\documentclass[3p,times,authoryear]{elsarticle}

\usepackage{ecrc}
\usepackage{epstopdf}
\usepackage{subfigure}
\usepackage{dsfont}
\usepackage{natbib}
\usepackage{amssymb}
\usepackage{amsthm}
\usepackage{amsmath}
\usepackage[utf8]{inputenc}
\usepackage{multirow}

\pdfmapfile{+txfonts.map}

\usepackage{xcolor}

\pdfmapfile{+txfonts.map}

\usepackage{epsfig}

\volume{00}

\firstpage{1}

\journalname{Physica A}

\runauth{Till Massing et al.}

\jid{procs}

\jnltitlelogo{Physica A}

\CopyrightLine{2021}{Published by Elsevier Ltd.}

\usepackage{amssymb}

\usepackage[figuresright]{rotating}

 \theoremstyle{remark}

\newcommand{\Levy}{L\'{e}vy }

\begin{document}

\begin{frontmatter}







\title{Student's $t$ mixture models for stock indices. A comparative study}

 \author[label1]{Till Massing\fnref{label4}}
  \author[label3]{Arturo Ramos\corref{cor1}}
 \address[label1]{Universit\"{a}t Duisburg-Essen,
 Fakult\"{a}t f\"{u}r Wirtschaftswissenschaften,
 Lehrstuhl für \"{O}konometrie, Essen, Germany}
\address[label3]{Departmento de An\'alisis Econ\'omico, Universidad de Zaragoza, Zaragoza, Spain}

\cortext[cor1]{Corresponding author\\
E-mail address: aramos@unizar.es}

\fntext[label4]{till.massing@uni-due.de (T. Massing)}

\begin{abstract}
We perform a comparative study for multiple equity indices of different countries using different models to determine the best fit using the Kolmogorov-Smirnov statistic, the Anderson-Darling statistic, the Akaike information criterion and the Bayesian information criteria as goodness-of-fit measures. We fit models both to daily and to hourly log-returns. The main result is the excellent performance of a mixture of three Student's $t$ distributions with the numbers of degrees of freedom fixed a priori (3St). In addition,
we find that the different components of the 3St mixture with small/moderate/high degree of freedom parameter describe the  extreme/moderate/small log-returns of the studied equity indices.
\end{abstract}


\end{frontmatter}

\section{Introduction}\label{sec:introappl}
This paper investigates which distribution gives the best fit for asset log-returns in a large class of parametric models. Asset price modeling goes back to \cite{bachelier1900theorie}, who proposed the normal distribution for log-returns. The distributional assumption is crucial, especially for option pricing. The famous \cite{black1973pricing} formula relies on the log-normality assumption. However, it is now well known that the normal distribution yields a poor fit for heavy-tailed returns. Several authors have proposed other more appropriate distributions. An important strand of the literature recommended the use of \Levy models (i.e., models with underlying infinitely divisible distributions \cite[]{Sato1999levy}). Among these, \cite{mandelbrot1961stable} recommended stable non-normal distributions. \cite{praetz1972distribution} suggested the Student's $t$ distribution because it allows a finite variance for a degree of freedom higher than two. Other suggestions include the variance gamma distribution \cite[]{madan1990variance}, the hyperbolic distribution \cite[]{eberlein1995hyperbolic}, the normal inverse Gaussian distribution \cite[]{barndorff1997normal}, the Meixner distribution \cite[]{schoutens2001meixner}, the generalized hyperbolic distribution \cite[]{eberlein2002generalized}, and the skew Student's $t$ distribution \cite[]{aas2006generalized}.
Another strand of the literature considered finite mixtures of distributions to model financial returns. As one of the first, \cite{kon1984models} proposed mixtures of normal distributions.
Let us mention that Student's $t$ distribution can be regarded as well as a weighted sum (integral) of normal distributions, as pointed out by \cite{praetz1972distribution}.
Empirical studies include \cite{peiro1994distribution} and \cite{behr2009alternatives}.

Since the distributional assumption is crucial for option pricing,
there is considerable interest in the choice of distribution. Many studies investigate this question. For example, \cite{gray1990empirical}, \cite{peiro1994distribution}, and \cite{aparicio2001empirical} compared different distributions for the daily log-returns of equity indices in different countries;
see \cite{corlu2016empirical}, \cite{goncu2016comparative} or \cite{Massing2019b} for more recent articles. \cite{corlu2015modelling} and \cite{nadarajah2015note} investigated foreign exchange rate returns. \cite{Chu2015} and \cite{ChaChuNadOst17,ZhaChuChaCha19} performed a similar analysis for cryptocurrencies. The results of the studies cited differ depending on the countries and time periods considered. The recent studies often favor the generalized hyperbolic, the normal inverse Gaussian or the variance gamma distribution.

This paper makes the following contributions. First, we analyze daily and hourly data for the multiple equity indices of different countries using different models to determine which yields the best fit, using the Kolmogorov-Smirnov statistic, the Anderson-Darling statistic, and the Akaike and  Bayesian information criteria as goodness-of-fit measures.
Second, we obtain a model which is excellent for describing the data at the body and above all, at the tails. This model is a finite mixture of three Student's $t$ distributions with numbers of degrees of freedom fixed a priori, so that some numerical issues of ML estimation have been avoided, namely numerical instability and the estimation algorithm may fail to converge. The models with two Student's $t$ distributions in the mixture also play a relevant role (see \cite{MasPueRam20}, where these mixtures have been introduced).
Let us note that Student's $t$ has been used previously on, for example, \cite{CasHamOuy10}.
A unified model for price return distributions in econophysics has been treated in \cite{BucJovSch11}.
Stock market return distributions have been treated, e.g., in \cite{GilToyKerKas00,DroForKwaOswRak07,SieHol08,SuaGom13,EomKaiSca19}.
Our proposed models are able to deal as well with the kurtosis of the data, as in \cite{LopGarGar12}.

The paper is organized as follows: Section~\ref{sec:themodels} introduces the different models and highlights some of their important properties and differences. Section~\ref{sec:Data} gives a brief overview of the data. Section~\ref{sec:Goodnessoffit} presents the goodness-of-fit results. Section~\ref{assessment3St} offers a further assessment of the 3-Student mixture. The last section concludes.

\section{The Models}\label{sec:themodels}

Throughout this paper, we use numerical maximum likelihood (ML) estimation unless there is a closed-form maximum likelihood estimator (MLE). We use the MLE because (under certain conditions) it enjoys asymptotic efficiency properties. Unreported results show that method-of-moments estimation generally performs much worse. For some models there may be more sophisticated estimation methods. However, we limit ourselves to ML estimation since, first, other methods are more involved and, second, so as to ensure comparability. Furthermore, numerical ML estimation works well in most settings and converges in a few minutes (or less).

We have also tried some other models as the stable distribution, the GH skew Student's $t$ distribution by \cite{aas2006generalized}, finite mixtures of normal distributions, and mixtures of logistic distributions. We omit the results because they have not been competitive compared to the distributions presented below.

\subsection{The Generalized Hyperbolic Model}\label{subsec:ghmodel}
The \emph{generalized hyperbolic (GH) distribution} $GH(\lambda,\alpha,\beta,\delta,\mu)$ is one of the most flexible distributions used to model asset returns. It contains many other models as special or limiting cases. Its density function is
\begin{equation*}
f_{GH}(x;\lambda,\alpha,\beta,\delta,\mu)=\frac{(\alpha^2-\beta^2)^{\lambda/2}K_{\lambda-1/2}\left(\alpha\sqrt{\delta^2+(x-\mu)^2}\right)\exp(\beta(x-\mu))}{\sqrt{2\pi}\alpha^{\lambda-1/2}\delta^{\lambda}K_{\lambda}\left(\delta\sqrt{\alpha^2-\beta^2}\right)\left(|\delta|+(x-\mu)^2\right)^{1/2-\lambda}},
\end{equation*}
with $K_{\nu}(x)$ the modified Bessel function of the second kind, shape parameters $\lambda\in\mathbb{R}$, $\alpha\in\mathbb{R}_0^+$, skewness parameter $\beta\in\mathbb{R}$, scale parameter $\delta$, and location parameter $\mu\in\mathbb{R}$ such that
\begin{align*}
\delta\ge0,\;0\le|\beta|<\alpha&\qquad\text{if } \lambda>0,\\
\delta>0,\;0\le|\beta|<\alpha&\qquad\text{if } \lambda=0,\\
\delta>0,\;0\le|\beta|\le\alpha&\qquad\text{if } \lambda<0.
\end{align*}
\cite{barndorff1977exponentially} introduced the GH distribution (as a model for sand movement) and \cite{Barndorff-Nielsen1977} proved its infinite divisibility. \cite{eberlein2002generalized} proposed using the GH distribution for asset price returns. \cite{eberlein2004generalized} discussed the special and limiting cases of the GH distribution that we introduce below. The GH distribution is fitted to data using numerical maximum likelihood estimation. Unfortunately, there is no closed-form maximum likelihood estimator and the likelihood function has a very complicated form and depends on five parameters. Thus, ML estimation may find only a local maximum. To address this issue, we use the numerical ML estimation algorithm of \cite{breymann2013ghyp} and our own Nelder-Mead-based approach for numerical maximization.

\subsubsection{The Normal Model}\label{subsubsec:normalmodel}
\cite{bachelier1900theorie} proposed Brownian motion as a model for log-returns. Although numerous authors have stressed that asset returns are too heavy tailed to be normal, the assumption features prominently in the frequently used \cite{black1973pricing} model for option pricing. Although there is overwhelming evidence against this model, we use it for the purposes of comparison. The normal distribution $N(\mu,\sigma^2)$ is the weak limit of the GH distribution $f_{N}(x;\mu,\sigma)=\lim f_{GH}(x;\lambda,\alpha,\beta,\delta,\mu)$ as $\alpha,\delta\to\infty$ and $\delta/\alpha\to\sigma^2$ for each $x\in \mathbb{R}$.

\subsubsection{The Student's $t$ Model}\label{subsec:studentmodel}
\cite{praetz1972distribution} and \cite{blattberg1974comparison} were among the first to propose the \emph{Student's $t$ distribution} $t(\nu,\mu,\sigma^2)$ for asset price returns. The Student's $t$ distribution has density function
\begin{equation*}
f_{St}(x;\mu,\sigma,\nu)=\frac{\Gamma\left(\frac{\nu+1}{2}\right)}{\Gamma\left(\frac{\nu}{2}\right)\sqrt{\pi\nu\sigma^2}}\left(1+\frac{1}{\nu}\left(\frac{x-\mu}{\sigma}\right)^2\right)^{-\frac{\nu+1}{2}},
\end{equation*}
with $\nu>0$ degrees of freedom, location parameter $\mu\in\mathbb{R}$, and scale parameter $\sigma>0$.
\cite{Cassidy20105736} and \cite{Cassidy20112794} used this for option pricing.
ML estimation for Student's $t$ random variables can be performed by using an expectation-conditional maximization either (ECME) algorithm \cite[]{liu1994ecme}.
The Student's $t$ distribution is the weak limit of the GH distribution $f_{St}(x;\mu,\sigma,\nu)=\lim_{\alpha,\beta\to0}f_{GH}(x;\lambda,\alpha,\beta,\delta,\mu)$ for each $x\in \mathbb{R}$.

\subsubsection{The Variance Gamma Model}\label{subsubsec:vgmodel}
\cite{madan1990variance} introduced the \emph{variance gamma distribution} $V\Gamma(\lambda,\alpha,\beta,\mu)$ to model market returns. It has density function
\begin{equation*}
f_{V\Gamma}(x;\lambda,\alpha,\beta,\mu)=\frac{(\alpha^2-\beta^2)^{\lambda}|x-\mu|^{\lambda-1/2}K_{\lambda-1/2}(\alpha|x-\mu|)\exp(\beta(x-\mu))}{\sqrt{\pi}\Gamma(\lambda)(2\alpha)^{\lambda-1/2}},
\end{equation*}
with $\alpha,\lambda>0$, $\beta\in\mathbb{R}$ such that $-\alpha<\beta<\alpha$, and $\mu\in \mathbb{R}$. It takes its name from the normal mean-variance mixture with a gamma distributed variable. The variance gamma distribution is the weak limit of the GH distribution $f_{V\Gamma}(x;\lambda,\alpha,\beta,\mu)=\lim_{\delta\downarrow0}f_{GH}(x;\lambda,\alpha,\beta,\delta,\mu)$ for each $x\in\mathbb{R}$.

\subsubsection{The Normal Inverse Gaussian Model}\label{subsubsec:nigmodel}
\cite{barndorff1977exponentially} introduced the \emph{normal inverse Gaussian (NIG) distribution} $NIG(\alpha,\beta,\delta,\mu)$. It has density function
\begin{equation*}
f_{NIG}(x;\alpha,\beta,\delta,\mu)=\frac{\alpha\delta K_1\left(\alpha\sqrt{\delta^2+(x-\mu)^2}\right)}{\pi\sqrt{\delta^2(x-\mu)^2}}\exp\left(\delta\sqrt{\alpha^2-\beta^2}+\beta(x-\mu)\right),
\end{equation*}
with $\alpha,\delta>0$ and $\beta,\mu\in\mathbb{R}$.
\cite{barndorff1997normal} used the NIG distribution in the context of asset returns. The NIG distribution is a special case of the GH distribution $NIG(\alpha,\beta,\delta,\mu)=GH(-1/2,\alpha,\beta,\delta,\mu)$.

\subsection{The Meixner Model}\label{subsec:meixnermodel}
The \emph{Meixner distribution} $M(\alpha,\beta,\mu,\delta)$ is not included in the GH family. \cite{schoutens1998levy,schoutens2001meixner} introduced the distribution for asset price returns as an alternative to the hyperbolic family. (It is named after Josef Meixner (1908--1994) to honor his work on so-called Meixner polynomials.) It has density function
\begin{equation*}
f_{M}(x;\alpha,\beta,\mu,\delta)=\frac{(2\cos(\beta/2))^{2\delta}}{2\alpha\pi\Gamma(2\delta)}\exp\left(\frac{\beta(x-\mu)}{\alpha}\right)
\left|\Gamma\left(\delta+\frac{\mathrm{i}(x-\mu)}{\alpha}\right)\right|,
\end{equation*}
with scale parameter $\alpha>0,$ shape parameter $\delta>0$, skewness parameter $-\pi<\beta<\pi$, and location parameter $\mu\in\mathbb{R}$.
The MLE can be found numerically using Newton methods since the derivative of the log-likelihood is explicitly available.

\subsection{Mixtures of Student's $t$}\label{subsec:mixtures}

The article of \cite{peiro1994distribution} suggested the use of mixtures of two normal distributions when studying log-returns. Here we will adapt this to mixtures of Student's $t$ distribution.
Mixtures of $m\geq 2$ non-standardized Student's $t$ distribution have the density
\begin{equation*}
f_{m{St}}(x;\mu_1,\sigma_1,\nu_1,
\ldots,\mu_m,\sigma_m,\nu_m,p_1,\ldots,p_{m-1})
=\sum_{j=1}^{m-1}p_jf_{St}(x;\mu_j,\sigma_j,\nu_j) +\left(1-\sum_{j=1}^{m-1}p_j\right)
f_{St}(x;\mu_m,\sigma_m,\nu_m)
\end{equation*}
where $\mu_1,\ldots,\mu_m\in\mathbb{R}$, $\sigma_1,\ldots,\sigma_m>0$, $\nu_1,\ldots,\nu_m>0$, and $0\le p_1,\ldots,p_{m-1},p_1+\dots+p_{m-1}\le1$. In this paper, we will consider 2-mixtures ($m=2$) and 3-mixtures ($m=3$).

For ML estimation, we use an expectation maximization algorithm proposed by \cite{Peel2000}.
There is an issue with the estimation of these mixtures because in their general form is numerically unstable. To make the estimation feasible we fix a priori the degrees of freedom $\nu_1,\nu_2$ for $m=2$ or $\nu_1,\nu_2,\nu_3$ for $m=3$. In particular, we estimate the remaining parameters for the three scenarios \citep{MasPueRam20}
\begin{eqnarray}
&&f_{2{St}}(x;\mu_1,\sigma_1,4,\mu_2,\sigma_2,12,p_1),\nonumber\\
&&f_{2{St}}(x;\mu_1,\sigma_1,4,\mu_2,\sigma_2,39,p_1),\nonumber\\
&&f_{3{St}}(x;\mu_1,\sigma_1,4,\mu_2,\sigma_2,12,\mu_3,\sigma_3,39,p_1,p_2)\nonumber
\end{eqnarray}
and call the distributions for brevity 2St12, 2St39 and 3St, respectively.
The intuition behind this is, e.g., for the 3-mixture, that a log-return $x$ is drawn from a Student's $t$ distribution with a small degree of freedom with \emph{posterior probability} (see, e.g., \cite{McLPee00}) $\tau_1(x)$ given by
$$
\tau _{1}(x)=p_{1}f_{{St}}(x;\mu _{1},\sigma _{1},4)/f_{{3St}}(x;\mu _{1},\sigma _{1},4,\mu _{2},\sigma _{2},12,\mu _{3},\sigma _{3},39,p_{1},p_{2})\,,
$$
from a distribution with moderately high degree of freedom with posterior probability
$$
\tau _{2}(x)=p_{2}f_{{St}}(x;\mu _{2},\sigma _{2},12)/f_{{3St}}(x;\mu _{1},\sigma _{1},4,\mu _{2},\sigma _{2},12,\mu _{3},\sigma _{3},39,p_{1},p_{2})\,,
$$
and with a high degree of freedom with posterior probability
$$
\tau _{3}(x)=(1-p_{1}-p_{2})f_{{St}}(x;\mu _{3},\sigma _{3},39)/f_{{3St}}(x;\mu _{1},\sigma _{1},4,\mu _{2},\sigma _{2},12,\mu _{3},\sigma _{3},39,p_{1},p_{2})\,.
$$
In Section~\ref{assessment3St}, we discuss which component of the 3St describes which part of the data best.

Of course, the particular values of the $\nu$'s are arbitrary and other choices can yield to estimates with higher log-likelihoods. However, some decision has to be taken (in order to avoid numerical problems when estimating $\nu$'s) and we have experienced good results with these choices compared to other choices.

\section{Data}\label{sec:Data}
This section gives a brief overview of the data we use in the following study. Data were provided by the \emph{Thomson Reuters Eikon} database. We consider the 78 equity indices from 70 countries for which hourly data are available. We observe daily closing prices from 01/02/1997 until 11/02/2017 (or shorter for some countries depending on availability). We compute daily log-returns for trading days. We observe hourly closing prices from 11/02/2016 12pm until 11/02/2017 12pm and compute hourly log-returns for trading hours. In other words, for the goodness-of-fit analysis of daily log-returns we can use a long sample of almost 20 years, whereas hourly data is restricted to one year. We compute all statistics for daily returns in this section for the full period.

Table \ref{tab:Deskd} reports on the countries and indices considered, the number of daily returns for the long period, and empirical mean, standard deviation, skewness, kurtosis, and minimal and maximal values. The index returns typically have a mean close to and usually larger than zero. There is some skewness in the data. The empirical kurtosis is greater than three, indicating heavy tails.
We list the countries alphabetically by country name.

\begin{table}[htbp]
  \hspace{-2.5cm}
\begin{tiny}
    \begin{tabular}{rrrrrrrrr}
        \hline
    Countries & Index & $n$ & Mean  & Sd    & Skewness & Kurtosis & Min   & Max \\
    \hline
       Argentina & MERVAL & 5121  & 0.000738 & 0.02168 & -0.30154 & 7.803876 & -0.14765 & 0.161165 \\
    Australia & S\&P/ASX 200 & 5269  & 0.000175 & 0.009857 & -0.46959 & 8.667021 & -0.08704 & 0.057244 \\
    Australia & All Ordinaries & 5268  & 0.000175 & 0.009582 & -0.54831 & 8.996051 & -0.08554 & 0.057361 \\
    Austria & ATX   & 5156  & 0.000215 & 0.014157 & -0.40108 & 9.821463 & -0.10253 & 0.12021 \\
    Bahrain & All Share & 3643  & 5.7E-05 & 0.005634 & -0.38586 & 9.328441 & -0.0492 & 0.036132 \\
    Belgium & BEL 20 & 5313  & 0.000148 & 0.012398 & -0.02947 & 8.577035 & -0.08319 & 0.09334 \\
    Brazil & Bovespa & 5158  & 0.000458 & 0.020565 & 0.287261 & 16.24525 & -0.17208 & 0.288325 \\
    Bulgaria & SOFIX & 4188  & 0.000453 & 0.015194 & -0.60166 & 37.19798 & -0.20899 & 0.210733 \\
    Canada & S\&P/TSX 60 & 4727  & 0.000195 & 0.011821 & -0.61546 & 12.44824 & -0.10327 & 0.09826 \\
    Canada & S\&P/TSX Composite & 5235  & 0.000191 & 0.010994 & -0.68304 & 12.09018 & -0.09788 & 0.093703 \\
    Chile & IPSA  & 5191  & 0.000351 & 0.01058 & 0.115497 & 11.47017 & -0.07658 & 0.118034 \\
    China & CSI 300 & 3056  & 0.000452 & 0.017978 & -0.53981 & 6.694908 & -0.09695 & 0.08931 \\
    China & SSE Composite Index & 5045  & 0.000258 & 0.016322 & -0.40072 & 7.849175 & -0.09335 & 0.09401 \\
    Colombia & IGBC  & 3967  & 0.000599 & 0.012723 & -0.17429 & 15.78783 & -0.11052 & 0.14688 \\
    Croatia & CROBEX & 4844  & 0.000135 & 0.014485 & 0.198162 & 19.14509 & -0.11092 & 0.174715 \\
    Cyprus & Cyprus Main Market Index & 3219  & -0.00095 & 0.026562 & 0.033523 & 9.42208 & -0.16704 & 0.174871 \\
    Czech Republic & PX    & 5209  & 0.00013 & 0.013762 & -0.4608 & 14.72246 & -0.16185 & 0.123641 \\
    Egypt & EGX 30 Index & 4838  & 0.00055 & 0.017055 & -0.32224 & 11.68218 & -0.17992 & 0.183692 \\
    Estonia & OMXT  & 5092  & 0.000396 & 0.015025 & -1.08837 & 28.48488 & -0.21577 & 0.128667 \\
    EuroStoxx & Euro Stoxx 50 & 5334  & 0.000104 & 0.013153 & -0.05247 & 8.441807 & -0.09001 & 0.102188 \\
    Finland & OMXH25 & 5229  & 0.000262 & 0.018097 & -0.35836 & 10.29514 & -0.17425 & 0.145631 \\
    France & CAC 40 & 5313  & 0.000168 & 0.014584 & -0.05873 & 7.475534 & -0.09472 & 0.105946 \\
    GB    & FTSE 100 & 5262  & 0.000116 & 0.011904 & -0.15033 & 8.546953 & -0.09266 & 0.093843 \\
    Germany & DAX   & 5288  & 0.000295 & 0.015216 & -0.09171 & 6.96101 & -0.08875 & 0.107975 \\
    Greece & Athex Composite Share Price Index & 5172  & -4.2E-05 & 0.019529 & -0.28642 & 8.330497 & -0.17713 & 0.134311 \\
    Hong Kong & Hang Seng & 5138  & 0.00015 & 0.016496 & 0.095766 & 13.14455 & -0.14735 & 0.172471 \\
    Hungary & Budapest SE & 5202  & 0.00043 & 0.01704 & -0.604 & 14.09001 & -0.18033 & 0.136157 \\
    India & Nifty 50 & 5180  & 0.000465 & 0.015348 & -0.21529 & 10.68258 & -0.13054 & 0.163343 \\
    India & BSE Sensex & 5180  & 0.00045 & 0.015367 & -0.16642 & 9.371377 & -0.11809 & 0.1599 \\
    Indonesia & IDX Composite & 5078  & 0.000442 & 0.015841 & -0.20109 & 11.08056 & -0.12732 & 0.131277 \\
    Ireland & ISEQ Overall Index & 5262  & 0.000179 & 0.013515 & -0.67045 & 11.15808 & -0.13964 & 0.097331 \\
    Israel & TA 35 & 5099  & 0.000366 & 0.012351 & -0.28931 & 7.344565 & -0.09884 & 0.092257 \\
    Italy & FTSE MIB & 5034  & -1.1E-05 & 0.015674 & -0.19843 & 7.493357 & -0.13331 & 0.108742 \\
    Japan & Topix & 5117  & 3.75E-05 & 0.013872 & -0.29215 & 8.500356 & -0.10007 & 0.128646 \\
    Kazakhstan & KASE Index & 4131  & 0.000735 & 0.026745 & 0.618863 & 67.29964 & -0.48644 & 0.487587 \\
    Kuwait & Kuwait 15 & 1352  & -2.4E-05 & 0.007646 & -0.02269 & 7.866281 & -0.04992 & 0.050591 \\
    Latvia & OMXR  & 4411  & 0.000518 & 0.014196 & -0.38008 & 19.73704 & -0.14705 & 0.115963 \\
    Lithuania & OMXV  & 4396  & 0.000428 & 0.010224 & -0.51677 & 24.18546 & -0.11938 & 0.110015 \\
    Luxembourg & LuxX Index & 4745  & 0.000104 & 0.012964 & -0.34199 & 9.525523 & -0.11159 & 0.091043 \\
    Malaysia & FTSE Bursa Malaysia KLCI & 5128  & 6.77E-05 & 0.013013 & 0.506517 & 65.72185 & -0.24153 & 0.208174 \\
    Mauritius & SEMDEX & 5097  & 0.000358 & 0.006224 & 0.336542 & 26.86229 & -0.06383 & 0.076546 \\
    Mexico & IPC   & 5240  & 0.000509 & 0.014181 & 0.01753 & 11.0234 & -0.14314 & 0.121536 \\
    Morocco & MASI  & 3950  & 0.000307 & 0.007633 & -0.42954 & 9.723467 & -0.06817 & 0.044635 \\
    Namibia & NSX Overall Index & 3773  & 0.000306 & 0.015552 & -0.41815 & 8.087981 & -0.14833 & 0.086961 \\
    Netherlands & AEX   & 5317  & 0.000123 & 0.014472 & -0.14407 & 8.600619 & -0.0959 & 0.100283 \\
    New Zealand & NZX 50 Index & 4232  & 0.000372 & 0.006938 & -0.5181 & 8.47095 & -0.05247 & 0.058146 \\
    Norway & OBX Index & 4555  & 0.000382 & 0.015299 & -0.53623 & 9.665639 & -0.11273 & 0.110198 \\
    Oman  & MSM 30 & 5083  & 0.000182 & 0.013809 & 0.554794 & 444.2203 & -0.43979 & 0.454222 \\
    Pakistan & KSE 100 Index & 5093  & 0.000667 & 0.01513 & -0.35097 & 9.260944 & -0.13213 & 0.127622 \\
    Peru  & S\&P Lima General Index & 5201  & 0.000506 & 0.013581 & -0.42559 & 14.02426 & -0.13291 & 0.128156 \\
    Philippines & PSEi Composite & 5119  & 0.000194 & 0.014254 & 0.184929 & 14.29941 & -0.13089 & 0.161776 \\
    Poland & WIG   & 5218  & 0.000287 & 0.013732 & -0.3973 & 7.196922 & -0.10286 & 0.078933 \\
    Portugal & PSI 20 & 5286  & 1.1E-05 & 0.012324 & -0.3426 & 9.10618 & -0.10379 & 0.101959 \\
    Qatar & QE 20 Index & 4877  & 0.000373 & 0.024117 & -0.55297 & 642.7196 & -0.85805 & 0.844237 \\
    Romania & BET 10 & 5024  & 0.000411 & 0.016471 & -0.35891 & 10.74084 & -0.13117 & 0.105645 \\
    Russia & MICEX & 5010  & 0.000605 & 0.025959 & 0.122738 & 19.34431 & -0.23336 & 0.275005 \\
    Russia & RTSI  & 3154  & -8.8E-05 & 0.019877 & -0.70406 & 32.3405 & -0.25957 & 0.221082 \\
    Saudi Arabia & Tadawul All Share & 5061  & 0.000302 & 0.014101 & -0.88529 & 13.5072 & -0.10328 & 0.093907 \\
    Serbia & BELEX & 3042  & -0.0001 & 0.012613 & 0.131727 & 18.90383 & -0.10861 & 0.121576 \\
    Singapore & STI Index & 4563  & 9.56E-05 & 0.011502 & -0.26366 & 8.375257 & -0.08696 & 0.075305 \\
    South Africa & FTSE/JSE All-Share Index & 5203  & 0.000438 & 0.012382 & -0.45179 & 8.874737 & -0.12626 & 0.07268 \\
    South Korea & KOSPI & 5236  & 0.00026 & 0.017339 & -0.3124 & 7.808938 & -0.12805 & 0.112844 \\
    Spain & IBEX 35 & 5272  & 0.00014 & 0.015083 & -0.13652 & 8.304619 & -0.13185 & 0.134836 \\
    Sri Lanka & CSE All-Share & 4969  & 0.000481 & 0.011142 & 0.259282 & 35.48838 & -0.13893 & 0.182881 \\
    Sweden & OMXS30 & 5230  & 0.000245 & 0.015189 & 0.041153 & 6.746038 & -0.088 & 0.110228 \\
    Switzerland & SMI   & 5241  & 0.000163 & 0.012148 & -0.19088 & 8.913466 & -0.0907 & 0.107876 \\
    Taiwan & Taiwan Weighted & 5261  & 8.72E-05 & 0.013925 & -0.16951 & 5.680875 & -0.06912 & 0.065246 \\
    Thailand & SET   & 5094  & 0.000147 & 0.015608 & 0.052839 & 10.97588 & -0.16063 & 0.113495 \\
    Tunesia & Tunindex & 4878  & 0.000371 & 0.007808 & -0.41858 & 561.8742 & -0.26694 & 0.265408 \\
    Turkey & BIST 100 & 5206  & 0.000911 & 0.023936 & -0.04054 & 9.59741 & -0.19979 & 0.177736 \\
    United Arab Emirates & DFM   & 3587  & 0.000359 & 0.017485 & -0.03534 & 9.110389 & -0.12157 & 0.122045 \\
    United Arab Emirates & Abu Dhabi & 4236  & 0.000349 & 0.011008 & -0.08549 & 11.5268 & -0.08679 & 0.076295 \\
    USA   & Dow 30 & 5241  & 0.000246 & 0.011443 & -0.15218 & 10.8003 & -0.08201 & 0.105083 \\
    USA   & S\&P 500 & 5241  & 0.000239 & 0.012151 & -0.23601 & 10.94939 & -0.0947 & 0.109572 \\
    USA   & Nasdaq & 5241  & 0.000389 & 0.018276 & 0.095177 & 8.848534 & -0.11115 & 0.17203 \\
    Venezuela & IBC   & 4768  & 0.002418 & 0.02006 & 1.087122 & 19.02831 & -0.20658 & 0.200618 \\
    Vietnam & HNX 30 & 2967  & 1.13E-05 & 0.019851 & 0.010562 & 7.257007 & -0.12862 & 0.09731 \\
    Zambia & All Share Index & 3919  & 0.001005 & 0.020664 & 1.219612 & 36.14717 & -0.2086 & 0.311243 \\
    \hline
    \end{tabular}%
				\end{tiny}
		  \caption{Countries with equity indices, number of nonzero daily log-returns from 01/03/1997 until 11/02/2017 (if available), mean, standard deviation, skewness, kurtosis, minimal and maximal log-return.}
  \label{tab:Deskd}%
\end{table}%

\begin{table}[htbp]
  \hspace{-2.5cm}
\begin{tiny}
    \begin{tabular}{rrrrrrrrr}
    \hline
    Countries & Index & $n$ & Mean  & Sd    & Skewness & Kurtosis & Min   & Max \\
    \hline
       Argentina & MERVAL & 1735  & 0.000303 & 0.004495 & -0.5839 & 18.70735 & -0.04464 & 0.032523 \\
    Australia & S\&P/ASX 200 & 1800  & 7.06E-05 & 0.002452 & 0.438839 & 21.35383 & -0.01809 & 0.029653 \\
    Australia & All Ordinaries & 1814  & 6.82E-05 & 0.002313 & 0.483321 & 23.6521 & -0.01767 & 0.029105 \\
    Austria & ATX   & 2242  & 0.000154 & 0.002566 & 0.238824 & 9.537616 & -0.01816 & 0.0191 \\
    Bahrain & All Share & 885   & 0.000128 & 0.002729 & 0.787911 & 36.21098 & -0.02807 & 0.031461 \\
    Belgium & BEL 20 & 2309  & 7.37E-05 & 0.002231 & 0.871004 & 20.22314 & -0.01262 & 0.028891 \\
    Brazil & Bovespa & 2077  & 7.46E-05 & 0.004382 & -3.99318 & 91.69129 & -0.09006 & 0.025154 \\
    Bulgaria & SOFIX & 1931  & 0.00012 & 0.002928 & 1.033648 & 28.36035 & -0.02875 & 0.036838 \\
    Canada & S\&P/TSX 60 & 1995  & 4.54E-05 & 0.001939 & -0.23815 & 9.073434 & -0.01321 & 0.011302 \\
    Canada & S\&P/TSX Composite & 2007  & 4.19E-05 & 0.001786 & -0.24981 & 9.370868 & -0.01242 & 0.010158 \\
    Chile & IPSA  & 2070  & 0.000129 & 0.002808 & 0.115631 & 300.5028 & -0.06562 & 0.066762 \\
    China & CSI 300 & 1709  & 0.000106 & 0.00249 & 0.121935 & 7.705952 & -0.01202 & 0.017058 \\
    China & SSE Composite Index & 1711  & 5.06E-05 & 0.002335 & -0.40249 & 8.522975 & -0.01567 & 0.011719 \\
    Colombia & IGBC  & 1837  & 3.11E-05 & 0.002074 & -0.18905 & 8.000927 & -0.01336 & 0.011554 \\
    Croatia & CROBEX & 2000  & -2.2E-05 & 0.002684 & -2.56038 & 50.92942 & -0.04391 & 0.019164 \\
    Cyprus & Cyprus Main Market Index & 1683  & 2.8E-05 & 0.007103 & -0.19354 & 7.770919 & -0.04725 & 0.039425 \\
    Czech Republic & PX    & 2008  & 8.42E-05 & 0.001962 & -0.17722 & 12.05851 & -0.0195 & 0.013509 \\
    Egypt & EGX 30 Index & 1200  & 0.000432 & 0.005 & 2.344607 & 20.73532 & -0.02916 & 0.045011 \\
    Estonia & OMXT  & 1629  & 0.000105 & 0.002358 & 0.486173 & 13.90778 & -0.01965 & 0.022084 \\
    EuroStoxx & Euro Stoxx 50 & 2320  & 6.63E-05 & 0.00207 & 0.564016 & 12.49036 & -0.01347 & 0.01768 \\
    Finland & OMXH25 & 2275  & 7.55E-05 & 0.002163 & 0.120082 & 12.57391 & -0.01588 & 0.015789 \\
    France & CAC 40 & 2311  & 9.37E-05 & 0.002505 & 1.785311 & 32.81083 & -0.01374 & 0.039058 \\
    GB    & FTSE 100 & 2269  & 3.73E-05 & 0.002032 & 0.177792 & 10.6633 & -0.01259 & 0.016327 \\
    Germany & DAX   & 2283  & 0.000111 & 0.002358 & 1.03419 & 16.45285 & -0.01467 & 0.025565 \\
    Greece & Athex Composite Share Price Index & 2019  & 0.000135 & 0.003812 & -0.05364 & 8.077142 & -0.02328 & 0.02532 \\
    Hong Kong & Hang Seng & 1927  & 0.000115 & 0.002497 & -0.17322 & 12.15748 & -0.0173 & 0.020471 \\
    Hungary & Budapest SE & 2279  & 0.000131 & 0.002347 & 0.223531 & 8.18847 & -0.01303 & 0.018177 \\
    India & Nifty 50 & 1734  & 0.000117 & 0.002343 & -0.48607 & 15.28521 & -0.02381 & 0.015614 \\
    India & BSE Sensex & 1738  & 0.000114 & 0.002309 & -0.16957 & 13.84181 & -0.02238 & 0.016214 \\
    Indonesia & IDX Composite & 1819  & 6.16E-05 & 0.002252 & -1.13553 & 45.12261 & -0.02988 & 0.023499 \\
    Ireland & ISEQ Overall Index & 2434  & 6.97E-05 & 0.002368 & 0.737489 & 12.47689 & -0.01401 & 0.021423 \\
    Israel & TA 35 & 2127  & 1.6E-05 & 0.001809 & 0.484082 & 11.65909 & -0.01246 & 0.013599 \\
    Italy & FTSE MIB & 2304  & 0.000142 & 0.003202 & 0.571569 & 12.44307 & -0.02082 & 0.033496 \\
    Japan & Topix & 1723  & 0.000168 & 0.003038 & 2.46633 & 52.58563 & -0.02219 & 0.049794 \\
    Kazakhstan & KASE Index & 1673  & 0.000289 & 0.003738 & 0.337259 & 7.405869 & -0.02292 & 0.024228 \\
    Kuwait & Kuwait 15 & 992   & 0.000152 & 0.004233 & 0.572548 & 6.444483 & -0.01426 & 0.025138 \\
    Latvia & OMXR  & 1421  & 0.00025 & 0.003896 & 6.959837 & 142.2951 & -0.01941 & 0.081791 \\
    Lithuania & OMXV  & 1580  & 0.000111 & 0.020067 & -0.20905 & 778.6937 & -0.56311 & 0.560521 \\
    Luxembourg & LuxX Index & 2270  & 1.79E-05 & 0.003866 & -0.37488 & 12.02111 & -0.03 & 0.026192 \\
    Malaysia & FTSE Bursa Malaysia KLCI & 1694  & 2.88E-05 & 0.001332 & -0.5679 & 13.66364 & -0.01171 & 0.00798 \\
    Mauritius & SEMDEX & 998   & 0.0002 & 0.001327 & 0.468473 & 11.66196 & -0.00826 & 0.008729 \\
    Mexico & IPC   & 1879  & 1.2E-05 & 0.002656 & -0.14613 & 11.91045 & -0.02062 & 0.021735 \\
    Morocco & MASI  & 1738  & 9.34E-05 & 0.002759 & 2.205793 & 36.822 & -0.01319 & 0.042122 \\
    Namibia & NSX Overall Index & 2221  & 5.16E-05 & 0.004033 & -0.30366 & 13.49559 & -0.03105 & 0.030742 \\
    Netherlands & AEX   & 2293  & 9.52E-05 & 0.002167 & 0.32895 & 14.67397 & -0.01479 & 0.02102 \\
    New Zealand & NZX 50 Index & 2107  & 7.83E-05 & 0.001662 & -0.47359 & 61.10236 & -0.02405 & 0.026364 \\
    Norway & OBX Index & 2019  & 0.000138 & 0.002465 & -0.27087 & 9.421002 & -0.01672 & 0.014571 \\
    Oman  & MSM 30 & 916   & -9E-05 & 0.001858 & 0.045612 & 7.814568 & -0.01033 & 0.011824 \\
    Pakistan & KSE 100 Index & 1727  & -1.7E-05 & 0.003888 & -0.52699 & 10.26866 & -0.03324 & 0.020818 \\
    Peru  & S\&P Lima General Index & 2043  & 0.000129 & 0.001938 & 0.163475 & 7.402088 & -0.01301 & 0.011891 \\
    Philippines & PSEi Composite & 1706  & 9.81E-05 & 0.003185 & 0.353588 & 9.700206 & -0.02127 & 0.020625 \\
    Poland & WIG   & 2259  & 0.000128 & 0.002353 & 0.395163 & 8.85866 & -0.01433 & 0.018426 \\
    Portugal & PSI 20 & 2311  & 7.49E-05 & 0.002464 & 0.112628 & 17.27194 & -0.02522 & 0.022203 \\
    Qatar & QE 20 Index & 1240  & -0.00017 & 0.004219 & -5.37658 & 108.2169 & -0.07975 & 0.026686 \\
    Romania & BET 10 & 2496  & 5.62E-05 & 0.001996 & -0.53437 & 19.27596 & -0.0186 & 0.020123 \\
    Russia & MICEX & 2277  & 1.84E-05 & 0.002748 & -0.37679 & 7.273298 & -0.02085 & 0.013887 \\
    Russia & RTSI  & 2223  & 8.47E-05 & 0.003753 & -0.58974 & 15.16277 & -0.04335 & 0.023125 \\
    Saudi Arabia & Tadawul All Share & 1303  & 0.000115 & 0.003409 & 0.753284 & 13.04813 & -0.02712 & 0.026833 \\
    Serbia & BELEX & 1409  & 6.44E-05 & 0.003369 & 0.066989 & 5.635979 & -0.01338 & 0.016387 \\
    Singapore & STI Index & 2249  & 8.35E-05 & 0.001801 & -0.14562 & 10.6594 & -0.01097 & 0.01229 \\
    South Africa & FTSE/JSE All-Share Index & 2230  & 7.4E-05 & 0.00235 & 0.39339 & 8.575543 & -0.01273 & 0.015363 \\
    South Korea & KOSPI & 1711  & 0.000147 & 0.00219 & 0.186581 & 14.7022 & -0.01573 & 0.018311 \\
    Spain & IBEX 35 & 2564  & 6.32E-05 & 0.00287 & -0.14603 & 28.57183 & -0.04008 & 0.031577 \\
    Sri Lanka & CSE All-Share & 1435  & 2.05E-05 & 0.001214 & 0.553361 & 6.271024 & -0.00517 & 0.007002 \\
    Sweden & OMXS30 & 2259  & 7.09E-05 & 0.002207 & 0.199227 & 11.2113 & -0.01338 & 0.016976 \\
    Switzerland & SMI   & 2276  & 7.89E-05 & 0.002023 & 0.618643 & 12.47614 & -0.0123 & 0.018293 \\
    Taiwan & Taiwan Weighted & 1233  & 0.000137 & 0.002502 & 0.324691 & 15.90517 & -0.0203 & 0.023795 \\
    Thailand & SET   & 1699  & 7.46E-05 & 0.00178 & 0.003001 & 6.706329 & -0.00847 & 0.009746 \\
    Tunesia & Tunindex & 1203  & 8.53E-05 & 0.001687 & 0.119257 & 4.40262 & -0.00684 & 0.006901 \\
    Turkey & BIST 100 & 2526  & 0.000154 & 0.002684 & -1.29678 & 21.84384 & -0.0373 & 0.017405 \\
    United Arab Emirates & DFM   & 1003  & 9.55E-05 & 0.003599 & 0.104579 & 8.433005 & -0.02659 & 0.020223 \\
    United Arab Emirates & Abu Dhabi & 1255  & 3.66E-05 & 0.003243 & -0.73731 & 15.2062 & -0.03358 & 0.021359 \\
    USA   & Dow 30 & 2008  & 0.000131 & 0.001537 & 1.094688 & 14.24498 & -0.01052 & 0.014671 \\
    USA   & S\&P 500 & 2000  & 0.0001 & 0.001592 & 0.640582 & 13.48703 & -0.01023 & 0.015537 \\
    USA   & Nasdaq & 2009  & 0.000135 & 0.002336 & -0.36333 & 16.4007 & -0.02325 & 0.019281 \\
    Venezuela & IBC   & 890   & 0.004317 & 0.017545 & 0.712966 & 8.708001 & -0.08353 & 0.107082 \\
    Vietnam & HNX 30 & 1264  & 0.000197 & 0.003173 & 0.126409 & 4.49775 & -0.0115 & 0.014023 \\
    Zambia & All Share Index & 470   & 0.00033 & 0.025373 & 0.032765 & 81.63779 & -0.29734 & 0.298635 \\
    \hline
    \end{tabular}%
				\end{tiny}
		  \caption{Countries with equity indices, number of nonzero hourly log-returns from 11/03/2016 1pm until 11/02/2017 12pm, mean, standard deviation, skewness, kurtosis, minimal and maximal log-return.}
  \label{tab:Deskh}%
\end{table}%

\section{Goodness-of-Fit}\label{sec:Goodnessoffit}

The aim of this section is to discover which model is the best fit for daily and intraday returns. We compare the models $B\in\{N,St,NIG,V\Gamma,GH,M,2St12,2St39,3St\}$ presented in Section \ref{sec:themodels}. We fit the distributions to the log-returns by (possibly numerically) maximizing the log-likelihood $\ell_B(\theta)=\sum_{i=1}^n\log f_{B}(x_i;\theta)$, where $\theta=(\theta_1,\ldots,\theta_k)$ is the vector of parameters. For the sake of brevity, we offer the ML estimates of all parameters for all models and samples upon request.
Note that $k$ depends on the specific model $B$. Let $\hat{\theta}$ denote the ML estimate given the log-returns $x_1,\ldots,x_n$.

We use four different measures of goodness-of-fit.

\begin{itemize}
\item The \emph{Kolmogorov-Smirnov (KS) statistic} \cite[]{KolmogorovAN1933} compares the empirical distribution function
$F_n(x)=\frac{1}{n}\sum_{i=1}^n\mathds{1}_{(-\infty,x]}(x_i)$
with the distribution function of the fitted model $F_{B}(x;\hat{\theta})$ and is given by the maximal deviance
\begin{equation*}
KS=\sup_{x\in\mathbb{R}}|F_n(x)-F_B(x;\hat{\theta})|.
\end{equation*}
\item The \emph{Anderson-Darling (AD) statistic} \cite[]{anderson1954test} is defined by
\begin{align*}
AD&=n\int_{-\infty}^{\infty}\frac{(F_n(x)-F_{B}(x;\hat{\theta}))^2}{F_{B}(x;\hat{\theta})(1-F_{B}(x;\hat{\theta}))}\mathrm{d}F_{B}(x;\hat{\theta})\\
&=-n-\sum_{i=1}^n\frac{2i-1}{n}\left(\log F_{B}(x_{(i)};\hat{\theta})+\log\left(1-F_{B}(x_{(n+1-i)};\hat{\theta})\right)\right),
\end{align*}
where $x_{(1)}\leq\ldots\leq x_{(n)}$ are the observed ordered data.
\item The \emph{Akaike information criterion (AIC)} \cite[]{Aka74} is defined by
\begin{equation*}
AIC=2k-2\ell_B(\hat{\theta}).
\end{equation*}
\item The \emph{Bayesian information criterion (BIC)} \cite[]{schwarz1978estimating} is defined by
\begin{equation*}
BIC=k\log(n)-2\ell_B(\hat{\theta}).
\end{equation*}
\end{itemize}
For each criterion, the model with the smallest statistic is considered to be the best fit among the models investigated.
The KS statistic better reflects the deviance between the empirical and the fitted distribution close to the center and the AD statistic better reflects the deviance in the tails \cite[]{razali2011power}. Generally, the KS and AD statistics do not account for overfitting. However, it is possible that a special or limiting case may have a lower distance due to the nature of the distances and the distributions. The AIC and BIC statistics adjust the log-likelihood by penalizing models that are too large to avoid overfitting, more heavily the BIC than the AIC.

Tables \ref{tab:dKS}--\ref{tab:dBIC} report KS, AD, AIC and BIC for daily log-returns for the last 20 years. For each country we compare all models presented here. The minimal statistic is in bold to indicate the best fit. The different criteria do not always lead to the same conclusion. This is reasonable, as they focus on different features. In terms of KS (Table \ref{tab:dKS}) and AD (Table \ref{tab:dAD}) statistics, the 3St model is clearly the most selected one, pointing out to a really good fit.

The AIC information criterion accounts for overfitting of the models, but not so heavily as the BIC does. There are models that are selected very often by AIC, namely, the GH, the 2St12 and the 3St.
The BIC favors small models, making it attractive since models with many parameters can lead to overfitting. The BIC for daily returns in the long period is minimal a number of times
for the Student's $t$, the NIG and the 2St12.

Table~\ref{tab:minsdaily} summarizes how many times each model is selected to each statistical criterion, according to
Tables~\ref{tab:dKS}--\ref{tab:dBIC}. We see that the model that is more often selected according to three criteria is clearly the 3St. Only for the BIC criterion the 3St is not the most often selected model, and then the Student's $t$, the NIG and the 2St12 are outstanding choices.

\begin{table}[htbp]
  \hspace{-2.5cm}
\begin{tiny}
\begin{tabular}{rrrrrrrrrrr}
    \hline
    Country & Index & N     & St    & NIG  & V$\Gamma$  & GH    & Meix & 2St12 & 2St39 & 3St \\
    \hline
     Argentina  &  MERVAL  &       0.0718    &       0.0114    &       0.0052    &       0.0111    & \textbf{      0.0049   } &       0.0055    &          0.0055    &          0.0054    &             0.0049    \\
     Australia  &  S\&P/ASX 200  &       0.0589    &       0.0125    &       0.0069    &       0.0090    &       0.0089    &       0.0079    &          0.0111    &          0.0112    & \textbf{            0.0053   } \\
     Australia  &  All Ordinaries  &       0.0611    &       0.0130    &       0.0077    &       0.0090    &       0.0099    &       0.0083    &          0.0132    &          0.0142    & \textbf{            0.0065   } \\
     Austria  &  ATX  &       0.0813    &       0.0154    &       0.0090    &       0.0181    &       0.0075    &       0.0115    &          0.0078    &          0.0091    & \textbf{            0.0072   } \\
     Bahrain  &  All Share  &       0.0835    &       0.0162    &       0.0111    &       0.0103    &       0.0103    &       0.0101    & \textbf{         0.0057   } &          0.0057    &             0.0059    \\
     Belgium  &  BEL 20  &       0.0731    &       0.0168    &       0.0072    &       0.0123    &       0.0071    &       0.0077    &          0.0071    &          0.0072    & \textbf{            0.0057   } \\
     Brazil  &  Bovespa  &       0.0599    &       0.0100    &       0.0090    &       0.0130    & \textbf{      0.0065   } &       0.0106    &          0.0079    &          0.0093    &             0.0076    \\
     Bulgaria  &  SOFIX  &       0.1393    &       0.0092    &       0.0095    &       0.0302    &       0.0076    &       0.0113    &          0.0076    &          0.0080    & \textbf{            0.0071   } \\
     Canada  &  S\&P/TSX 60  &       0.0776    &       0.0155    &       0.0084    &       0.0143    &       0.0090    &       0.0100    &          0.0091    &          0.0100    & \textbf{            0.0080   } \\
     Canada  &  S\&P/TSX Composite  &       0.0809    &       0.0163    &       0.0098    &       0.0165    &       0.0075    &       0.0111    &          0.0056    &          0.0061    & \textbf{            0.0055   } \\
     Chile  &  IPSA  &       0.0610    &       0.0084    &       0.0072    &       0.0115    &       0.0081    &       0.0079    &          0.0098    &          0.0073    & \textbf{            0.0072   } \\
     China  &  CSI 300  &       0.0867    &       0.0281    &       0.0182    &       0.0167    &       0.0158    &       0.0191    &          0.0115    &          0.0117    & \textbf{            0.0098   } \\
     China  &  SSE Composite Index  &       0.0842    &       0.0185    &       0.0115    &       0.0144    &       0.0095    &       0.0106    &          0.0077    &          0.0075    & \textbf{            0.0074   } \\
     Colombia  &  IGBC  &       0.0830    &       0.0137    &       0.0098    &       0.0142    &       0.0130    &       0.0099    &          0.0143    &          0.0098    & \textbf{            0.0084   } \\
     Croatia  &  CROBEX  &       0.1226    &       0.0095    &       0.0096    &       0.0258    &       0.0091    &       0.0113    &          0.0093    &          0.0101    & \textbf{            0.0079   } \\
     Cyprus  &  Cyprus Main Market Index  &       0.1042    &       0.0210    &       0.0144    &       0.0174    &       0.0101    &       0.0140    &          0.0084    &          0.0086    & \textbf{            0.0056   } \\
     Czech Republic  &  PX   &       0.0721    &       0.0127    &       0.0092    &       0.0153    &       0.0080    &       0.0107    &          0.0084    &          0.0082    & \textbf{            0.0071   } \\
     Egypt  &  EGX 30 Index  &       0.0691    &       0.0103    &       0.0072    &       0.0136    &       0.0072    &       0.0082    &          0.0076    &          0.0079    & \textbf{            0.0054   } \\
     Estonia  &  OMXT  &       0.1310    &       0.0146    &       0.0079    &       0.0208    &       0.0074    &       0.0063    &          0.0070    &          0.0072    & \textbf{            0.0056   } \\
     EuroStoxx  &  Euro Stoxx 50  &       0.0704    &       0.0135    &       0.0060    &       0.0097    &       0.0055    &       0.0052    &          0.0055    &          0.0055    & \textbf{            0.0046   } \\
     Finland  &  OMXH25  &       0.0787    &       0.0156    &       0.0069    &       0.0131    &       0.0095    &       0.0061    &          0.0052    & \textbf{         0.0049   } &             0.0049    \\
     France  &  CAC 40  &       0.0656    &       0.0111    &       0.0050    &       0.0111    & \textbf{      0.0050   } &       0.0064    &          0.0075    &          0.0061    &             0.0058    \\
     GB   &  FTSE 100  &       0.0685    &       0.0099    &       0.0057    &       0.0100    &       0.0052    &       0.0074    & \textbf{         0.0048   } &          0.0052    &             0.0051    \\
     Germany  &  DAX  &       0.0671    &       0.0160    &       0.0107    &       0.0055    &       0.0055    &       0.0095    &          0.0083    &          0.0085    & \textbf{            0.0044   } \\
     Greece  &  Athex Composite Share Price Index  &       0.0711    &       0.0119    &       0.0076    &       0.0113    &       0.0068    &       0.0073    &          0.0058    & \textbf{         0.0054   } &             0.0057    \\
     Hong Kong  &  Hang Seng  &       0.0801    &       0.0165    &       0.0107    &       0.0109    &       0.0118    &       0.0095    &          0.0086    &          0.0090    & \textbf{            0.0052   } \\
     Hungary  &  Budapest SE  &       0.0682    &       0.0070    &       0.0102    &       0.0172    &       0.0062    &       0.0124    &          0.0105    &          0.0063    & \textbf{            0.0062   } \\
     India  &  Nifty 50  &       0.0684    &       0.0111    &       0.0058    &       0.0091    &       0.0073    &       0.0058    &          0.0095    &          0.0061    & \textbf{            0.0044   } \\
     India  &  BSE Sensex  &       0.0661    &       0.0120    &       0.0060    &       0.0096    &       0.0065    &       0.0062    &          0.0082    &          0.0082    & \textbf{            0.0058   } \\
     Indonesia  &  IDX Composite  &       0.0863    &       0.0119    &       0.0111    &       0.0185    &       0.0091    &       0.0130    &          0.0089    &          0.0090    & \textbf{            0.0087   } \\
     Ireland  &  ISEQ Overall Index  &       0.0790    &       0.0101    &       0.0076    &       0.0160    & \textbf{      0.0055   } &       0.0088    &          0.0076    &          0.0087    &             0.0067    \\
     Israel  &  TA 35  &       0.0618    &       0.0125    &       0.0094    &       0.0093    &       0.0084    &       0.0086    &          0.0073    &          0.0070    & \textbf{            0.0056   } \\
     Italy  &  FTSE MIB  &       0.0668    &       0.0164    &       0.0093    &       0.0082    &       0.0060    &       0.0074    &          0.0074    &          0.0075    & \textbf{            0.0059   } \\
     Japan  &  Topix  &       0.0575    &       0.0131    &       0.0079    &       0.0101    &       0.0110    &       0.0080    &          0.0143    &          0.0154    & \textbf{            0.0063   } \\
     Kazakhstan  &  KASE Index  &       0.1625    &       0.0375    &       0.0333    &       0.0237    &       0.0392    &       0.0326    &          0.0351    &          0.0361    & \textbf{            0.0108   } \\
     Kuwait  &  Kuwait 15  &       0.0546    &       0.0146    &       0.0164    &       0.0170    &       0.0148    &       0.0173    &          0.0165    &          0.0163    & \textbf{            0.0089   } \\
     Latvia  &  OMXR  &       0.1191    &       0.0083    &       0.0069    &       0.0224    &       0.0059    &       0.0086    &          0.0070    &          0.0075    & \textbf{            0.0051   } \\
     Lithuania  &  OMXV  &       0.1140    &       0.0075    &       0.0083    &       0.0232    & \textbf{      0.0070   } &       0.0109    &          0.0071    &          0.0070    &             0.0083    \\
     Luxembourg  &  LuxX Index  &       0.0639    &       0.0078    &       0.0064    &       0.0126    &       0.0062    &       0.0076    &          0.0055    & \textbf{         0.0047   } &             0.0050    \\
     Malaysia  &  FTSE Bursa Malaysia KLCI  &       0.1392    &       0.0107    &       0.0092    &       0.0254    &       0.0081    &       0.0113    &          0.0096    &          0.0101    & \textbf{            0.0047   } \\
     Mauritius  &  SEMDEX  &       0.1330    &       0.0103    &       0.0126    &       0.0284    &       0.0079    &       0.0152    &          0.0086    &          0.0095    & \textbf{            0.0059   } \\
     Mexico  &  IPC  &       0.0716    &       0.0110    &       0.0071    &       0.0134    &       0.0075    &       0.0074    &          0.0068    &          0.0065    & \textbf{            0.0056   } \\
     Morocco  &  MASI  &       0.0813    &       0.0122    &       0.0068    &       0.0120    &       0.0068    &       0.0071    &          0.0075    &          0.0079    & \textbf{            0.0058   } \\
     Namibia  &  NSX Overall Index  &       0.0577    &       0.0095    &       0.0080    &       0.0116    &       0.0120    &       0.0092    &          0.0088    &          0.0073    & \textbf{            0.0071   } \\
     Netherlands  &  AEX  &       0.0731    &       0.0231    &       0.0064    &       0.0112    &       0.0064    &       0.0066    &          0.0083    &          0.0082    & \textbf{            0.0042   } \\
     New Zealand  &  NZX 50 Index  &       0.0526    &       0.0163    &       0.0102    &       0.0106    &       0.0130    &       0.0088    &          0.0128    &          0.0109    & \textbf{            0.0062   } \\
     Norway  &  OBX Index  &       0.0717    &       0.0134    &       0.0070    &       0.0147    &       0.0064    &       0.0092    &          0.0065    & \textbf{         0.0058   } &             0.0062    \\
     Oman  &  MSM 30  &       0.1805    &       0.0100    &       0.0101    &       0.0365    &       0.0099    &       0.0136    &          0.0113    &          0.0115    & \textbf{            0.0075   } \\
     Pakistan  &  KSE 100 Index  &       0.0970    &       0.0176    &       0.0113    &       0.0227    &       0.0120    &       0.0114    &          0.0130    &          0.0137    & \textbf{            0.0102   } \\
     Peru  &  S\&P Lima General Index  &       0.0927    &       0.0116    &       0.0096    &       0.0204    &       0.0108    &       0.0106    &          0.0114    &          0.0116    & \textbf{            0.0092   } \\
     Philippines  &  PSEi Composite  &       0.0711    &       0.0082    &       0.0097    &       0.0151    &       0.0062    &       0.0111    &          0.0059    & \textbf{         0.0058   } &             0.0063    \\
     Poland  &  WIG  &       0.0615    &       0.0126    &       0.0077    &       0.0090    &       0.0059    &       0.0060    &          0.0058    &          0.0058    & \textbf{            0.0057   } \\
     Portugal  &  PSI 20  &       0.0692    &       0.0118    &       0.0075    &       0.0108    &       0.0077    &       0.0062    &          0.0070    &          0.0071    & \textbf{            0.0056   } \\
     Qatar  &  QE 20 Index  &       0.2211    &       0.0139    &       0.0140    &       0.0378    &       0.0138    &       0.0140    &          0.0171    &          0.0183    & \textbf{            0.0082   } \\
     Romania  &  BET 10  &       0.0988    &       0.0106    &       0.0080    &       0.0206    &       0.0083    &       0.0087    &          0.0063    &          0.0066    & \textbf{            0.0059   } \\
     Russia  &  MICEX  &       0.1090    &       0.0106    &       0.0116    &       0.0207    &       0.0078    &       0.0132    &          0.0106    &          0.0114    & \textbf{            0.0056   } \\
     Russia  &  RTSI  &       0.1308    &       0.0190    &       0.0118    &       0.0220    &       0.0157    &       0.0110    &          0.0106    &          0.0106    & \textbf{            0.0083   } \\
     Saudi Arabia  &  Tadawul All Share  &       0.1371    &       0.0157    &       0.0090    &       0.0287    &       0.0090    &       0.0097    &          0.0128    &          0.0136    & \textbf{            0.0070   } \\
     Serbia  &  BELEX  &       0.1118    &       0.0099    &       0.0084    &       0.0209    &       0.0064    &       0.0096    & \textbf{         0.0064   } &          0.0064    &             0.0068    \\
     Singapore  &  STI Index  &       0.0715    &       0.0124    & \textbf{      0.0058   } &       0.0111    &       0.0059    &       0.0069    &          0.0065    &          0.0074    &             0.0067    \\
     South Africa  &  FTSE/JSE All-Share Index  &       0.0569    &       0.0124    &       0.0079    &       0.0112    &       0.0062    &       0.0089    &          0.0067    &          0.0071    & \textbf{            0.0054   } \\
     South Korea  &  KOSPI  &       0.0919    &       0.0199    &       0.0124    &       0.0158    & \textbf{      0.0073   } &       0.0107    &          0.0091    &          0.0093    &             0.0083    \\
     Spain  &  IBEX 35  &       0.0569    &       0.0143    &       0.0100    &       0.0078    &       0.0091    &       0.0088    & \textbf{         0.0055   } &          0.0055    &             0.0058    \\
     Sri Lanka  &  CSE All-Share  &       0.1168    &       0.0243    &       0.0119    &       0.0242    &       0.0134    &       0.0119    &          0.0068    &          0.0071    & \textbf{            0.0044   } \\
     Sweden  &  OMXS30  &       0.0580    &       0.0125    &       0.0087    &       0.0101    &       0.0079    &       0.0074    &          0.0075    & \textbf{         0.0061   } &             0.0080    \\
     Switzerland  &  SMI  &       0.0724    &       0.0126    &       0.0067    &       0.0120    &       0.0078    &       0.0065    &          0.0110    &          0.0113    & \textbf{            0.0058   } \\
     Taiwan  &  Taiwan Weighted  &       0.0720    &       0.0164    &       0.0111    &       0.0108    &       0.0074    &       0.0676    &          0.0075    & \textbf{         0.0074   } &             0.0078    \\
     Thailand  &  SET  &       0.0754    &       0.0150    &       0.0106    &       0.0128    &       0.0104    &       0.0098    &          0.0071    & \textbf{         0.0071   } &             0.0073    \\
     Tunesia  &  Tunindex  &       0.1642    &       0.0150    &       0.0162    &       0.0267    &       0.0145    &       0.0494    &          0.0116    &          0.0084    & \textbf{            0.0075   } \\
     Turkey  &  BIST 100  &       0.0735    &       0.0086    &       0.0065    &       0.0120    &       0.0060    &       0.0066    &          0.0060    &          0.0060    & \textbf{            0.0050   } \\
     United Arab Emirates  &  DFM  &       0.0894    &       0.0138    &       0.0100    &       0.0199    &       0.0101    &       0.0107    &          0.0106    &          0.0106    & \textbf{            0.0074   } \\
     United Arab Emirates  &  Abu Dhabi  &       0.1056    &       0.0169    &       0.0112    &       0.0193    &       0.0088    &       0.0096    &          0.0090    &          0.0094    & \textbf{            0.0070   } \\
     USA  &  Dow 30  &       0.0812    &       0.0151    &       0.0097    &       0.0130    &       0.0085    &       0.0086    &          0.0082    &          0.0086    & \textbf{            0.0069   } \\
     USA  &  S\&P 500  &       0.0827    &       0.0170    &       0.0095    &       0.0122    &       0.0077    &       0.0080    &          0.0073    &          0.0077    & \textbf{            0.0053   } \\
     USA  &  Nasdaq  &       0.0865    &       0.0233    &       0.0166    &       0.0126    &       0.0085    &       0.0150    & \textbf{         0.0066   } &          0.0067    &             0.0074    \\
     Venezuela  &  IBC  &       0.1329    &       0.0242    &       0.0142    &       0.0225    &       0.0118    &       0.0128    &          0.0182    &          0.0190    & \textbf{            0.0068   } \\
     Vietnam  &  HNX 30  &       0.0998    &       0.0144    &       0.0097    &       0.0212    &       0.0103    &       0.1071    &          0.0082    &          0.0078    & \textbf{            0.0070   } \\
     Zambia  &  All Share Index  &       0.1870    &       0.0539    &       0.0525    &       0.0285    &       0.0265    &       0.0516    &          0.0266    &          0.0291    & \textbf{            0.0146   } \\
    \hline
    \end{tabular}%
		\end{tiny}
		  \caption{KS distance between the empirical and fitted distributions for daily log-returns, from 01/03/1997 until 11/02/2017.}
  \label{tab:dKS}%
\end{table}%

\begin{table}[htbp]
    \hspace{-2.5cm}
		\begin{tiny}
		\begin{tabular}{rrrrrrrrrrr}
    \hline
    Country & Index & N     & St    & NIG  & V$\Gamma$  & GH    & Meix & 2St12 & 2St39 & 3St \\
    \hline
        Argentina  &  MERVAL  &       61.4018    &         1.4315    & \textbf{        0.1071   } &         1.1051    &         0.1073    &         0.1574    &         0.1577    &         0.1728    &              0.1150    \\
     Australia  &  S\&P/ASX 200  &       43.5157    &         1.7596    &         0.4156    &         0.9159    &         0.4632    &         0.5626    &         0.5793    &         0.5933    & \textbf{             0.1443   } \\
     Australia  &  All Ordinaries  &       44.9225    &         2.1225    &         0.3609    &         0.8719    &         0.3849    &         0.5056    &         0.4544    &         0.5889    & \textbf{             0.1664   } \\
     Austria  &  ATX  &       71.4871    &         2.7950    &         0.5604    &         2.4360    &         0.3067    &         0.9490    &         0.3457    &         0.4231    & \textbf{             0.1602   } \\
     Bahrain  &  All Share  &       57.7486    &         1.5727    &         0.7218    &         0.7982    &         0.4826    &         0.5959    &         0.1672    &         0.1760    & \textbf{             0.1552   } \\
     Belgium  &  BEL 20  &       59.9251    &         2.4349    &         0.3143    &         1.1152    &         0.3109    &         0.3095    &         0.4900    &         0.5241    & \textbf{             0.2327   } \\
     Brazil  &  Bovespa  &       51.9114    &         0.9076    &         0.7759    &         1.7570    &         0.3297    &         1.1442    &         0.2454    &         0.3354    & \textbf{             0.2420   } \\
     Bulgaria  &  SOFIX  &     190.9200    &         0.4959    &         0.4356    &         6.5727    &         0.2769    &         0.8130    &         0.2817    &         0.3330    & \textbf{             0.1321   } \\
     Canada  &  S\&P/TSX 60  &       70.2820    &         2.8171    &         0.4534    &         1.8278    &         0.3953    &         0.7275    &         0.1987    &         0.2398    & \textbf{             0.1727   } \\
     Canada  &  S\&P/TSX Composite  &       81.5348    &         3.8481    &         0.4928    &         2.3741    &         0.3290    &         0.8748    &         0.1205    &         0.1631    & \textbf{             0.1081   } \\
     Chile  &  IPSA  &       51.4843    &         0.3826    &         0.3852    &         1.2470    &         0.2433    &         0.6280    &         0.3369    &         0.2060    & \textbf{             0.1864   } \\
     China  &  CSI 300  &       46.2091    &         2.5890    &         1.4802    &         1.1695    &         1.0113    &         1.3146    &         0.9017    &         0.9135    & \textbf{             0.2768   } \\
     China  &  SSE Composite Index  &       75.9015    &         2.2158    &         0.8366    &         1.0343    &         0.5890    &         0.6984    &         0.5676    &         0.5858    & \textbf{             0.2584   } \\
     Colombia  &  IGBC  &       66.1748    &         0.7093    &         0.3938    &         1.4865    &         0.3580    &         0.6218    &         0.5009    &         0.4662    & \textbf{             0.0949   } \\
     Croatia  &  CROBEX  &     190.9279    &         0.5575    &         0.3959    &         8.0189    &         0.2535    &         0.7441    &         0.4296    &         0.5392    & \textbf{             0.1674   } \\
     Cyprus  &  Cyprus Main Market Index  &       83.5638    &         2.2376    &         0.9360    &         1.2512    &         0.2991    &         0.6936    &         0.2729    &         0.3098    & \textbf{             0.0975   } \\
     Czech Republic  &  PX   &       62.7585    &         0.9750    &         0.6136    &         2.0849    &         0.2203    &         1.0301    &         0.2807    &         0.3022    & \textbf{             0.1948   } \\
     Egypt  &  EGX 30 Index  &       49.9205    &         0.7505    &         0.3358    &         1.2463    &         0.3322    &         0.4573    &         0.4180    &         0.4677    & \textbf{             0.1331   } \\
     Estonia  &  OMXT  &     221.5153    &         1.7963    &         0.4095    &         5.3429    &         0.3744    &         0.4263    &         0.4766    &         0.5470    & \textbf{             0.1210   } \\
     EuroStoxx  &  Euro Stoxx 50  &       63.1276    &         1.8154    &         0.2345    &         0.6848    &         0.1751    &         0.1721    &         0.2331    &         0.2433    & \textbf{             0.0933   } \\
     Finland  &  OMXH25  &       76.8740    &         2.1777    &         0.2916    &         1.1807    &         0.4975    &         0.1931    &         0.1134    &         0.1116    & \textbf{             0.0855   } \\
     France  &  CAC 40  &       49.0040    &         1.4135    &         0.1711    &         0.7670    &         0.1798    &         0.2086    &         0.4414    &         0.4962    & \textbf{             0.1043   } \\
     GB   &  FTSE 100  &       56.6985    &         1.2483    &         0.1300    &         0.9116    &         0.1374    &         0.1891    &         0.1751    &         0.1807    & \textbf{             0.1058   } \\
     Germany  &  DAX  &       50.7628    &         3.5610    &         1.0327    &         0.2967    &         0.2967    &         0.7677    &         0.7819    &         0.8433    & \textbf{             0.0993   } \\
     Greece  &  Athex Composite Share Price Index  &       57.8528    &         1.6080    &         0.2490    &         0.6981    &         0.1792    &         0.1601    &         0.1820    &         0.1905    & \textbf{             0.1244   } \\
     Hong Kong  &  Hang Seng  &       79.7223    &         2.0783    &         0.6629    &         0.8426    &         0.6733    &         0.5975    &         0.5540    &         0.5791    & \textbf{             0.1269   } \\
     Hungary  &  Budapest SE  &       67.5311    &         0.3042    &         0.7701    &         2.1833    &         0.2666    &         1.2236    &         0.4818    &         0.2087    & \textbf{             0.2050   } \\
     India  &  Nifty 50  &       58.5942    &         1.4940    &         0.1959    &         0.7941    &         0.2495    &         0.2605    &         0.4221    &         0.4318    & \textbf{             0.0998   } \\
     India  &  BSE Sensex  &       57.2633    &         1.4038    &         0.1644    &         0.8563    &         0.1762    &         0.2256    &         0.3226    &         0.3272    & \textbf{             0.1348   } \\
     Indonesia  &  IDX Composite  &       95.9611    &         1.4356    &         0.3977    &         2.7623    &         0.3138    &         0.6708    &         0.2111    &         0.2346    & \textbf{             0.1984   } \\
     Ireland  &  ISEQ Overall Index  &       75.7232    &         1.7031    &         0.4120    &         2.4842    &         0.1230    &         0.8240    &         0.1646    &         0.2257    & \textbf{             0.0871   } \\
     Israel  &  TA 35  &       40.7644    &         1.2899    &         0.4726    &         0.5717    &         0.3669    &         0.3877    &         0.4870    &         0.5446    & \textbf{             0.1364   } \\
     Italy  &  FTSE MIB  &       47.8938    &         2.3304    &         0.3825    &         0.3231    &         0.1636    &         0.2301    &         0.4458    &         0.4943    & \textbf{             0.1392   } \\
     Japan  &  Topix  &       35.5084    &         1.3942    &         0.5419    &         0.8529    &         0.6177    &         0.6292    &         1.0712    &         1.3218    & \textbf{             0.1760   } \\
     Kazakhstan  &  KASE Index  &     271.2909    &         6.0586    &         4.5664    &         3.3495    &         9.6257    &         4.2544    &         4.7109    &         5.1635    & \textbf{             0.4200   } \\
     Kuwait  &  Kuwait 15  &         8.5261    &         0.3296    &         0.4819    &         0.7261    &         0.3409    &         0.5729    &         0.3208    &         0.3001    & \textbf{             0.1073   } \\
     Latvia  &  OMXR  &     153.6989    &         0.3680    &         0.3253    &         4.8440    &         0.1092    &         0.7715    &         0.1987    &         0.2405    & \textbf{             0.1000   } \\
     Lithuania  &  OMXV  &     146.3636    &         0.4593    &         0.3895    &         4.6047    &         0.2338    &         0.7769    &         0.3025    &         0.3462    & \textbf{             0.1773   } \\
     Luxembourg  &  LuxX Index  &       48.7480    &         0.4079    &         0.2576    &         1.2032    &         0.1529    &         0.4730    &         0.1715    & \textbf{        0.1115   } &              0.1185    \\
     Malaysia  &  FTSE Bursa Malaysia KLCI  &     279.5591    &         0.8532    &         0.6847    &         7.1045    &         0.4008    &         1.3513    &         0.4747    &         0.5600    & \textbf{             0.1269   } \\
     Mauritius  &  SEMDEX  &     241.1763    &         1.1387    &         0.9492    &         8.5619    &         0.2976    &         1.8280    &         0.4014    &         0.5045    & \textbf{             0.1319   } \\
     Mexico  &  IPC  &       71.1773    &         1.0533    &         0.3015    &         1.4603    &         0.3130    &         0.4498    &         0.1430    &         0.1267    & \textbf{             0.1139   } \\
     Morocco  &  MASI  &       67.4822    &         0.6225    &         0.2241    &         1.4898    &         0.2233    &         0.3015    &         0.2566    &         0.2833    & \textbf{             0.1707   } \\
     Namibia  &  NSX Overall Index  &       29.0520    &         0.2739    &         0.2097    &         0.7937    &         0.4121    &         0.3419    &         0.2077    &         0.1966    & \textbf{             0.1178   } \\
     Netherlands  &  AEX  &       73.8303    &         3.9438    &         0.1918    &         1.3243    &         0.1971    &         0.2968    &         0.2418    &         0.2671    & \textbf{             0.0972   } \\
     New Zealand  &  NZX 50 Index  &       30.7972    &         2.1691    &         0.3677    &         0.6692    &         0.4015    &         0.4651    &         0.4941    &         0.4071    & \textbf{             0.1160   } \\
     Norway  &  OBX Index  &       56.7607    &         1.6070    &         0.3081    &         1.6389    &         0.1454    &         0.5942    &         0.2223    &         0.2483    & \textbf{             0.1413   } \\
     Oman  &  MSM 30  &     424.0793    &         0.5342    &         0.9099    &       13.5977    &         0.4528    &         1.8845    &         0.6962    &         0.7862    & \textbf{             0.1938   } \\
     Pakistan  &  KSE 100 Index  &       92.7593    &         3.3135    &         0.9706    &         3.0725    &         0.9689    &         0.9245    &         1.2103    &         1.2760    & \textbf{             0.4369   } \\
     Peru  &  S\&P Lima General Index  &     107.3345    &         0.3809    &         0.5982    &         3.8051    &         0.3235    &         1.0404    &         0.2795    &         0.2919    & \textbf{             0.1767   } \\
     Philippines  &  PSEi Composite  &       70.0531    &         0.5201    &         0.6322    &         2.4278    &         0.1849    &         1.0816    &         0.1514    &         0.1646    & \textbf{             0.1467   } \\
     Poland  &  WIG  &       49.2788    &         1.2862    &         0.2753    &         0.5120    & \textbf{        0.1987   } &         0.2135    &         0.3149    &         0.2644    &              0.2029    \\
     Portugal  &  PSI 20  &       54.4711    &         1.8733    &         0.3175    &         0.9093    &         0.3292    &         0.3117    &         0.6014    &         0.6193    & \textbf{             0.1880   } \\
     Qatar  &  QE 20 Index  &     567.4108    &         1.4914    &         0.9137    &       15.1353    &         0.9505    &         1.2925    &         1.8682    &         2.1362    & \textbf{             0.3293   } \\
     Romania  &  BET 10  &     107.7361    &         0.9509    &         0.2574    &         2.5199    &         0.2765    &         0.3135    &         0.1912    &         0.2021    & \textbf{             0.1080   } \\
     Russia  &  MICEX  &     153.1403    &         1.1384    &         0.6646    &         4.6802    &         0.3677    &         1.1908    &         0.5455    &         0.6184    & \textbf{             0.0975   } \\
     Russia  &  RTSI  &     130.4714    &         1.8457    &         0.7024    &         2.9798    &         0.8545    &         0.8784    &         0.7600    &         0.8112    & \textbf{             0.1119   } \\
     Saudi Arabia  &  Tadawul All Share  &     201.0783    &         2.8584    &         0.5337    &         7.3013    &         0.5559    &         0.6470    &         1.3092    &         1.5547    & \textbf{             0.2671   } \\
     Serbia  &  BELEX  &       99.6180    &         0.3806    &         0.1806    &         2.7045    &         0.1442    &         0.3792    &         0.1294    &         0.1388    & \textbf{             0.0927   } \\
     Singapore  &  STI Index  &       53.2461    &         1.0570    &         0.1878    &         0.9559    &         0.1933    &         0.2275    &         0.2249    &         0.2266    & \textbf{             0.1573   } \\
     South Africa  &  FTSE/JSE All-Share Index  &       43.0165    &         1.3711    &         0.4221    &         1.3931    &         0.2035    &         0.6975    &         0.1490    &         0.1728    & \textbf{             0.1258   } \\
     South Korea  &  KOSPI  &       97.5514    &         4.2503    &         1.0463    &         2.1092    &         0.3051    &         0.6633    &         0.4027    &         0.4328    & \textbf{             0.2836   } \\
     Spain  &  IBEX 35  &       43.8417    &         2.1678    &         0.4713    &         0.3981    &         0.3951    &         0.3668    &         0.2970    &         0.2806    & \textbf{             0.1194   } \\
     Sri Lanka  &  CSE All-Share  &     164.8302    &         3.0701    &         0.9446    &         4.7761    &         0.9799    &         1.2512    &         0.3159    &         0.3255    & \textbf{             0.1118   } \\
     Sweden  &  OMXS30  &       44.2956    &         1.3522    &         0.2349    &         0.5904    &         0.2024    &         0.2029    &         0.1676    &         0.1394    & \textbf{             0.1080   } \\
     Switzerland  &  SMI  &       62.1912    &         1.6151    &         0.2955    &         1.2924    &         0.2739    &         0.4733    &         0.5377    &         0.5703    & \textbf{             0.1323   } \\
     Taiwan  &  Taiwan Weighted  &       58.6538    &         3.0170    &         0.7470    &         0.6716    &         0.2834    &       48.0961    &         0.5635    &         0.5889    & \textbf{             0.2777   } \\
     Thailand  &  SET  &       72.1878    &         1.6181    &         0.6543    &         0.9743    &         0.6235    &         0.5977    & \textbf{        0.2069   } &         0.2124    &              0.2148    \\
     Tunesia  &  Tunindex  &     335.9487    &         1.9712    &         2.8552    &         7.4779    &         1.7832    &       31.0061    &         0.7370    &         0.5006    & \textbf{             0.1864   } \\
     Turkey  &  BIST 100  &       73.0789    &         0.4407    &         0.1642    &         1.4645    &         0.1464    &         0.3344    &         0.1639    &         0.1528    & \textbf{             0.1072   } \\
     United Arab Emirates  &  DFM  &       67.3002    &         0.7892    &         0.5082    &         2.0386    &         0.4991    &         0.6304    &         0.5545    &         0.6030    & \textbf{             0.1681   } \\
     United Arab Emirates  &  Abu Dhabi  &     114.3571    &         1.8512    &         0.5395    &         2.6397    &         0.3752    &         0.4450    &         0.4698    &         0.5167    & \textbf{             0.1921   } \\
     USA  &  Dow 30  &       78.3204    &         2.5815    &         0.6680    &         0.9541    &         0.5189    &         0.5430    &         0.5311    &         0.5574    & \textbf{             0.1868   } \\
     USA  &  S\&P 500  &       85.2271    &         3.0328    &         0.6245    &         1.0956    &         0.4356    &         0.4799    &         0.4670    &         0.5005    & \textbf{             0.1826   } \\
     USA  &  Nasdaq  &       91.1063    &         6.2771    &         1.9766    &         0.7419    &         0.4397    &         1.4235    &         0.3121    &         0.3321    & \textbf{             0.2491   } \\
     Venezuela  &  IBC  &     210.9153    &         7.6051    &         0.7853    &         3.8059    &         0.5438    &         0.6433    &         1.4585    &         1.6996    & \textbf{             0.1985   } \\
     Vietnam  &  HNX 30  &       61.0065    &         1.1820    &         0.3806    &         1.7214    &         0.2726    &       63.4368    &         0.2021    &         0.2036    & \textbf{             0.1264   } \\
     Zambia  &  All Share Index  &     317.7578    &       16.6685    &       16.8868    &         4.5607    &         3.1559    &       16.4118    &         5.9229    &         6.5435    & \textbf{             0.6230   } \\
    \hline
    \end{tabular}%
		\end{tiny}
		  \caption{AD distance between the empirical and fitted distributions for daily log-returns, from 01/03/1997 until 11/02/2017.}
  \label{tab:dAD}%
\end{table}%

\begin{table}[htbp]
  \hspace{-2.5cm}
\begin{tiny}
    \begin{tabular}{rrrrrrrrrrr}
    \hline
    Country & Index & N     & St    & NIG  & V$\Gamma$  & GH    & Meix & 2St12 & 2St39 & 3St \\
    \hline
        Argentina & MERVAL & -24705 & -25622 & \textbf{-25650} & -25608 & -25648 & -25648 & -25643 & -25643 & -25636 \\
    Australia & S\&P/ASX 200 & -33725 & -34510 & -34514 & -34480 & -34521 & -34503 & -34521 & -34520 & \textbf{-34525} \\
    Australia & All Ordinaries & -34017 & -34821 & -34830 & -34794 & -34836 & -34818 & \textbf{-34837} & -34833 & -34837 \\
    Austria & ATX   & -29269 & -30368 & -30389 & -30313 & -30395 & -30374 & \textbf{-30399} & -30398 & -30397 \\
    Bahrain & All Share & -27392 & -28234 & -28260 & -28251 & -28264 & -28262 & \textbf{-28273} & -28272 & -28265 \\
    Belgium & BEL 20 & -31570 & -32495 & -32523 & -32485 & -32521 & -32519 & -32521 & -32521 & \textbf{-32524} \\
    Brazil & Bovespa & -25428 & -26420 & -26390 & -26326 & -26421 & -26367 & \textbf{-26430} & -26428 & -26422 \\
    Bulgaria & SOFIX & -23181 & -25862 & -25868 & -25647 & \textbf{-25873} & -25853 & -25865 & -25860 & -25871 \\
    Canada & S\&P/TSX 60 & -28538 & -29683 & -29693 & -29622 & -29701 & -29678 & \textbf{-29710} & -29709 & -29701 \\
    Canada & S\&P/TSX Composite & -32365 & -33655 & -33676 & -33586 & -33685 & -33658 & \textbf{-33694} & -33693 & -33686 \\
    Chile & IPSA  & -32491 & -33438 & -33422 & -33368 & -33435 & -33407 & -33434 & \textbf{-33438} & -33430 \\
    China & CSI 300 & -15886 & -16459 & -16501 & \textbf{-16524} & -16523 & -16508 & -16514 & -16513 & -16521 \\
    China & SSE Composite Index & -27202 & -28276 & -28320 & -28309 & \textbf{-28325} & -28323 & -28322 & -28322 & -28325 \\
    Colombia & IGBC  & -23366 & -24478 & -24470 & -24409 & -24479 & -24455 & -24475 & -24478 & \textbf{-24484} \\
    Croatia & CROBEX & -27275 & -29812 & -29834 & \textbf{-29609} & -29835 & -29824 & -29829 & -29823 & -29830 \\
    Cyprus & Cyprus Main Market Index & -14221 & -15309 & -15363 & \textbf{-15349} & -15382 & -15371 & -15374 & -15372 & -15375 \\
    Czech Republic & PX    & -29865 & -31021 & -31001 & \textbf{-30918} & -31025 & -30978 & -31025 & -31022 & -31024 \\
    Egypt & EGX 30 Index & -25662 & -26501 & -26505 & -26462 & -26506 & -26497 & -26504 & -26507 & \textbf{-26512} \\
    Estonia & OMXT  & -28299 & -31432 & -31475 & -31260 & -31473 & -31466 & -31454 & -31449 & \textbf{-31482} \\
    EuroStoxx & Euro Stoxx 50 & -31064 & -32035 & -32063 & -32035 & -32061 & -32061 & \textbf{-32064} & -32064 & -32061 \\
    Finland & OMXH25 & -27116 & -28242 & -28282 & -28241 & -28281 & -28283 & -28287 & \textbf{-28288} & -28282 \\
    France & CAC 40 & -29844 & -30625 & \textbf{-30643} & -30614 & -30641 & -30639 & -30634 & -30637 & -30638 \\
    GB    & FTSE 100 & -31694 & -32594 & \textbf{-32612} & -32574 & -32611 & -32608 & -32610 & -32611 & -32608 \\
    Germany & DAX   & -29255 & -29995 & -30034 & -30053 & -30051 & -30038 & -30044 & -30049 & \textbf{-30054} \\
    Greece & Athex Composite Share Price Index & -26032 & -26901 & -26931 & -26907 & -26931 & \textbf{-26932} & -26930 & -26931 & -26923 \\
    Hong Kong & Hang Seng & -27595 & -28868 & -28885 & -28852 & -28882 & -28878 & \textbf{-28906} & -28906 & -28905 \\
    Hungary & Budapest SE & -27601 & \textbf{-28875} & -28841 & -28754 & -28873 & -28815 & -28869 & -28871 & -28874 \\
    India & Nifty 50 & -28568 & -29527 & -29540 & -29505 & -29540 & -29532 & -29536 & \textbf{-29541} & -29540 \\
    India & BSE Sensex & -28555 & -29451 & \textbf{-29470} & -29439 & -29468 & -29465 & -29464 & -29464 & -29461 \\
    Indonesia & IDX Composite & -27684 & -29128 & -29144 & -29048 & -29145 & -29133 & \textbf{-29149} & -29147 & -29141 \\
    Ireland & ISEQ Overall Index & -30359 & -31564 & -31576 & -31488 & \textbf{-31584} & -31560 & -31583 & -31582 & -31577 \\
    Israel & TA 35 & -30337 & -30979 & -30998 & -30990 & -30997 & -30998 & -30995 & -30992 & \textbf{-30999} \\
    Italy & FTSE MIB & -27551 & -28263 & -28296 & -28290 & -28298 & \textbf{-28298} & -28291 & -28292 & -28296 \\
    Japan & Topix & -29255 & -29920 & -29919 & -29893 & -29927 & -29908 & -29924 & -29917 & \textbf{-29939} \\
    Kazakhstan & KASE Index & -18194 & -21812 & -21894 & -22127 & \textbf{-22157} & -21900 & -21837 & -21818 & -22133 \\
    Kuwait & Kuwait 15 & -9338 & \textbf{-9506} & -9499 & -9485 & -9502 & -9494 & -9505 & -9506 & -9501 \\
    Latvia & OMXR  & -25015 & -27271 & -27271 & -27085 & \textbf{-27279} & -27254 & -27269 & -27265 & -27272 \\
    Lithuania & OMXV  & -27816 & -29951 & -29951 & -29777 & \textbf{-29958} & -29934 & -29952 & -29949 & -29955 \\
    Luxembourg & LuxX Index & -27771 & -28631 & -28625 & -28574 & -28632 & -28613 & -28630 & \textbf{-28634} & -28626 \\
    Malaysia & FTSE Bursa Malaysia KLCI & -29974 & -33593 & -33573 & -33297 & \textbf{-33595} & -33543 & -33581 & -33573 & -33593 \\
    Mauritius & SEMDEX & -37310 & -40646 & -40629 & -40317 & \textbf{-40654} & -40601 & -40632 & -40625 & -40649 \\
    Mexico & IPC   & -29728 & -30865 & -30873 & -30813 & -30873 & -30863 & -30879 & \textbf{-30879} & -30871 \\
    Morocco & MASI  & -27302 & -28293 & \textbf{-28312} & -28263 & -28310 & -28309 & -28306 & -28304 & -28303 \\
    Namibia & NSX Overall Index & -20708 & -21210 & \textbf{-21211} & -21184 & -21210 & -21205 & -21207 & -21209 & -21202 \\
    Netherlands & AEX   & -29949 & -31032 & \textbf{-31094} & -31040 & -31092 & -31088 & -31090 & -31088 & -31086 \\
    New Zealand & NZX 50 Index & -30059 & -30607 & -30618 & -30594 & -30623 & -30609 & -30621 & -30619 & \textbf{-30624} \\
    Norway & OBX Index & -25150 & -26088 & -26097 & -26035 & \textbf{-26103} & -26084 & -26099 & -26098 & -26098 \\
    Oman  & MSM 30 & -29107 & -35123 & -35077 & -34670 & -35120 & -35035 & -35085 & -35076 & -35114 \\
    Pakistan & KSE 100 Index & -28234 & \textbf{-29443} & -29506 & -29462 & -29508 & -29511 & -29498 & -29498 & -29517 \\
    Peru  & S\&P Lima General Index & -29956 & -31627 & -31622 & -31504 & \textbf{-31632} & -31604 & -31630 & -31629 & -31626 \\
    Philippines & PSEi Composite & -28989 & -30159 & -30141 & -30051 & -30158 & -30120 & \textbf{-30162} & -30162 & -30156 \\
    Poland & WIG   & -29939 & -30687 & -30713 & -30699 & -30712 & \textbf{-30713} & -30703 & -30708 & -30703 \\
    Portugal & PSI 20 & -31473 & -32343 & -32367 & -32334 & -32365 & -32363 & -32356 & -32362 & \textbf{-32373} \\
    Qatar & QE 20 Index & -22489 & -30301 & -30309 & -29807 & -30314 & -30287 & -30244 & -30229 & \textbf{-30314} \\
    Romania & BET 10 & -26998 & -28554 & \textbf{-28587} & -28501 & -28585 & -28583 & -28584 & -28584 & -28580 \\
    Russia & MICEX & -22365 & -24550 & -24545 & -24372 & -24558 & -24524 & -24555 & -24551 & \textbf{-24560} \\
    Russia & RTSI  & -15762 & -17640 & -17646 & -17546 & -17648 & -17632 & -17636 & -17634 & \textbf{-17673} \\
    Saudi Arabia & Tadawul All Share & -28769 & -31409 & -31478 & -31304 & -31476 & \textbf{-31478} & -31439 & -31430 & -31470 \\
    Serbia & BELEX & -17969 & -19330 & -19334 & -19223 & \textbf{-19336} & -19324 & -19335 & -19334 & -19334 \\
    Singapore & STI Index & -27798 & -28628 & \textbf{-28645} & -28605 & -28643 & -28641 & -28642 & -28644 & -28636 \\
    South Africa & FTSE/JSE All-Share Index & -30930 & -31699 & -31701 & -31649 & -31709 & -31688 & \textbf{-31710} & -31709 & -31703 \\
    South Korea & KOSPI & -27600 & -28831 & -28917 & -28908 & \textbf{-28942} & -28933 & -28931 & -28930 & -28923 \\
    Spain & IBEX 35 & -29259 & -29966 & -29984 & -29976 & -29983 & -29983 & \textbf{-29992} & -29990 & -29990 \\
    Sri Lanka & CSE All-Share & -30587 & -32975 & -32993 & -32818 & -33000 & -32974 & -33012 & -33011 & \textbf{-33014} \\
    Sweden & OMXS30 & -28953 & -29625 & -29647 & -29631 & -29646 & \textbf{-29647} & -29640 & -29643 & -29642 \\
    Switzerland & SMI   & -31355 & -32347 & -32360 & -32316 & -32362 & -32350 & -32358 & -32355 & \textbf{-32362} \\
    Taiwan & Taiwan Weighted & -30038 & -30751 & -30814 & -30827 & \textbf{-30833} & -30126 & -30808 & -30808 & -30826 \\
    Thailand & SET   & -27923 & -29070 & -29087 & -29056 & -29085 & -29082 & -29102 & \textbf{-29102} & -29094 \\
    Tunesia & Tunindex & -33496 & -38617 & -38483 & -38224 & -38613 & -37936 & -38604 & -38601 & \textbf{-38631} \\
    Turkey & BIST 100 & -24084 & -25246 & -25255 & -25193 & \textbf{-25255} & -25245 & -25254 & -25254 & -25248 \\
    United Arab Emirates & DFM   & -18847 & -19806 & -19824 & -19770 & -19823 & -19821 & -19818 & -19817 & \textbf{-19825} \\
    United Arab Emirates & Abu Dhabi & -26177 & -27774 & -27817 & -27747 & -27818 & -27818 & -27809 & -27807 & \textbf{-27820} \\
    USA   & Dow 30 & -31982 & -33152 & -33179 & -33155 & -33178 & -33175 & -33194 & \textbf{-33194} & -33193 \\
    USA   & S\&P 500 & -31352 & -32595 & -32630 & -32599 & -32630 & -32628 & \textbf{-32642} & -32642 & -32639 \\
    USA   & Nasdaq & -27074 & -28256 & -28340 & -28366 & -28379 & -28355 & \textbf{-28383} & -28382 & -28376 \\
    Venezuela & IBC   & -23742 & -26362 & -26447 & -26361 & -26448 & -26446 & -26408 & -26397 & \textbf{-26457} \\
    Vietnam & HNX 30 & -14835 & -15593 & -15636 & -15601 & \textbf{-15641} & -14788 & -15639 & -15639 & -15635 \\
    Zambia & All Share Index & -19282 & -23381 & -23466 & -23886 & \textbf{-23903} & -23481 & -23657 & -23640 & -23888 \\
    \hline
    \end{tabular}%
		\end{tiny}
		  \caption{AIC of fitted distributions for daily log-returns, from 01/03/1997 until 11/02/2017.}
  \label{tab:dAIC}%
\end{table}%

\begin{table}[htbp]
  \hspace{-2.5cm}
\begin{tiny}
    \begin{tabular}{rrrrrrrrrrr}
    \hline
    Country & Index & N     & St    & NIG  & V$\Gamma$  & GH    & Meix & 2St12 & 2St39 & 3St \\
    \hline
       Argentina & MERVAL & -24692 & -25602 & \textbf{-25624} & -25582 & -25615 & -25621 & -25617 & -25617 & -25584 \\
    Australia & S\&P/ASX 200 & -33712 & -34490 & -34488 & -34454 & -34488 & -34477 & \textbf{-34494} & -34494 & -34472 \\
    Australia & All Ordinaries & -34004 & -34801 & -34804 & -34768 & -34803 & -34792 & \textbf{-34811} & -34807 & -34784 \\
    Austria & ATX   & -29256 & -30348 & -30363 & -30286 & -30363 & -30348 & \textbf{-30373} & -30372 & -30345 \\
    Bahrain & All Share & -27380 & -28216 & -28236 & -28226 & -28233 & -28237 & \textbf{-28248} & -28247 & -28215 \\
    Belgium & BEL 20 & -31557 & -32475 & \textbf{-32497} & -32459 & -32488 & -32493 & -32495 & -32495 & -32471 \\
    Brazil & Bovespa & -25415 & -26401 & -26364 & -26300 & -26388 & -26341 & \textbf{-26404} & -26402 & -26370 \\
    Bulgaria & SOFIX & -23168 & \textbf{-25843} & -25842 & -25622 & -25841 & -25827 & -25839 & -25835 & -25820 \\
    Canada & S\&P/TSX 60 & -28525 & -29664 & -29667 & -29597 & -29669 & -29652 & \textbf{-29684} & -29683 & -29650 \\
    Canada & S\&P/TSX Composite & -32352 & -33636 & -33650 & -33559 & -33652 & -33632 & \textbf{-33668} & -33667 & -33634 \\
    Chile & IPSA  & -32478 & \textbf{-33418} & -33396 & -33342 & -33402 & -33380 & -33407 & -33412 & -33377 \\
    China & CSI 300 & -15874 & -16441 & -16477 & \textbf{-16500} & -16492 & -16484 & -16490 & -16489 & -16473 \\
    China & SSE Composite Index & -27189 & -28256 & -28293 & -28283 & -28292 & \textbf{-28297} & -28296 & -28296 & -28272 \\
    Colombia & IGBC  & -23353 & \textbf{-24459} & -24445 & -24384 & -24448 & -24430 & -24450 & -24453 & -24433 \\
    Croatia & CROBEX & -27262 & -29793 & \textbf{-29808} & -29583 & -29803 & -29798 & -29803 & -29797 & -29779 \\
    Cyprus & Cyprus Main Market Index & -14209 & -15290 & -15338 & -15324 & \textbf{-15351} & -15347 & -15350 & -15348 & -15327 \\
    Czech Republic & PX    & -29852 & \textbf{-31001} & -30975 & -30891 & -30993 & -30952 & -30999 & -30996 & -30972 \\
    Egypt & EGX 30 Index & -25649 & -26481 & -26480 & -26436 & -26474 & -26471 & -26478 & \textbf{-26481} & -26460 \\
    Estonia & OMXT  & -28286 & -31412 & \textbf{-31448} & -31234 & -31440 & -31440 & -31428 & -31423 & -31429 \\
    EuroStoxx & Euro Stoxx 50 & -31051 & -32015 & -32036 & -32009 & -32028 & -32035 & \textbf{-32038} & -32038 & -32008 \\
    Finland & OMXH25 & -27102 & -28223 & -28256 & -28215 & -28248 & -28257 & -28261 & \textbf{-28261} & -28229 \\
    France & CAC 40 & -29831 & -30605 & \textbf{-30617} & -30588 & -30608 & -30613 & -30607 & -30610 & -30585 \\
    GB    & FTSE 100 & -31681 & -32574 & \textbf{-32586} & -32548 & -32578 & -32581 & -32583 & -32585 & -32555 \\
    Germany & DAX   & -29242 & -29975 & -30007 & \textbf{-30026} & -30018 & -30012 & -30017 & -30023 & -30002 \\
    Greece & Athex Composite Share Price Index & -26019 & -26881 & -26905 & -26881 & -26898 & \textbf{-26906} & -26904 & -26905 & -26871 \\
    Hong Kong & Hang Seng & -27582 & -28849 & -28859 & -28826 & -28850 & -28852 & \textbf{-28880} & -28879 & -28853 \\
    Hungary & Budapest SE & -27588 & \textbf{-28856} & -28815 & -28728 & -28841 & -28789 & -28843 & -28845 & -28822 \\
    India & Nifty 50 & -28555 & -29507 & -29514 & -29479 & -29507 & -29506 & -29510 & \textbf{-29515} & -29488 \\
    India & BSE Sensex & -28542 & -29432 & \textbf{-29444} & -29413 & -29436 & -29439 & -29438 & -29438 & -29408 \\
    Indonesia & IDX Composite & -27671 & -29108 & -29118 & -29022 & -29113 & -29107 & \textbf{-29122} & -29121 & -29089 \\
    Ireland & ISEQ Overall Index & -30346 & -31544 & -31550 & -31462 & -31551 & -31534 & \textbf{-31557} & -31556 & -31524 \\
    Israel & TA 35 & -30324 & -30959 & \textbf{-30972} & -30964 & -30965 & -30972 & -30969 & -30966 & -30947 \\
    Italy & FTSE MIB & -27538 & -28243 & -28270 & -28264 & -28266 & \textbf{-28272} & -28264 & -28266 & -28244 \\
    Japan & Topix & -29242 & \textbf{-29901} & -29892 & -29866 & -29894 & -29882 & -29897 & -29891 & -29887 \\
    Kazakhstan & KASE Index & -18181 & -21793 & -21868 & -22102 & \textbf{-22125} & -21874 & -21811 & -21793 & -22082 \\
    Kuwait & Kuwait 15 & -9328 & \textbf{-9490} & -9478 & -9464 & -9476 & -9473 & -9484 & -9485 & -9460 \\
    Latvia & OMXR  & -25002 & \textbf{-27252} & -27245 & -27060 & -27247 & -27228 & -27243 & -27240 & -27220 \\
    Lithuania & OMXV  & -27803 & \textbf{-29932} & -29925 & -29751 & -29926 & -29908 & -29926 & -29924 & -29904 \\
    Luxembourg & LuxX Index & -27758 & \textbf{-28612} & -28599 & -28548 & -28599 & -28587 & -28604 & -28608 & -28574 \\
    Malaysia & FTSE Bursa Malaysia KLCI & -29961 & \textbf{-33573} & -33547 & -33271 & -33563 & -33517 & -33554 & -33547 & -33541 \\
    Mauritius & SEMDEX & -37297 & \textbf{-40627} & -40603 & -40291 & -40621 & -40575 & -40606 & -40599 & -40597 \\
    Mexico & IPC   & -29715 & -30845 & -30846 & -30787 & -30840 & -30837 & -30853 & \textbf{-30853} & -30818 \\
    Morocco & MASI  & -27289 & -28274 & \textbf{-28287} & -28238 & -28278 & -28284 & -28281 & -28279 & -28253 \\
    Namibia & NSX Overall Index & -20695 & \textbf{-21191} & -21186 & -21160 & -21179 & -21180 & -21182 & -21184 & -21152 \\
    Netherlands & AEX   & -29935 & -31013 & \textbf{-31067} & -31014 & -31059 & -31061 & -31063 & -31062 & -31033 \\
    New Zealand & NZX 50 Index & -30046 & -30588 & -30592 & -30569 & -30591 & -30584 & \textbf{-30595} & -30594 & -30573 \\
    Norway & OBX Index & -25137 & -26069 & -26071 & -26009 & -26071 & -26058 & \textbf{-26073} & -26072 & -26047 \\
    Oman  & MSM 30 & -29094 & \textbf{-35103} & -35051 & -34644 & -35087 & -35009 & -35059 & -35050 & -35062 \\
    Pakistan & KSE 100 Index & -28221 & -29423 & -29480 & -29436 & -29475 & \textbf{-29485} & -29472 & -29472 & -29465 \\
    Peru  & S\&P Lima General Index & -29943 & \textbf{-31607} & -31596 & -31478 & -31599 & -31578 & -31604 & -31603 & -31574 \\
    Philippines & PSEi Composite & -28976 & \textbf{-30139} & -30114 & -30024 & -30126 & -30094 & -30136 & -30136 & -30103 \\
    Poland & WIG   & -29926 & -30667 & -30687 & -30673 & -30680 & \textbf{-30687} & -30677 & -30682 & -30650 \\
    Portugal & PSI 20 & -31460 & -32323 & \textbf{-32341} & -32308 & -32333 & -32337 & -32329 & -32335 & -32321 \\
    Qatar & QE 20 Index & -22476 & -30281 & \textbf{-30283} & -29781 & -30281 & -30261 & -30218 & -30203 & -30262 \\
    Romania & BET 10 & -26985 & -28534 & \textbf{-28560} & -28475 & -28552 & -28557 & -28558 & -28558 & -28528 \\
    Russia & MICEX & -22351 & \textbf{-24530} & -24519 & -24346 & -24526 & -24498 & -24528 & -24525 & -24508 \\
    Russia & RTSI  & -15750 & -17622 & -17622 & -17522 & -17618 & -17608 & -17612 & -17610 & \textbf{-17625} \\
    Saudi Arabia & Tadawul All Share & -28756 & -31389 & -31452 & -31278 & -31444 & \textbf{-31452} & -31413 & -31404 & -31418 \\
    Serbia & BELEX & -17957 & \textbf{-19312} & -19310 & -19199 & -19305 & -19300 & -19311 & -19310 & -19286 \\
    Singapore & STI Index & -27785 & -28609 & \textbf{-28619} & -28579 & -28611 & -28615 & -28616 & -28618 & -28585 \\
    South Africa & FTSE/JSE All-Share Index & -30917 & -31679 & -31675 & -31623 & -31677 & -31662 & \textbf{-31684} & -31683 & -31650 \\
    South Korea & KOSPI & -27587 & -28812 & -28891 & -28882 & \textbf{-28909} & -28907 & -28905 & -28904 & -28870 \\
    Spain & IBEX 35 & -29246 & -29946 & -29958 & -29949 & -29950 & -29956 & \textbf{-29966} & -29964 & -29937 \\
    Sri Lanka & CSE All-Share & -30574 & -32956 & -32967 & -32792 & -32968 & -32948 & \textbf{-32986} & -32985 & -32962 \\
    Sweden & OMXS30 & -28939 & -29605 & -29621 & -29604 & -29613 & \textbf{-29621} & -29614 & -29617 & -29590 \\
    Switzerland & SMI   & -31342 & -32327 & \textbf{-32334} & -32289 & -32329 & -32324 & -32331 & -32329 & -32310 \\
    Taiwan & Taiwan Weighted & -30025 & -30732 & -30788 & \textbf{-30801} & -30800 & -30099 & -30782 & -30782 & -30773 \\
    Thailand & SET   & -27910 & -29051 & -29061 & -29029 & -29052 & -29056 & -29076 & \textbf{-29076} & -29042 \\
    Tunesia & Tunindex & -33483 & \textbf{-38597} & -38457 & -38198 & -38581 & -37910 & -38578 & -38575 & -38579 \\
    Turkey & BIST 100 & -24071 & -25227 & \textbf{-25228} & -25167 & -25222 & -25219 & -25228 & -25228 & -25195 \\
    United Arab Emirates & DFM   & -18834 & -19788 & \textbf{-19800} & -19745 & -19792 & -19797 & -19794 & -19792 & -19776 \\
    United Arab Emirates & Abu Dhabi & -26165 & -27755 & -27792 & -27722 & -27787 & \textbf{-27792} & -27784 & -27781 & -27769 \\
    USA   & Dow 30 & -31969 & -33132 & -33153 & -33128 & -33145 & -33149 & -33168 & \textbf{-33168} & -33140 \\
    USA   & S\&P 500 & -31339 & -32575 & -32604 & -32573 & -32597 & -32602 & \textbf{-32616} & -32615 & -32587 \\
    USA   & Nasdaq & -27061 & -28237 & -28314 & -28339 & -28346 & -28329 & \textbf{-28357} & -28356 & -28323 \\
    Venezuela & IBC   & -23729 & -26343 & \textbf{-26421} & -26336 & -26416 & -26420 & -26382 & -26372 & -26405 \\
    Vietnam & HNX 30 & -14823 & -15575 & -15612 & -15577 & -15611 & -14764 & \textbf{-15615} & -15615 & -15587 \\
    Zambia & All Share Index & -19269 & -23362 & -23441 & -23861 & \textbf{-23872} & -23456 & -23631 & -23615 & -23838 \\
    \hline
    \end{tabular}%
		\end{tiny}
		  \caption{BIC of fitted distributions for daily log-returns, from 01/03/1997 until 11/02/2017.}
  \label{tab:dBIC}%
\end{table}%

\begin{table}[htbp]
  \centering
		\begin{tabular}{rrrrrrrrrr}
    \hline
    &     N     & St    & NIG  & V$\Gamma$  & GH    & Meix & 2St12 & 2St39 & 3St \\
    \hline
    KS  & 0 & 0 &  1 &  0 & 6 & 0 & 5 & 8 & 58 \\
    AD  & 0 &  0 & 1  & 0 & 1 & 0 &  1& 1  & 74 \\
    AIC  & 0 & 3 & 9  & 1 & 19 & 5 & 14 & 7  & 20 \\
    BIC  & 0 &  19 & 18  & 3 & 4 & 8 & 19  &  6 & 1 \\
    \hline
    \end{tabular}%
		  \caption{Number of lowest statistics per distribution, for daily returns.}
  \label{tab:minsdaily}%
\end{table}%

\begin{figure}
	\centering
\begin{tabular}{ccc}
{\includegraphics[width=.33\textwidth, height=5cm]{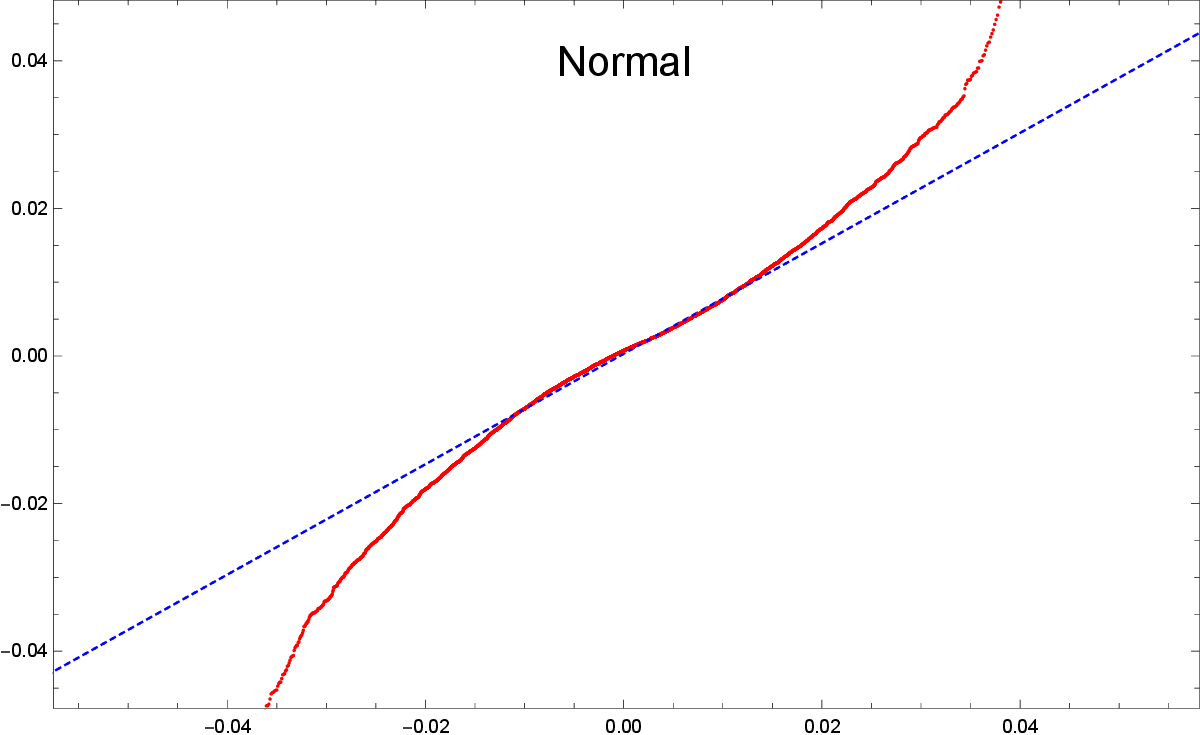}}&
  {\includegraphics[width=.33\textwidth,
  height=5cm]{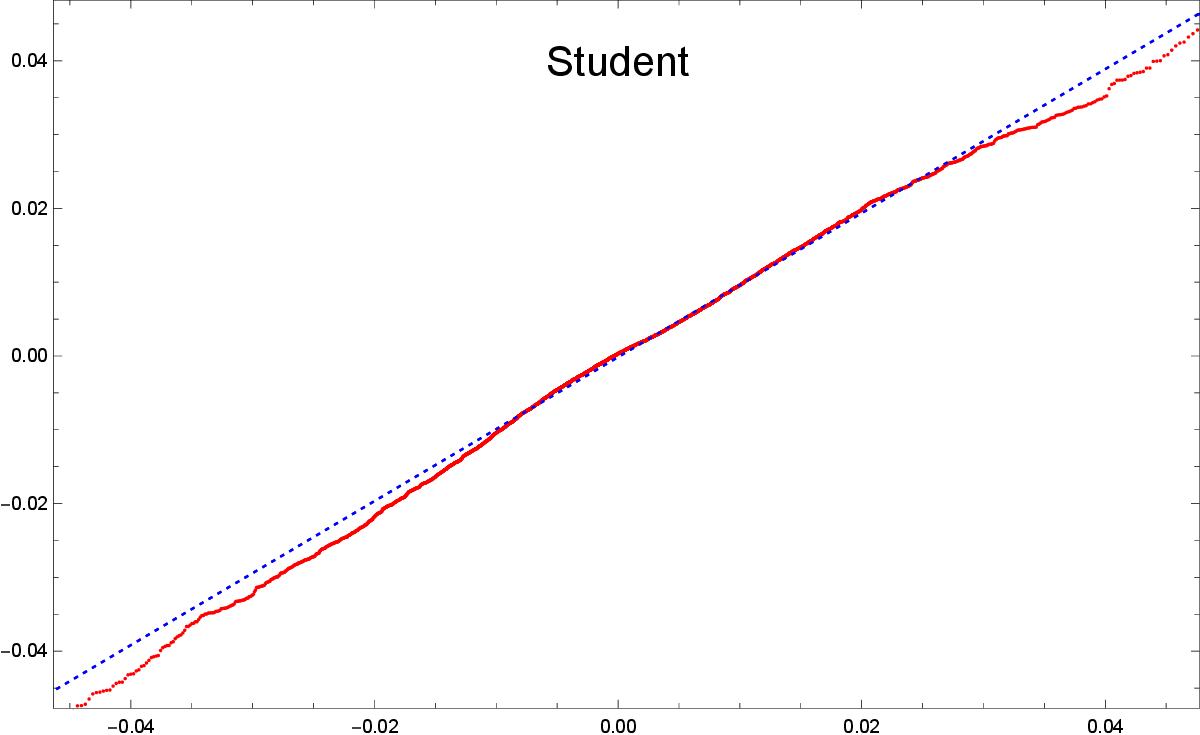}} &
  {\includegraphics[width=.33\textwidth,
  height=5cm]{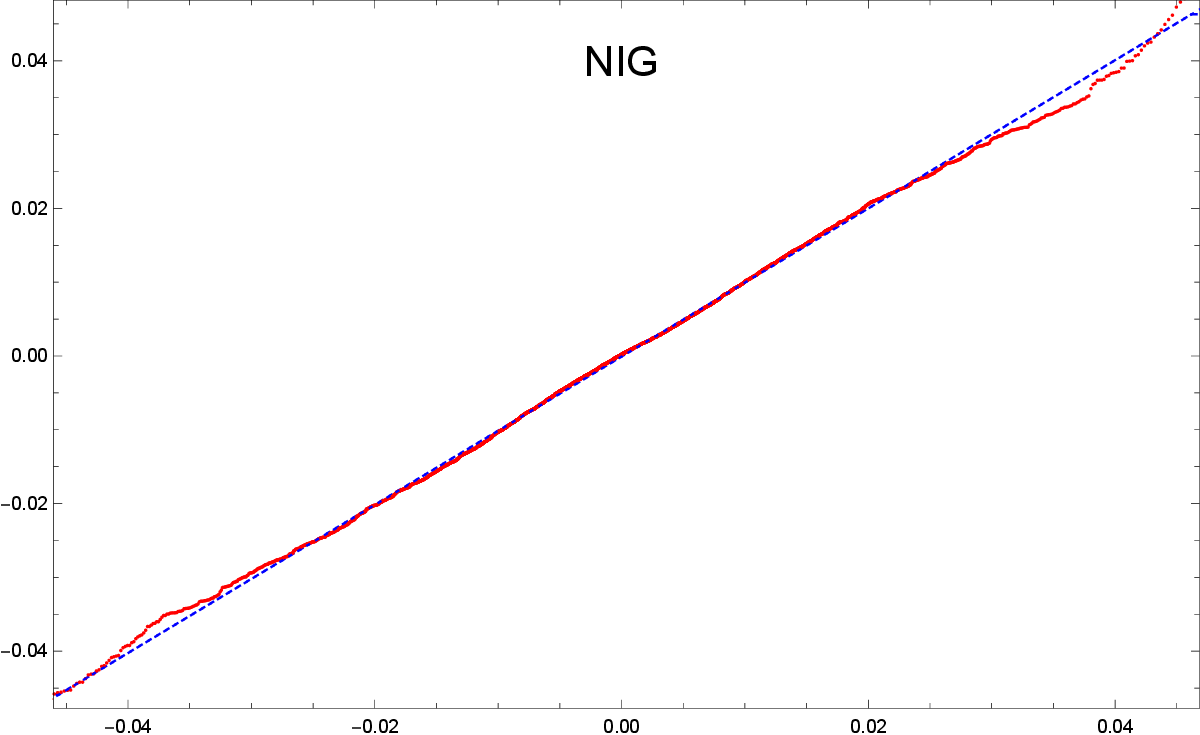}}\\
  {\includegraphics[width=.33\textwidth, height=5cm]{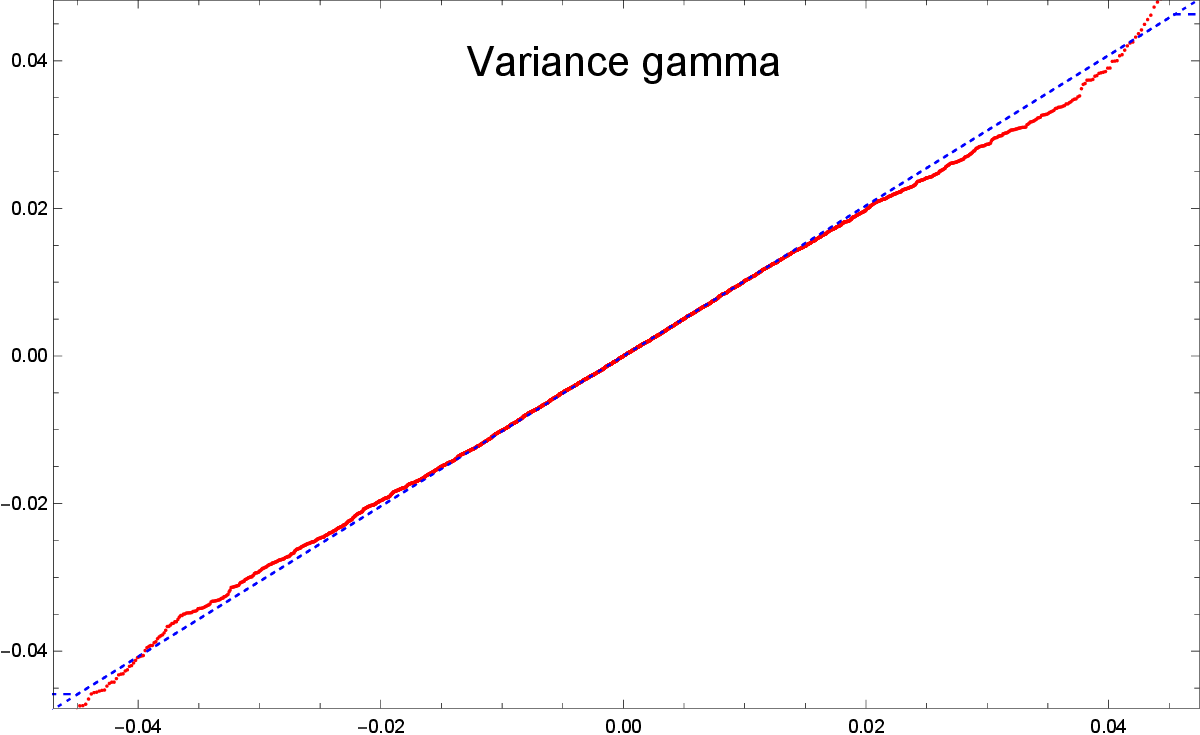}} &
  {\includegraphics[width=.33\textwidth,
  height=5cm]{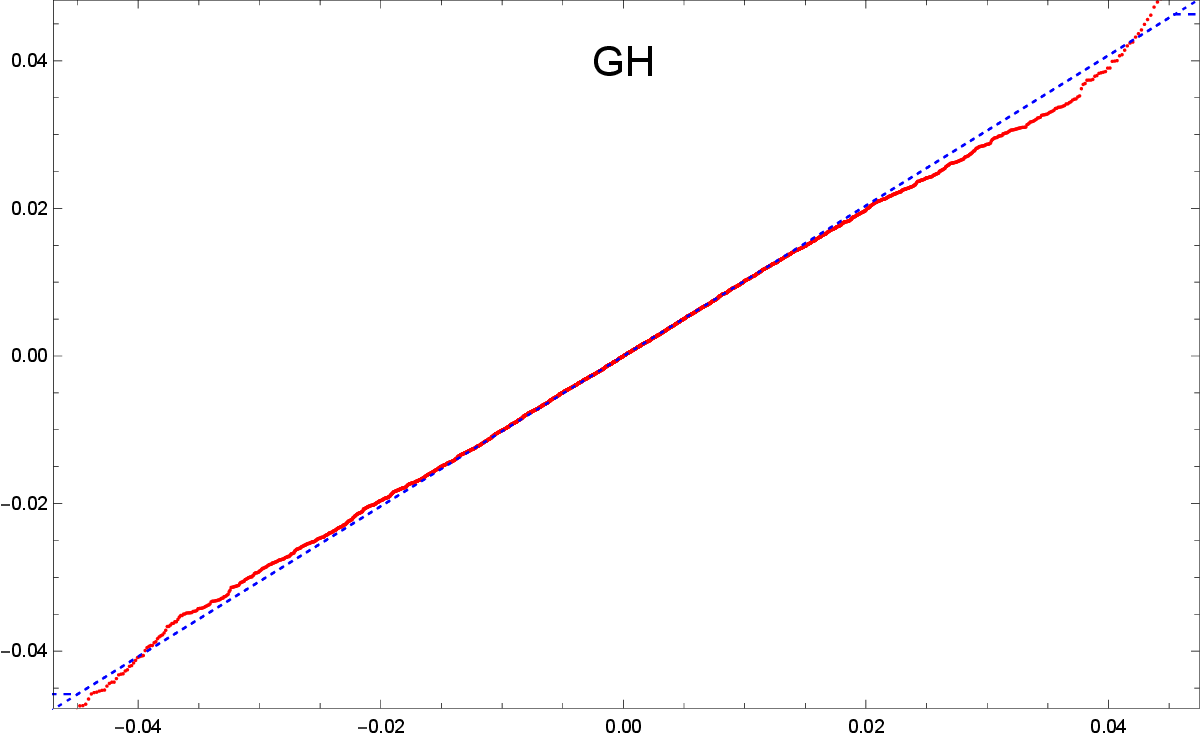}} &
  {\includegraphics[width=.33\textwidth,
  height=5cm]{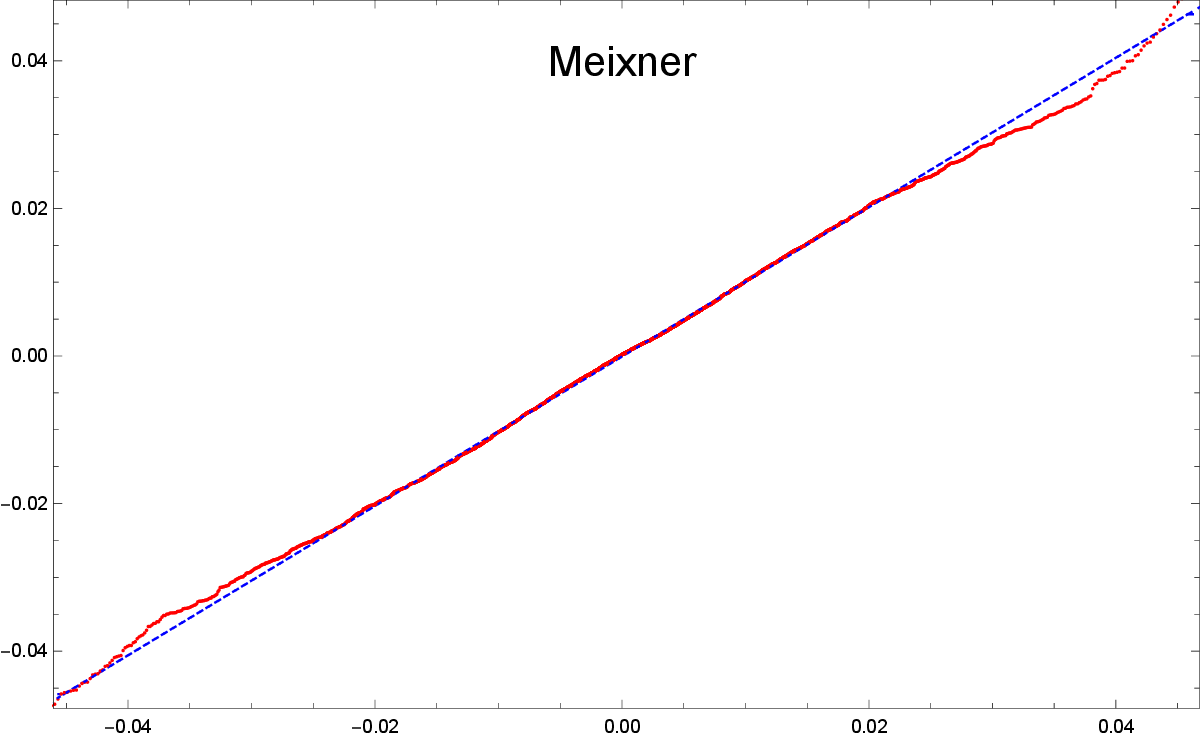}}\\
  {\includegraphics[width=.33\textwidth, height=5cm]{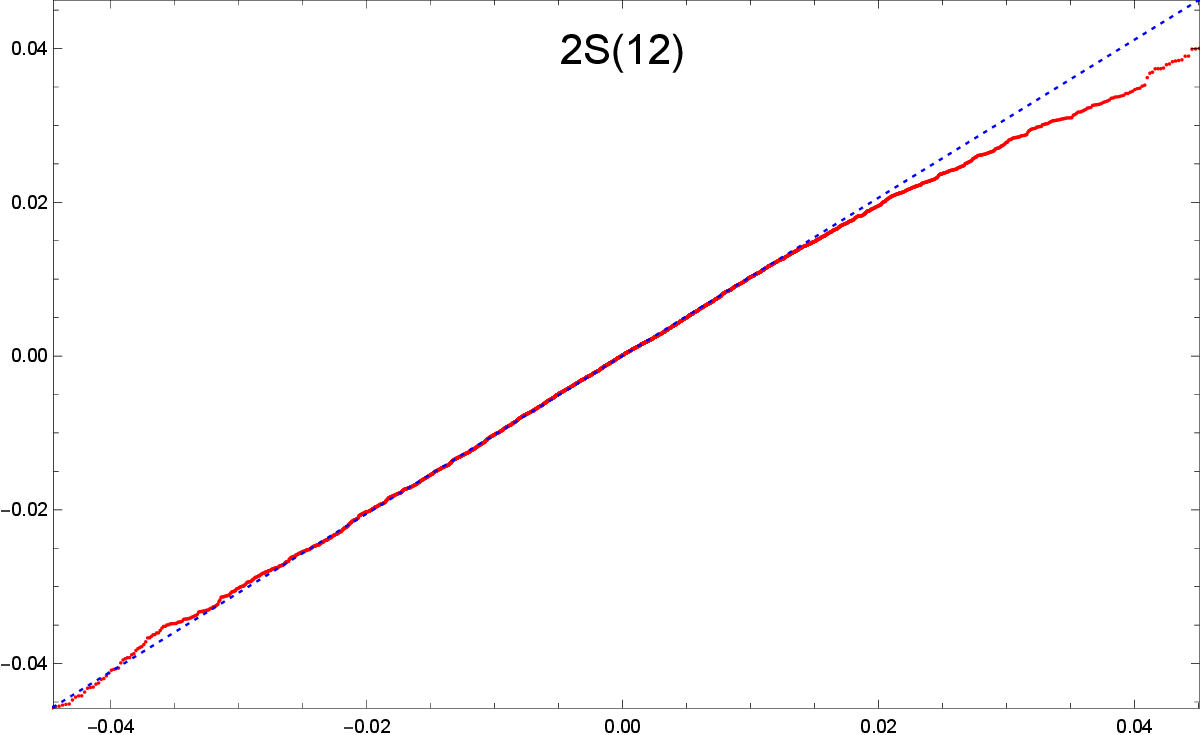}} &
  {\includegraphics[width=.33\textwidth,
  height=5cm]{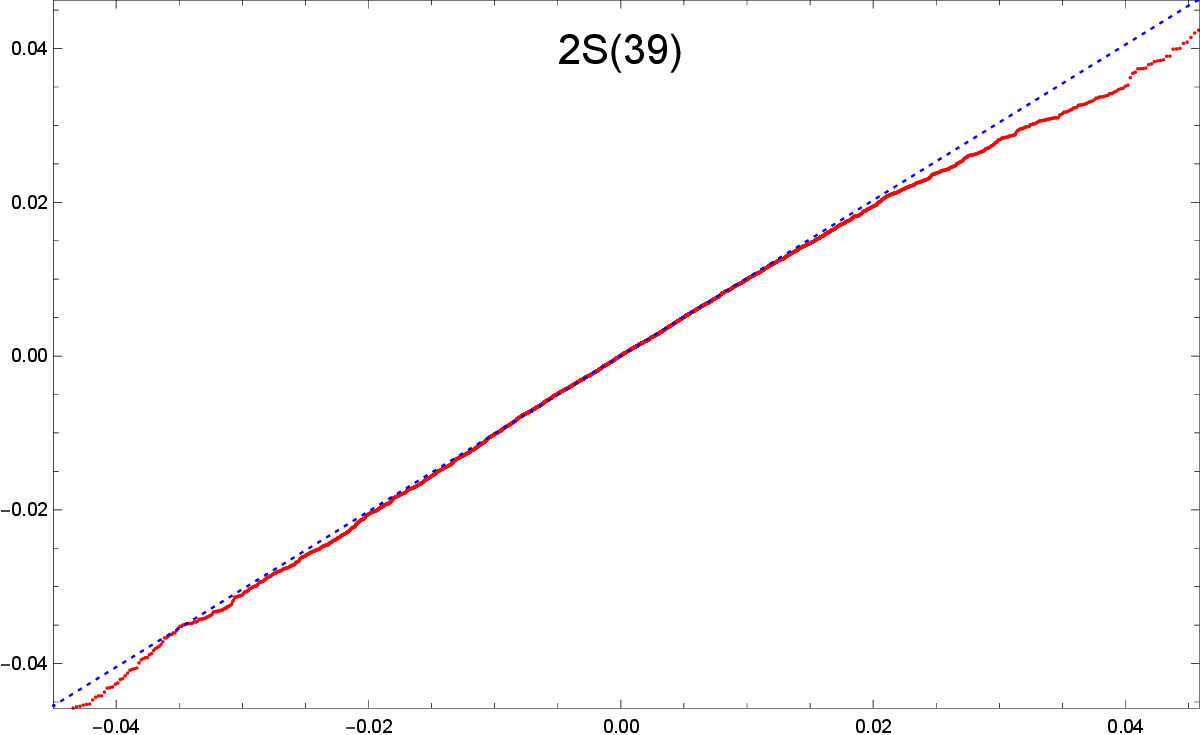}} &
  {\includegraphics[width=.33\textwidth,
  height=5cm]{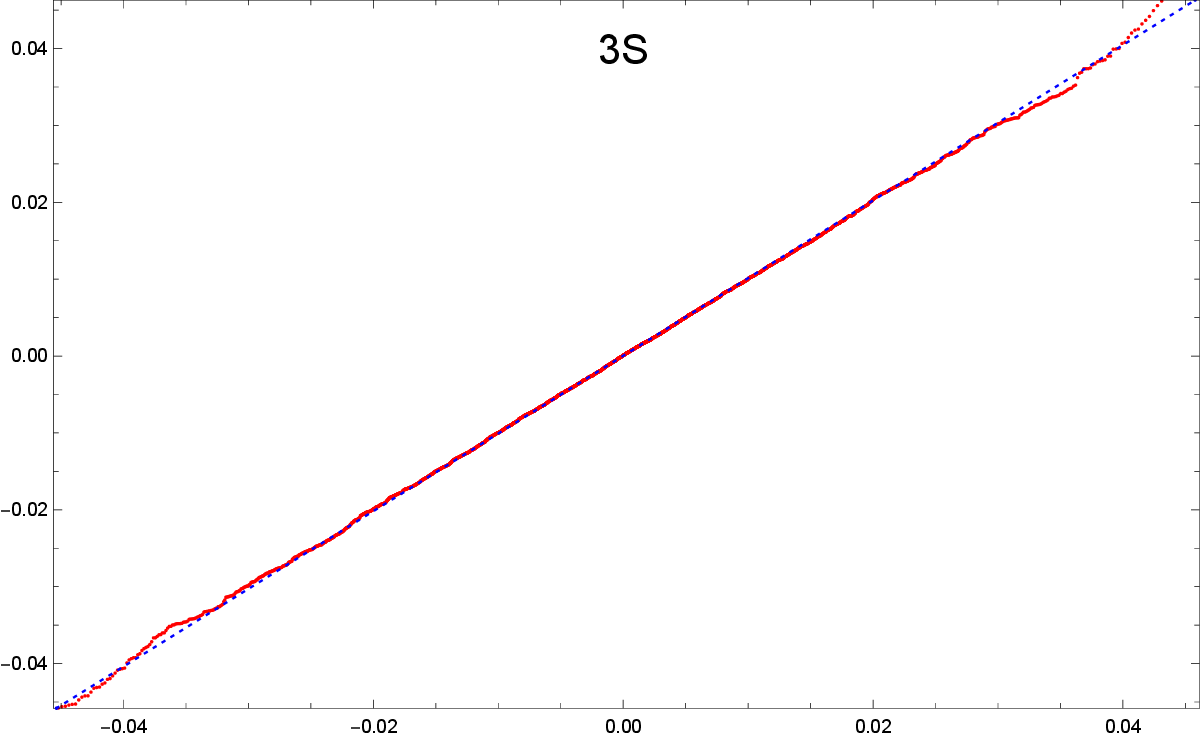}}\\
  \end{tabular}
  \caption{QQ plots of empirical quantiles for daily DAX log-returns from 01/03/1997 until 11/02/2017 versus model distributions.}
  \label{pic:qqDAXd}
\end{figure}

To this point, we have only investigated which of the models fits best, not whether the fits are qualitatively good. To analyze this, we use QQ plots. We discuss one example in some detail (others are available on request).
Figure~\ref{pic:qqDAXd} plots the empirical quantiles of the German DAX index against the theoretical quantiles, each panel corresponding to one model. The figure makes it obvious that the normal model yields a bad fit. The Student's $t$ model has some difficulty in capturing the skewness in the data. Each of the other distributions fits quite well, making it hard to find the best by visual inspection alone. For this index, the KS, AD and AIC select the 3St, but the BIC selects the $V\Gamma$. In fact, visually the 3St offers an excellent fit for this sample.

We now turn to the case of hourly data. We will proceed in parallel to what we have done already for the daily data.

\begin{table}[htbp]
  \hspace{-2.5cm}
\begin{tiny}
    \begin{tabular}{rrrrrrrrrrr}
    \hline
    Country & Index & N     & St    & NIG  & V$\Gamma$  & GH    & Meix & 2St12 & 2St39 & 3St \\
    \hline
   Argentina  &  MERVAL  &       0.1216    &    0.0143    &    0.0112    &    0.0243    &    0.0112    &    0.0111    &    0.0129    &    0.0142    & \textbf{   0.0084   } \\
     Australia  &  S\&P/ASX 200  &       0.1218    &    0.0159    &    0.0146    &    0.0329    &    0.0158    &    0.0149    &    0.0179    &    0.0190    & \textbf{   0.0098   } \\
     Australia  &  All Ordinaries  &       0.1236    &    0.0199    &    0.0187    &    0.0301    &    0.0198    &    0.0183    &    0.0205    &    0.0206    & \textbf{   0.0096   } \\
     Austria  &  ATX  &       0.0806    &    0.0353    &    0.0116    &    0.0186    &    0.0090    &    0.0128    &    0.0086    &    0.0098    & \textbf{   0.0079   } \\
     Bahrain  &  All Share  &       0.1135    &    0.0222    &    0.0187    &    0.0206    &    0.0222    &    0.0177    &    0.0171    &    0.0173    & \textbf{   0.0160   } \\
     Belgium  &  BEL 20  &       0.0866    &    0.0120    &    0.0132    &    0.0204    &    0.0122    &    0.0154    &    0.0106    &    0.0118    & \textbf{   0.0089   } \\
     Brazil  &  Bovespa  &       0.1021    &    0.0122    &    0.0129    &    0.0186    &    0.0116    &    0.0138    &    0.0099    &    0.0096    & \textbf{   0.0069   } \\
     Bulgaria  &  SOFIX  &       0.1015    &    0.0135    &    0.0099    &    0.0188    &    0.0130    &    0.0120    &    0.0147    &    0.0144    & \textbf{   0.0087   } \\
     Canada  &  S\&P/TSX 60  &       0.1132    &    0.0434    &    0.0232    &    0.0200    &    0.0311    &    0.0218    &    0.0151    &    0.0159    & \textbf{   0.0094   } \\
     Canada  &  S\&P/TSX Composite  &       0.1132    &    0.0420    &    0.0247    &    0.0177    &    0.0312    &    0.0233    &    0.0113    &    0.0120    & \textbf{   0.0088   } \\
     Chile  &  IPSA  &       0.1397    &    0.0084    &    0.0131    &    0.0205    &    0.0085    &    0.0162    &    0.0094    &    0.0090    & \textbf{   0.0076   } \\
     China  &  CSI 300  &       0.1293    &    0.0844    &    0.0793    &    0.0504    &    0.0848    &    0.0786    &    0.0141    &    0.0136    & \textbf{   0.0134   } \\
     China  &  SSE Composite Index  &       0.1321    &    0.0754    &    0.0696    &    0.0501    &    0.0749    &    0.0687    &    0.0101    & \textbf{   0.0098   } &    0.0117    \\
     Colombia  &  IGBC  &       0.0752    &    0.0104    &    0.0090    &    0.0182    &    0.0103    &    0.0094    &    0.0094    &    0.0094    & \textbf{   0.0086   } \\
     Croatia  &  CROBEX  &       0.1191    &    0.0141    &    0.0112    &    0.0227    &    0.0123    &    0.0130    &    0.0090    & \textbf{   0.0089   } &    0.0096    \\
     Cyprus  &  Cyprus Main Market Index  &       0.1039    &    0.0325    &    0.0273    &    0.0633    &    0.0327    & \textbf{   0.0272   } &    0.0324    &    0.0326    &    0.0321    \\
     Czech Republic  &  PX   &       0.0684    &    0.0160    &    0.0159    &    0.0150    &    0.0140    &    0.0162    &    0.0121    &    0.0125    & \textbf{   0.0078   } \\
     Egypt  &  EGX 30 Index  &       0.1860    &    0.0404    &    0.0176    &    0.0489    &    0.0117    &    0.0136    &    0.0152    &    0.0176    & \textbf{   0.0079   } \\
     Estonia  &  OMXT  &       0.0573    &    0.0214    &    0.0159    &    0.0108    &    0.0213    &    0.0139    &    0.0259    &    0.0272    & \textbf{   0.0092   } \\
     EuroStoxx  &  Euro Stoxx 50  &       0.0874    &    0.0195    &    0.0142    &    0.0215    &    0.0145    &    0.0167    &    0.0164    &    0.0177    & \textbf{   0.0089   } \\
     Finland  &  OMXH25  &       0.1050    &    0.0138    &    0.0183    &    0.0272    &    0.0137    &    0.0201    &    0.0128    & \textbf{   0.0123   } &    0.0125    \\
     France  &  CAC 40  &       0.0934    &    0.0106    &    0.0127    &    0.0232    &    0.0095    &    0.0151    &    0.0112    &    0.0121    & \textbf{   0.0076   } \\
     GB   &  FTSE 100  &       0.0859    &    0.0310    &    0.0108    &    0.0159    &    0.0079    &    0.0129    & \textbf{   0.0072   } &    0.0082    &    0.0074    \\
     Germany  &  DAX  &       0.1027    &    0.0106    &    0.0173    &    0.0305    &    0.0107    &    0.0198    &    0.0111    &    0.0103    & \textbf{   0.0097   } \\
     Greece  &  Athex Composite Share Price Index  &       0.0702    &    0.0124    &    0.0110    &    0.0141    &    0.0122    &    0.0126    &    0.0130    &    0.0096    & \textbf{   0.0088   } \\
     Hong Kong  &  Hang Seng  &       0.1223    &    0.0418    &    0.0382    &    0.0328    &    0.0415    &    0.0377    &    0.0231    &    0.0237    & \textbf{   0.0072   } \\
     Hungary  &  Budapest SE  &       0.0731    &    0.0113    &    0.0109    &    0.0188    &    0.0113    &    0.0122    & \textbf{   0.0085   } &    0.0088    &    0.0108    \\
     India  &  Nifty 50  &       0.0919    &    0.0130    &    0.0102    &    0.0171    &    0.0128    &    0.0118    &    0.0104    &    0.0107    & \textbf{   0.0086   } \\
     India  &  BSE Sensex  &       0.0957    &    0.0156    &    0.0161    &    0.0268    &    0.0155    &    0.0177    &    0.0123    &    0.0133    & \textbf{   0.0122   } \\
     Indonesia  &  IDX Composite  &       0.1205    &    0.0253    &    0.0137    &    0.0237    &    0.0174    &    0.0148    &    0.0106    &    0.0108    & \textbf{   0.0105   } \\
     Ireland  &  ISEQ Overall Index  &       0.0747    &    0.0235    &    0.0142    &    0.0218    &    0.0123    &    0.0165    &    0.0096    &    0.0088    & \textbf{   0.0088   } \\
     Israel  &  TA 35  &       0.1074    &    0.0164    &    0.0144    &    0.0268    &    0.0110    &    0.0156    &    0.0122    &    0.0135    & \textbf{   0.0074   } \\
     Italy  &  FTSE MIB  &       0.0842    &    0.0111    &    0.0105    &    0.0145    &    0.0111    &    0.0124    &    0.0086    &    0.0090    & \textbf{   0.0066   } \\
     Japan  &  Topix  &       0.1729    &    0.0414    &    0.0168    &    0.0405    &    0.0228    &    0.0157    &    0.0226    &    0.0228    & \textbf{   0.0119   } \\
     Kazakhstan  &  KASE Index  &       0.0765    &    0.0268    &    0.0149    &    0.0135    &    0.0261    &    0.0129    & \textbf{   0.0117   } &    0.0127    &    0.0120    \\
     Kuwait  &  Kuwait 15  &       0.0746    &    0.0286    &    0.0239    &    0.0199    &    0.0286    &    0.0216    &    0.0246    &    0.0215    & \textbf{   0.0091   } \\
     Latvia  &  OMXR  &       0.1579    &    0.0299    &    0.0184    &    0.0406    &    0.0295    &    0.0171    &    0.0126    &    0.0135    & \textbf{   0.0096   } \\
     Lithuania  &  OMXV  &       0.4093    &    0.0134    &    0.0476    &    0.0879    &    0.0134    &    0.0497    &    0.0145    & \textbf{   0.0129   } &    0.0207    \\
     Luxembourg  &  LuxX Index  &       0.1399    &    0.0413    &    0.0180    &    0.0303    &    0.0207    &    0.0172    &    0.0168    &    0.0187    & \textbf{   0.0083   } \\
     Malaysia  &  FTSE Bursa Malaysia KLCI  &       0.1009    &    0.0174    &    0.0131    &    0.0236    &    0.0171    &    0.0154    &    0.0149    &    0.0152    & \textbf{   0.0079   } \\
     Mauritius  &  SEMDEX  &       0.1276    &    0.0475    &    0.0302    &    0.0330    &    0.0420    &    0.5024    &    0.0150    &    0.0148    & \textbf{   0.0107   } \\
     Mexico  &  IPC  &       0.0826    &    0.0215    &    0.0106    &    0.0164    &    0.0165    &    0.0105    &    0.0118    &    0.0127    & \textbf{   0.0075   } \\
     Morocco  &  MASI  &       0.0972    &    0.0480    &    0.0215    &    0.0208    &    0.0275    &    0.0196    &    0.0106    &    0.0112    & \textbf{   0.0065   } \\
     Namibia  &  NSX Overall Index  &       0.1631    &    0.0382    &    0.0414    &    0.0354    &    0.0432    &    0.0412    &    0.0185    &    0.0198    & \textbf{   0.0110   } \\
     Netherlands  &  AEX  &       0.0966    &    0.0154    &    0.0124    &    0.0196    &    0.0153    &    0.0126    &    0.0155    &    0.0155    & \textbf{   0.0101   } \\
     New Zealand  &  NZX 50 Index  &       0.1084    &    0.0147    &    0.0198    &    0.0279    &    0.0147    &    0.0216    &    0.0170    &    0.0173    & \textbf{   0.0121   } \\
     Norway  &  OBX Index  &       0.0835    &    0.0202    &    0.0144    &    0.0211    &    0.0115    &    0.0154    & \textbf{   0.0095   } &    0.0102    &    0.0096    \\
     Oman  &  MSM 30  &       0.1040    &    0.0760    &    0.0718    &    0.0612    &    0.0762    &    0.0983    &    0.0247    &    0.0256    & \textbf{   0.0163   } \\
     Pakistan  &  KSE 100 Index  &       0.0816    &    0.0181    &    0.0154    &    0.0149    &    0.0181    &    0.0144    &    0.0139    &    0.0130    & \textbf{   0.0117   } \\
     Peru  &  S\&P Lima General Index  &       0.0902    &    0.0302    &    0.0236    &    0.0261    &    0.0302    &    0.0220    & \textbf{   0.0148   } &    0.0152    &    0.0154    \\
     Philippines  &  PSEi Composite  &       0.1194    &    0.0445    &    0.0358    &    0.0122    &    0.0442    &    0.0351    &    0.0175    &    0.0182    & \textbf{   0.0091   } \\
     Poland  &  WIG  &       0.0773    &    0.0096    &    0.0119    &    0.0193    &    0.0093    &    0.0142    &    0.0091    &    0.0092    & \textbf{   0.0070   } \\
     Portugal  &  PSI 20  &       0.0881    &    0.0113    &    0.0128    &    0.0200    &    0.0130    &    0.0137    &    0.0123    &    0.0127    & \textbf{   0.0064   } \\
     Qatar  &  QE 20 Index  &       0.1124    &    0.0158    &    0.0169    &    0.0247    &    0.0146    &    0.0199    &    0.0144    &    0.0134    & \textbf{   0.0116   } \\
     Romania  &  BET 10  &       0.0935    &    0.0350    &    0.0134    &    0.0203    &    0.0105    &    0.0153    &    0.0094    &    0.0100    & \textbf{   0.0084   } \\
     Russia  &  MICEX  &       0.0653    &    0.0118    &    0.0108    &    0.0145    &    0.0118    &    0.0111    &    0.0127    &    0.0122    & \textbf{   0.0100   } \\
     Russia  &  RTSI  &       0.0694    &    0.0274    &    0.0264    &    0.0395    &    0.0273    & \textbf{   0.0258   } &    0.0268    &    0.0265    &    0.0263    \\
     Saudi Arabia  &  Tadawul All Share  &       0.1116    &    0.0300    &    0.0269    &    0.0309    &    0.0302    &    0.0262    &    0.0292    &    0.0295    & \textbf{   0.0137   } \\
     Serbia  &  BELEX  &       0.0722    &    0.0195    & \textbf{   0.0169   } &    0.0305    &    0.0193    &    0.0813    &    0.0193    &    0.0189    &    0.0191    \\
     Singapore  &  STI Index  &       0.1079    &    0.0199    &    0.0167    &    0.0307    &    0.0138    &    0.0194    &    0.0134    &    0.0131    & \textbf{   0.0080   } \\
     South Africa  &  FTSE/JSE All-Share Index  &       0.0784    &    0.0106    &    0.0114    &    0.0205    &    0.0106    &    0.0135    &    0.0111    &    0.0116    & \textbf{   0.0096   } \\
     South Korea  &  KOSPI  &       0.1223    &    0.0135    &    0.0159    &    0.0302    &    0.0133    &    0.0169    &    0.0130    &    0.0130    & \textbf{   0.0128   } \\
     Spain  &  IBEX 35  &       0.0740    &    0.0338    &    0.0103    &    0.0145    &    0.0073    &    0.0117    &    0.0085    &    0.0085    & \textbf{   0.0071   } \\
     Sri Lanka  &  CSE All-Share  &       0.0604    &    0.0186    &    0.0144    &    0.0138    &    0.0206    &    0.0144    &    0.0163    &    0.0158    & \textbf{   0.0117   } \\
     Sweden  &  OMXS30  &       0.0902    &    0.0089    &    0.0108    &    0.0183    &    0.0089    &    0.0133    &    0.0100    &    0.0103    & \textbf{   0.0075   } \\
     Switzerland  &  SMI  &       0.0872    &    0.0170    &    0.0171    &    0.0242    &    0.0140    &    0.0183    &    0.0122    & \textbf{   0.0120   } &    0.0132    \\
     Taiwan  &  Taiwan Weighted  &       0.1110    &    0.0169    &    0.0172    &    0.0328    &    0.0166    &    0.0186    &    0.0126    &    0.0115    & \textbf{   0.0093   } \\
     Thailand  &  SET  &       0.0656    &    0.0109    & \textbf{   0.0084   } &    0.0136    &    0.0109    &    0.0096    &    0.0096    &    0.0100    &    0.0087    \\
     Tunesia  &  Tunindex  &       0.0497    &    0.0272    &    0.0247    &    0.0246    &    0.0273    &    0.0225    &    0.0309    &    0.0260    & \textbf{   0.0122   } \\
     Turkey  &  BIST 100  &       0.0928    &    0.0222    &    0.0201    &    0.0176    &    0.0222    &    0.0194    & \textbf{   0.0077   } &    0.0077    &    0.0078    \\
     United Arab Emirates  &  DFM  &       0.0665    &    0.0206    &    0.0189    &    0.0198    & \textbf{   0.0170   } &    0.0194    &    0.0172    &    0.0173    &    0.0173    \\
     United Arab Emirates  &  Abu Dhabi  &       0.0630    &    0.0247    &    0.0133    &    0.0161    &    0.0158    &    0.0139    &    0.0216    &    0.0225    & \textbf{   0.0113   } \\
     USA  &  Dow 30  &       0.1167    &    0.0476    &    0.0270    &    0.0314    &    0.0305    &    0.0258    &    0.0133    &    0.0140    & \textbf{   0.0076   } \\
     USA  &  S\&P 500  &       0.1053    &    0.0420    &    0.0370    &    0.0190    &    0.0419    &    0.0359    &    0.0149    &    0.0150    & \textbf{   0.0074   } \\
     USA  &  Nasdaq  &       0.1157    &    0.0371    &    0.0309    &    0.0291    &    0.0373    &    0.0302    &    0.0208    &    0.0208    & \textbf{   0.0061   } \\
     Venezuela  &  IBC  &       0.1638    &    0.0842    &    0.0595    &    0.0671    &    0.0437    &    0.0551    &    0.0302    &    0.0315    & \textbf{   0.0205   } \\
     Vietnam  &  HNX 30  &       0.0380    &    0.0154    &    0.0163    &    0.0170    &    0.0155    &    0.0163    &    0.0165    &    0.0158    & \textbf{   0.0133   } \\
     Zambia  &  All Share Index  &       0.1744    &    0.1505    &    0.1420    &    0.1635    &    0.1507    & \textbf{   0.1404   } &    0.1556    &    0.1583    &    0.1443    \\
     \hline
    \end{tabular}%
				\end{tiny}
		  \caption{KS distance between the empirical and fitted distributions for hourly log-returns, from 11/02/2016 1pm until 11/02/2017 12pm.}
  \label{tab:hKS}%
\end{table}%

\begin{table}[htbp]
  \hspace{-2.5cm}
\begin{tiny}
    \begin{tabular}{rrrrrrrrrrr}
    \hline
    Country & Index & N     & St    & NIG  & V$\Gamma$  & GH    & Meix & 2St12 & 2St39 & 3St \\
    \hline
   Argentina  &  MERVAL  &       64.6049    &         0.5361    &         0.1780    &         1.8260    &         0.1760    &         0.1986    &         0.3235    &         0.3895    & \textbf{        0.0821   } \\
     Australia  &  S\&P/ASX 200  &       63.5397    &         0.3335    &         0.3804    &         3.2506    &         0.3377    &         0.5287    &         0.3185    &         0.3604    & \textbf{        0.1491   } \\
     Australia  &  All Ordinaries  &       66.2863    &         0.4300    &         0.4706    &         2.7968    &         0.4304    &         0.6198    &         0.4544    &         0.5103    & \textbf{        0.1489   } \\
     Austria  &  ATX  &       34.2717    &         4.6665    &         0.3640    &         1.4352    &         0.2572    &         0.5715    &         0.1467    &         0.1534    & \textbf{        0.1192   } \\
     Bahrain  &  All Share  &       25.6256    &         0.4881    &         0.2797    &         0.6277    &         0.4773    &         0.3057    &         0.2821    &         0.2904    & \textbf{        0.1212   } \\
     Belgium  &  BEL 20  &       45.9048    &         0.5090    &         0.6062    &         1.7749    &         0.5361    &         0.8773    &         0.2371    &         0.2586    & \textbf{        0.0968   } \\
     Brazil  &  Bovespa  &       51.2750    &         0.3994    &         0.4146    &         1.3465    &         0.3724    &         0.5879    &         0.2536    &         0.2546    & \textbf{        0.1065   } \\
     Bulgaria  &  SOFIX  &       48.9907    &         0.5919    &         0.3189    &         1.0089    &         0.2980    &         0.4765    &         0.3773    &         0.3862    & \textbf{        0.1267   } \\
     Canada  &  S\&P/TSX 60  &       50.9177    &         6.6251    &         1.6520    &         0.7087    &         2.6457    &         1.4268    &         0.2805    &         0.3083    & \textbf{        0.1297   } \\
     Canada  &  S\&P/TSX Composite  &       53.2639    &         6.8481    &         2.0130    &         0.4552    &         3.1116    &         1.7701    &         0.2934    &         0.3242    & \textbf{        0.1203   } \\
     Chile  &  IPSA  &     108.9088    &         0.1597    &         0.7635    &         2.8535    &         0.1618    &         1.2275    &         0.1679    &         0.1625    & \textbf{        0.1436   } \\
     China  &  CSI 300  &       48.5645    &       13.2797    &       12.3525    &         3.1807    &       13.1634    &       12.2292    &         0.1958    &         0.1901    & \textbf{        0.1836   } \\
     China  &  SSE Composite Index  &       50.7337    &       10.8647    &       10.1643    &         4.0847    &       10.8032    &       10.0208    &         0.3495    &         0.3384    & \textbf{        0.2247   } \\
     Colombia  &  IGBC  &       23.7075    &         0.2651    &         0.1282    &         0.7257    &         0.2619    &         0.1705    &         0.1180    &         0.1153    & \textbf{        0.1094   } \\
     Croatia  &  CROBEX  &       72.5397    &         0.4067    &         0.4427    &         1.6846    &         0.3806    &         0.6906    &         0.2474    &         0.2563    & \textbf{        0.1219   } \\
     Cyprus  &  Cyprus Main Market Index  &       36.1398    &         2.0526    &         1.3229    &         3.6741    &         2.0573    &         1.1618    &         0.7778    &         0.7696    & \textbf{        0.7537   } \\
     Czech Republic  &  PX   &       24.5275    &         1.0372    &         0.4686    &         0.9475    &         0.3700    &         0.6176    &         0.2050    &         0.2266    & \textbf{        0.0997   } \\
     Egypt  &  EGX 30 Index  &       77.1603    &         4.6331    &         0.3427    &         4.0783    &         0.1542    &         0.2031    &         0.3574    &         0.4418    & \textbf{        0.1426   } \\
     Estonia  &  OMXT  &       12.6155    &         0.7130    &         0.5272    &         0.4141    &         0.7040    &         0.5084    &         0.9256    &         1.0470    & \textbf{        0.1363   } \\
     EuroStoxx  &  Euro Stoxx 50  &       46.0165    &         0.9544    &         0.3670    &         1.4842    &         0.2258    &         0.5711    &         0.2789    &         0.3238    & \textbf{        0.1316   } \\
     Finland  &  OMXH25  &       56.0978    &         0.4647    &         0.9469    &         3.3088    &         0.4622    &         1.3291    &         0.3367    & \textbf{        0.3218   } &         0.3336    \\
     France  &  CAC 40  &       54.4623    &         0.1968    &         0.4006    &         1.7717    &         0.1954    &         0.6354    &         0.1952    &         0.2309    & \textbf{        0.1273   } \\
     GB   &  FTSE 100  &       41.7813    &         4.6204    &         0.2726    &         1.5446    &         0.1695    &         0.4829    &         0.1246    &         0.1557    & \textbf{        0.0882   } \\
     Germany  &  DAX  &       56.9001    &         0.2909    &         0.6794    &         2.7711    &         0.2967    &         0.9876    &         0.2390    &         0.2506    & \textbf{        0.2249   } \\
     Greece  &  Athex Composite Share Price Index  &       21.9151    &         0.2808    &         0.2842    &         0.6892    &         0.2788    &         0.3703    &         0.3603    &         0.1838    & \textbf{        0.1422   } \\
     Hong Kong  &  Hang Seng  &       72.2368    &         3.0319    &         2.1814    &         2.6863    &         2.9800    &         2.0298    &         2.6042    &         2.6989    & \textbf{        0.0831   } \\
     Hungary  &  Budapest SE  &       28.2722    &         0.3097    &         0.2624    &         0.9169    &         0.3232    &         0.3519    &         0.1564    &         0.1580    & \textbf{        0.1141   } \\
     India  &  Nifty 50  &       34.8370    &         0.1822    &         0.1687    &         1.1093    &         0.1743    &         0.2902    &         0.1089    &         0.1160    & \textbf{        0.1038   } \\
     India  &  BSE Sensex  &       35.1337    &         0.4911    &         0.2689    &         1.4985    &         0.4673    &         0.3709    &         0.2186    &         0.2272    & \textbf{        0.2127   } \\
     Indonesia  &  IDX Composite  &       66.9248    &         2.2545    &         0.6128    &         1.6530    &         0.7851    &         0.7591    &         0.1360    &         0.1426    & \textbf{        0.1095   } \\
     Ireland  &  ISEQ Overall Index  &       37.5286    &         1.3755    &         0.8351    &         1.9910    &         0.3515    &         1.1746    &         0.1474    &         0.1271    & \textbf{        0.1230   } \\
     Israel  &  TA 35  &       50.5617    &         0.6329    &         0.4118    &         2.0069    &         0.3172    &         0.6102    &         0.2970    &         0.3357    & \textbf{        0.0955   } \\
     Italy  &  FTSE MIB  &       36.6913    &         0.3230    &         0.2739    &         1.1133    &         0.3225    &         0.3676    &         0.2112    &         0.2185    & \textbf{        0.1233   } \\
     Japan  &  Topix  &     118.4660    &         3.5287    &         0.3694    &         6.2719    &         0.7682    &         0.3464    &         0.8797    &         0.9919    & \textbf{        0.2314   } \\
     Kazakhstan  &  KASE Index  &       20.7294    &         2.7707    &         0.5738    &         0.2699    &         2.6296    &         0.4301    &         0.1969    &         0.2081    & \textbf{        0.1595   } \\
     Kuwait  &  Kuwait 15  &         7.8919    &         1.2858    &         0.7764    &         0.4412    &         1.2528    &         0.6817    &         0.8315    &         0.4692    & \textbf{        0.1158   } \\
     Latvia  &  OMXR  &       82.0664    &         1.9607    &         0.7745    &         1.9667    &         1.8824    &         0.6820    &         0.3254    &         0.3767    & \textbf{        0.1123   } \\
     Lithuania  &  OMXV  &     504.1008    &         0.3025    &         9.5807    &       31.0164    &         0.3035    &       10.3979    &         0.2721    & \textbf{        0.2358   } &         0.5430    \\
     Luxembourg  &  LuxX Index  &       97.8343    &         5.1473    &         1.1948    &         3.2033    &         1.7223    &         1.0442    &         0.6986    &         0.8220    & \textbf{        0.1653   } \\
     Malaysia  &  FTSE Bursa Malaysia KLCI  &       41.5234    &         0.6425    &         0.3607    &         1.4040    &         0.6176    &         0.4435    &         0.3538    &         0.3866    & \textbf{        0.1304   } \\
     Mauritius  &  SEMDEX  &       35.6296    &         3.8971    &         1.1538    &         1.3216    &         3.6253    &     385.5455    &         0.2525    &         0.2628    & \textbf{        0.0884   } \\
     Mexico  &  IPC  &       32.5629    &         0.9043    &         0.1705    &         0.6164    &         0.9673    &         0.2023    &         0.1947    &         0.2149    & \textbf{        0.1077   } \\
     Morocco  &  MASI  &       42.0081    &         6.5151    &         0.6334    &         0.5638    &         1.4094    &         0.5396    &         0.2136    &         0.2279    & \textbf{        0.0769   } \\
     Namibia  &  NSX Overall Index  &     130.7378    &         6.6396    &         8.2153    &         3.0141    &         7.5289    &         8.2290    &         1.1294    &         1.2868    & \textbf{        0.2998   } \\
     Netherlands  &  AEX  &       53.4257    &         0.3014    &         0.3483    &         1.7431    &         0.3106    &         0.5473    &         0.2671    &         0.2779    & \textbf{        0.1243   } \\
     New Zealand  &  NZX 50 Index  &       59.5363    &         0.3832    &         0.9617    &         2.4454    &         0.3884    &         1.3768    &         0.3982    &         0.4131    & \textbf{        0.2876   } \\
     Norway  &  OBX Index  &       34.5181    &         1.0883    &         0.7337    &         2.1675    &         0.2977    &         1.0504    &         0.1628    &         0.1509    & \textbf{        0.1447   } \\
     Oman  &  MSM 30  &       19.9979    &         4.7329    &         4.2594    &         4.2701    &         4.7415    &       20.9619    &         0.5974    &         0.6343    & \textbf{        0.1896   } \\
     Pakistan  &  KSE 100 Index  &       28.9548    &         0.5280    &         0.2680    &         0.5813    &         0.5292    &         0.2631    &         0.2466    &         0.2522    & \textbf{        0.1678   } \\
     Peru  &  S\&P Lima General Index  &       33.2462    &         2.6000    &         1.4011    &         0.8080    &         2.5800    &         1.1560    &         0.2747    &         0.2843    & \textbf{        0.2570   } \\
     Philippines  &  PSEi Composite  &       49.0890    &         4.5096    &         2.9632    &         0.3837    &         4.4266    &         2.7243    &         0.7093    &         0.7817    & \textbf{        0.1486   } \\
     Poland  &  WIG  &       31.6454    &         0.2955    &         0.2380    &         1.0904    &         0.2894    &         0.3881    &         0.1743    &         0.1918    & \textbf{        0.0989   } \\
     Portugal  &  PSI 20  &       44.0223    &         0.4065    &         0.5106    &         1.5788    &         0.5780    &         0.7349    &         0.3795    &         0.3913    & \textbf{        0.1117   } \\
     Qatar  &  QE 20 Index  &       37.2608    &         0.3026    &         0.6321    &         1.2982    &         0.2042    &         0.9147    &         0.1570    &         0.1409    & \textbf{        0.1345   } \\
     Romania  &  BET 10  &       55.5069    &         4.7910    &         0.7553    &         2.4330    &         0.2059    &         1.1783    &         0.1939    &         0.2049    & \textbf{        0.1245   } \\
     Russia  &  MICEX  &       18.4654    &         0.4060    &         0.2443    &         0.5760    &         0.4244    &         0.2818    &         0.3084    &         0.2934    & \textbf{        0.1969   } \\
     Russia  &  RTSI  &       27.2834    &         0.5510    &         0.5705    &         1.5853    &         0.5016    &         0.6980    &         0.5202    &         0.5005    & \textbf{        0.2449   } \\
     Saudi Arabia  &  Tadawul All Share  &       35.0071    &         0.6438    &         0.4680    &         1.5179    &         0.5840    &         0.5300    &         0.5440    &         0.5936    & \textbf{        0.2165   } \\
     Serbia  &  BELEX  &       15.2473    &         0.9288    &         0.4793    &         0.9361    &         0.9097    &       17.1000    &         0.4146    &         0.4093    & \textbf{        0.3925   } \\
     Singapore  &  STI Index  &       60.3014    &         1.0571    &         0.8641    &         3.5824    &         0.4167    &         1.2104    &         0.2534    &         0.2317    & \textbf{        0.1337   } \\
     South Africa  &  FTSE/JSE All-Share Index  &       32.4245    &         0.2222    &         0.4177    &         1.4105    &         0.2203    &         0.6028    &         0.2078    &         0.2200    & \textbf{        0.1359   } \\
     South Korea  &  KOSPI  &       62.3498    &         0.3180    &         0.3551    &         3.0698    &         0.3082    &         0.5055    &         0.2207    &         0.2460    & \textbf{        0.1841   } \\
     Spain  &  IBEX 35  &       36.1543    &         6.5471    &         0.4055    &         0.9703    &         0.2037    &         0.6101    &         0.3082    &         0.2976    & \textbf{        0.1346   } \\
     Sri Lanka  &  CSE All-Share  &         8.3816    &         0.7019    &         0.1863    &         0.3075    &         1.4909    &         0.2100    &         0.3087    &         0.3166    & \textbf{        0.1485   } \\
     Sweden  &  OMXS30  &       42.6429    &         0.2121    &         0.3527    &         1.5844    &         0.2207    &         0.5738    &         0.2498    &         0.2740    & \textbf{        0.0945   } \\
     Switzerland  &  SMI  &       41.1358    &         0.7891    &         0.5450    &         1.7875    &         0.2746    &         0.8125    &         0.2377    &         0.2432    & \textbf{        0.2271   } \\
     Taiwan  &  Taiwan Weighted  &       36.3549    &         0.3619    &         0.4332    &         2.0928    &         0.3488    &         0.5606    &         0.1708    &         0.1425    & \textbf{        0.0895   } \\
     Thailand  &  SET  &       18.1060    &         0.1596    &         0.1512    &         0.5235    &         0.1606    &         0.2149    &         0.1278    &         0.1332    & \textbf{        0.1211   } \\
     Tunesia  &  Tunindex  &         4.9198    &         0.8092    &         0.7186    &         0.6431    &         0.8076    &         0.6882    &         0.9220    &         0.7147    & \textbf{        0.2701   } \\
     Turkey  &  BIST 100  &       41.6775    &         2.3598    &         1.4413    &         0.9661    &         2.3289    &         1.2495    &         0.2441    &         0.2367    & \textbf{        0.0608   } \\
     United Arab Emirates  &  DFM  &         8.2957    &         0.5151    &         0.3108    &         0.5248    &         0.2489    &         0.3746    & \textbf{        0.2256   } &         0.2278    &         0.2319    \\
     United Arab Emirates  &  Abu Dhabi  &       10.1539    &         1.1049    &         0.2553    &         0.3701    &         0.2443    &         0.2996    &         0.3899    &         0.4231    & \textbf{        0.1714   } \\
     USA  &  Dow 30  &       65.3198    &         7.0260    &         1.5977    &         1.7143    &         2.7187    &         1.4125    &         0.8404    &         0.9135    & \textbf{        0.2589   } \\
     USA  &  S\&P 500  &       55.8784    &         3.7922    &         2.7317    &         0.6900    &         3.7964    &         2.5021    &         0.9082    &         0.9372    & \textbf{        0.1306   } \\
     USA  &  Nasdaq  &       66.5410    &         2.5418    &         1.7973    &         2.4980    &         2.5600    &         1.6896    &         1.5565    &         1.5986    & \textbf{        0.0945   } \\
     Venezuela  &  IBC  &       47.9978    &       13.1605    &         5.2416    &         3.4120    &         2.1086    &         4.2988    &         0.8694    &         0.9480    & \textbf{        0.2177   } \\
     Vietnam  &  HNX 30  &         3.6923    &         0.2365    &         0.2748    &         0.3756    &         0.2346    &         0.3087    &         0.2390    & \textbf{        0.2242   } &         0.2413    \\
     Zambia  &  All Share Index  &       24.9193    &       16.4023    &       15.4692    &       18.9661    &       16.4108    &       15.3148    &       17.1749    &       17.6813    & \textbf{      12.6451   } \\
    \hline
    \end{tabular}%
				\end{tiny}
		  \caption{AD distance between the empirical and fitted distributions for hourly log-returns, from 11/02/2016 1pm until 11/02/2017 12pm.}
  \label{tab:hAD}%
\end{table}%

\begin{table}[htbp]
  \hspace{-2.5cm}
\begin{tiny}
     \begin{tabular}{rrrrrrrrrrr}
    \hline
    Country & Index & N     & St    & NIG  & V$\Gamma$  & GH    & Meix & 2St12 & 2St39 & 3St \\
    \hline
   Argentina & MERVAL & -13828 & -14730 & \textbf{-14743} & -14692 & -14741 & -14741 & -14738 & -14736 & -14740 \\
    Australia & S\&P/ASX 200 & -16528 & -17401 & -17405 & -17359 & -17405 & -17401 & -17410 & -17408 & \textbf{-17470} \\
    Australia & All Ordinaries & -16868 & -17783 & -17787 & -17754 & -17787 & -17783 & -17790 & -17789 & \textbf{-17874} \\
    Austria & ATX   & -20383 & -20860 & -20915 & -20873 & -20918 & -20908 & -20921 & \textbf{-20921} & -20913 \\
    Bahrain & All Share & -7935 & -8336 & -8334 & -8314 & -8334 & -8330 & \textbf{-8339} & -8339 & -8335 \\
    Belgium & BEL 20 & -21640 & -22379 & -22366 & -22312 & -22376 & -22355 & \textbf{-22381} & -22381 & -22377 \\
    Brazil & Bovespa & -16660 & -17553 & -17540 & -17489 & -17549 & -17529 & \textbf{-17554} & -17554 & -17550 \\
    Bulgaria & SOFIX & -17046 & \textbf{-17807} & -17798 & -17755 & -17806 & -17787 & -17804 & -17803 & -17805 \\
    Canada & S\&P/TSX 60 & -19255 & -19822 & -19938 & -19971 & \textbf{-19979} & -19947 & -19975 & -19974 & -19973 \\
    Canada & S\&P/TSX Composite & -19700 & -20296 & -20415 & -20468 & \textbf{-20473} & -20425 & -20465 & -20463 & -20464 \\
    Chile & IPSA  & -18447 & \textbf{-20278} & -20214 & -20120 & -20274 & -20184 & -20273 & -20273 & -20267 \\
    China & CSI 300 & -15639 & -16136 & -16206 & -16551 & -16555 & -16223 & -16677 & \textbf{-16677} & -16668 \\
    China & SSE Composite Index & -15877 & -16435 & -16498 & -16736 & -16750 & -16512 & -16853 & \textbf{-16854} & -16850 \\
    Colombia & IGBC  & -17482 & -17838 & \textbf{-17844} & -17822 & -17842 & -17842 & -17843 & -17843 & -17835 \\
    Croatia & CROBEX & -18003 & \textbf{-19154} & -19138 & -19066 & -19151 & -19124 & -19144 & -19143 & -19151 \\
    Cyprus & Cyprus Main Market Index & -11873 & -12311 & -12343 & -12332 & -12356 & -12350 & -12362 & \textbf{-12363} & -12354 \\
    Czech Republic & PX    & -19334 & -19761 & -19756 & -19728 & -19761 & -19748 & \textbf{-19769} & -19769 & -19764 \\
    Egypt & EGX 30 Index & -9307 & -10217 & -10259 & -10192 & -10262 & \textbf{-10263} & -10257 & -10254 & -10254 \\
    Estonia & OMXT  & -15084 & -15344 & -15334 & -15329 & -15340 & -15330 & -15341 & -15339 & \textbf{-15354} \\
    EuroStoxx & Euro Stoxx 50 & -22089 & -22783 & -22789 & -22741 & -22793 & -22781 & \textbf{-22793} & -22792 & -22790 \\
    Finland & OMXH25 & -21461 & -22292 & -22283 & -22196 & -22292 & -22273 & -22300 & \textbf{-22300} & -22292 \\
    France & CAC 40 & -21122 & -21979 & -21971 & -21910 & -21977 & -21961 & \textbf{-21980} & -21979 & -21974 \\
    GB    & FTSE 100 & -21688 & -22272 & -22328 & -22276 & -22330 & -22321 & \textbf{-22332} & -22331 & -22326 \\
    Germany & DAX   & -21142 & -21988 & -21980 & -21907 & -21987 & -21970 & -21994 & \textbf{-21994} & -21987 \\
    Greece & Athex Composite Share Price Index & -16757 & -17122 & -17119 & -17099 & -17120 & -17115 & -17119 & \textbf{-17122} & -17117 \\
    Hong Kong & Hang Seng & -17625 & -18550 & -18588 & -18671 & -18669 & -18593 & -18636 & -18634 & \textbf{-18705} \\
    Hungary & Budapest SE & -21127 & -21561 & -21565 & -21542 & -21564 & -21562 & -21565 & \textbf{-21565} & -21558 \\
    India & Nifty 50 & -16079 & -16625 & -16624 & -16583 & -16624 & -16618 & \textbf{-16626} & -16626 & -16619 \\
    India & BSE Sensex & -16168 & -16696 & -16699 & -16659 & -16698 & -16695 & \textbf{-16702} & -16701 & -16694 \\
    Indonesia & IDX Composite & -17012 & -18040 & -18037 & -17971 & -18043 & -18027 & \textbf{-18053} & -18053 & -18051 \\
    Ireland & ISEQ Overall Index & -22519 & -23136 & -23129 & -23074 & -23143 & -23116 & -23152 & \textbf{-23153} & -23145 \\
    Israel & TA 35 & -20824 & -21548 & -21557 & -21501 & -21558 & -21551 & \textbf{-21558} & -21556 & -21555 \\
    Italy & FTSE MIB & -19926 & -20486 & -20489 & -20459 & -20488 & -20484 & \textbf{-20489} & -20489 & -20482 \\
    Japan & Topix & -15083 & -16610 & -16665 & -16543 & -16663 & \textbf{-16665} & -16640 & -16634 & -16663 \\
    Kazakhstan & KASE Index & -13951 & -14220 & -14244 & -14251 & -14249 & -14247 & \textbf{-14252} & -14251 & -14246 \\
    Kuwait & Kuwait 15 & -8024 & -8128 & -8135 & -8144 & -8142 & -8136 & -8128 & -8145 & \textbf{-8148} \\
    Latvia & OMXR  & -11731 & -12912 & -12932 & -12899 & -12936 & -12929 & -12944 & -12943 & \textbf{-12948} \\
    Lithuania & OMXV  & -7865 & -15820 & -15588 & -14858 & -15816 & -15584 & -15804 & -15805 & \textbf{-15836} \\
    Luxembourg & LuxX Index & -18777 & -19948 & -20050 & -20002 & -20068 & -20059 & -20065 & -20061 & \textbf{-20078} \\
    Malaysia & FTSE Bursa Malaysia KLCI & -17623 & -18233 & -18237 & -18196 & -18236 & -18233 & \textbf{-18239} & -18238 & -18238 \\
    Mauritius & SEMDEX & -10389 & -10838 & -10863 & -10852 & -10870 & 20533 & -10875 & -10874 & \textbf{-10876} \\
    Mexico & IPC   & -16954 & -17444 & \textbf{-17458} & -17439 & -17457 & -17456 & -17458 & -17458 & -17455 \\
    Morocco & MASI  & -15548 & -16096 & -16193 & -16191 & -16198 & -16192 & \textbf{-16212} & -16212 & -16209 \\
    Namibia & NSX Overall Index & -18183 & -19871 & -19931 & -20095 & \textbf{-20137} & -19944 & -20079 & -20072 & -20120 \\
    Netherlands & AEX   & -21622 & \textbf{-22430} & -22427 & -22370 & -22430 & -22419 & -22428 & -22427 & -22427 \\
    New Zealand & NZX 50 Index & -20986 & \textbf{-22041} & -22003 & -21933 & -22038 & -21981 & -22034 & -22033 & -22031 \\
    Norway & OBX Index & -18518 & -19040 & -19042 & -18985 & -19049 & -19033 & -19057 & \textbf{-19058} & -19051 \\
    Oman  & MSM 30 & -8917 & -9126 & -9150 & -9431 & -9388 & -8901 & -9496 & -9492 & \textbf{-9531} \\
    Pakistan & KSE 100 Index & -14265 & -14689 & \textbf{-14697} & -14692 & -14696 & -14697 & -14697 & -14696 & -14692 \\
    Peru  & S\&P Lima General Index & -19720 & -20128 & -20164 & \textbf{-20219} & -20217 & -20172 & -20197 & -20196 & -20187 \\
    Philippines & PSEi Composite & -14772 & -15367 & -15418 & \textbf{-15516} & -15516 & -15429 & -15496 & -15494 & -15511 \\
    Poland & WIG   & -20929 & -21417 & -21420 & -21388 & \textbf{-21420} & -21415 & -21419 & -21419 & -21414 \\
    Portugal & PSI 20 & -21199 & \textbf{-21886} & -21874 & -21821 & -21883 & -21862 & -21884 & -21884 & -21884 \\
    Qatar & QE 20 Index & -10039 & \textbf{-10718} & -10692 & -10647 & -10716 & -10678 & -10717 & -10717 & -10713 \\
    Romania & BET 10 & -23946 & -24804 & -24835 & -24757 & -24855 & -24818 & \textbf{-24858} & -24857 & -24852 \\
    Russia & MICEX & -20390 & -20689 & \textbf{-20693} & -20678 & -20691 & -20691 & -20687 & -20688 & -20684 \\
    Russia & RTSI  & -18520 & -19016 & -19007 & -18978 & -19013 & -18999 & -19013 & -19013 & \textbf{-19026} \\
    Saudi Arabia & Tadawul All Share & -11105 & -11586 & -11595 & -11604 & -11593 & -11593 & -11590 & -11589 & \textbf{-11716} \\
    Serbia & BELEX & -12042 & -12229 & \textbf{-12241} & -12239 & -12240 & -12017 & -12239 & -12240 & -12238 \\
    Singapore & STI Index & -22040 & -22875 & -22880 & -22789 & -22884 & -22872 & -22897 & \textbf{-22898} & -22893 \\
    South Africa & FTSE/JSE All-Share Index & -20667 & -21165 & -21165 & -21125 & -21166 & -21159 & \textbf{-21167} & -21167 & -21162 \\
    South Korea & KOSPI & -16098 & -16936 & -16942 & -16853 & -16942 & -16939 & \textbf{-16945} & -16943 & -16938 \\
    Spain & IBEX 35 & -22738 & -23379 & -23437 & -23392 & -23460 & -23421 & -23459 & -23458 & \textbf{-23463} \\
    Sri Lanka & CSE All-Share & -15193 & -15329 & \textbf{-15335} & -15328 & -15333 & -15333 & -15330 & -15329 & -15327 \\
    Sweden & OMXS30 & -21218 & \textbf{-21887} & -21882 & -21827 & -21886 & -21874 & -21886 & -21885 & -21883 \\
    Switzerland & SMI   & -21775 & -22424 & -22420 & -22363 & -22427 & -22410 & \textbf{-22429} & -22428 & -22422 \\
    Taiwan & Taiwan Weighted & -11271 & -11761 & -11763 & -11709 & -11763 & -11760 & -11773 & \textbf{-11775} & -11770 \\
    Thailand & SET   & -16689 & -16961 & \textbf{-16964} & -16952 & -16961 & -16962 & -16961 & -16962 & -16953 \\
    Tunesia & Tunindex & -11944 & -12001 & -12003 & -12007 & -12005 & -12004 & -11995 & -11997 & \textbf{-12010} \\
    Turkey & BIST 100 & -22738 & -23382 & -23400 & -23405 & -23408 & -23402 & -23435 & \textbf{-23435} & -23434 \\
    United Arab Emirates & DFM   & -8439 & -8582 & -8581 & -8571 & -8581 & -8578 & -8584 & \textbf{-8584} & -8576 \\
    United Arab Emirates & Abu Dhabi & -10821 & -11031 & -11033 & -11020 & -11039 & -11029 & \textbf{-11040} & -11038 & -11038 \\
    USA   & Dow 30 & -20313 & -21097 & -21206 & -21238 & -21245 & -21212 & -21231 & -21228 & \textbf{-21248} \\
    USA   & S\&P 500 & -20092 & -20825 & -20864 & -20951 & -20949 & -20871 & -20954 & -20953 & \textbf{-20967} \\
    USA   & Nasdaq & -18641 & -19551 & -19579 & -19620 & -19613 & -19581 & -19623 & -19622 & \textbf{-19657} \\
    Venezuela & IBC   & -4668 & -5189 & -5239 & -5304 & \textbf{-5328} & -5253 & -5311 & -5308 & -5326 \\
    Vietnam & HNX 30 & -10954 & \textbf{-11012} & -11011 & -11008 & -11008 & -11010 & -11006 & -11006 & -11004 \\
    Zambia & All Share Index & -2117 & -2432 & -2392 & -2433 & -2428 & -2378 & -2469 & -2484 & \textbf{-3005} \\
    \hline
    \end{tabular}%
				\end{tiny}
		  \caption{AIC of fitted distributions for hourly log-returns, from 11/02/2016 1pm until 11/02/2017 12pm.}
  \label{tab:hAIC}%
\end{table}%

\begin{table}[htbp]
  \hspace{-2.5cm}
\begin{tiny}
     \begin{tabular}{rrrrrrrrrrr}
    \hline
    Country & Index & N     & St    & NIG  & V$\Gamma$  & GH    & Meix & 2St12 & 2St39 & 3St \\
    \hline
   Argentina & MERVAL & -13817 & -14714 & \textbf{-14721} & -14670 & -14714 & -14719 & -14716 & -14714 & -14696 \\
    Australia & S\&P/ASX 200 & -16517 & -17384 & -17383 & -17337 & -17378 & -17379 & -17388 & -17386 & \textbf{-17426} \\
    Australia & All Ordinaries & -16857 & -17767 & -17765 & -17732 & -17760 & -17761 & -17768 & -17767 & \textbf{-17830} \\
    Austria & ATX   & -20372 & -20842 & -20892 & -20850 & -20889 & -20886 & -20898 & \textbf{-20899} & -20867 \\
    Bahrain & All Share & -7926 & \textbf{-8321} & -8315 & -8295 & -8310 & -8311 & -8320 & -8320 & -8296 \\
    Belgium & BEL 20 & -21628 & \textbf{-22361} & -22343 & -22290 & -22347 & -22332 & -22358 & -22358 & -22331 \\
    Brazil & Bovespa & -16649 & \textbf{-17536} & -17517 & -17467 & -17521 & -17507 & -17532 & -17532 & -17505 \\
    Bulgaria & SOFIX & -17035 & \textbf{-17790} & -17775 & -17733 & -17778 & -17765 & -17782 & -17780 & -17760 \\
    Canada & S\&P/TSX 60 & -19244 & -19805 & -19915 & -19949 & -19951 & -19924 & \textbf{-19953} & -19951 & -19928 \\
    Canada & S\&P/TSX Composite & -19689 & -20279 & -20393 & \textbf{-20446} & -20445 & -20403 & -20443 & -20441 & -20420 \\
    Chile & IPSA  & -18436 & \textbf{-20261} & -20192 & -20097 & -20246 & -20161 & -20250 & -20251 & -20221 \\
    China & CSI 300 & -15629 & -16119 & -16185 & -16529 & -16528 & -16201 & -16655 & \textbf{-16656} & -16625 \\
    China & SSE Composite Index & -15867 & -16419 & -16476 & -16714 & -16723 & -16490 & -16832 & \textbf{-16833} & -16806 \\
    Colombia & IGBC  & -17471 & \textbf{-17822} & -17822 & -17800 & -17814 & -17820 & -17821 & -17821 & -17791 \\
    Croatia & CROBEX & -17991 & \textbf{-19137} & -19116 & -19043 & -19123 & -19102 & -19122 & -19121 & -19107 \\
    Cyprus & Cyprus Main Market Index & -11862 & -12295 & -12322 & -12310 & -12329 & -12329 & -12340 & \textbf{-12341} & -12311 \\
    Czech Republic & PX    & -19323 & -19744 & -19733 & -19705 & -19733 & -19726 & \textbf{-19747} & -19747 & -19719 \\
    Egypt & EGX 30 Index & -9297 & -10201 & -10239 & -10172 & -10237 & \textbf{-10243} & -10236 & -10234 & -10214 \\
    Estonia & OMXT  & -15073 & \textbf{-15328} & -15313 & -15308 & -15313 & -15308 & -15320 & -15317 & -15311 \\
    EuroStoxx & Euro Stoxx 50 & -22077 & -22766 & -22766 & -22718 & -22764 & -22758 & \textbf{-22770} & -22769 & -22744 \\
    Finland & OMXH25 & -21449 & -22275 & -22260 & -22174 & -22263 & -22250 & -22277 & \textbf{-22277} & -22246 \\
    France & CAC 40 & -21111 & \textbf{-21962} & -21948 & -21887 & -21949 & -21938 & -21957 & -21956 & -21928 \\
    GB    & FTSE 100 & -21676 & -22255 & -22305 & -22253 & -22302 & -22298 & \textbf{-22309} & -22308 & -22280 \\
    Germany & DAX   & -21131 & -21971 & -21957 & -21884 & -21958 & -21947 & -21971 & \textbf{-21971} & -21941 \\
    Greece & Athex Composite Share Price Index & -16746 & \textbf{-17105} & -17097 & -17076 & -17092 & -17093 & -17097 & -17100 & -17072 \\
    Hong Kong & Hang Seng & -17613 & -18533 & -18565 & -18649 & -18641 & -18571 & -18613 & -18612 & \textbf{-18661} \\
    Hungary & Budapest SE & -21115 & \textbf{-21544} & -21542 & -21519 & -21535 & -21539 & -21542 & -21542 & -21512 \\
    India & Nifty 50 & -16069 & \textbf{-16609} & -16602 & -16561 & -16597 & -16596 & -16605 & -16604 & -16575 \\
    India & BSE Sensex & -16157 & -16679 & -16677 & -16637 & -16671 & -16673 & \textbf{-16680} & -16679 & -16650 \\
    Indonesia & IDX Composite & -17001 & -18023 & -18015 & -17949 & -18015 & -18005 & \textbf{-18031} & -18031 & -18007 \\
    Ireland & ISEQ Overall Index & -22508 & -23119 & -23106 & -23051 & -23114 & -23093 & -23129 & \textbf{-23130} & -23098 \\
    Israel & TA 35 & -20813 & -21531 & -21534 & -21478 & -21529 & -21529 & \textbf{-21535} & -21534 & -21510 \\
    Italy & FTSE MIB & -19915 & \textbf{-20468} & -20466 & -20436 & -20459 & -20461 & -20466 & -20466 & -20436 \\
    Japan & Topix & -15072 & -16593 & -16643 & -16521 & -16636 & \textbf{-16643} & -16619 & -16612 & -16620 \\
    Kazakhstan & KASE Index & -13940 & -14204 & -14222 & -14229 & -14222 & -14226 & \textbf{-14230} & -14230 & -14202 \\
    Kuwait & Kuwait 15 & -8014 & -8113 & -8115 & -8124 & -8117 & -8117 & -8108 & \textbf{-8126} & -8109 \\
    Latvia & OMXR  & -11720 & -12897 & -12911 & -12878 & -12910 & -12908 & \textbf{-12923} & -12922 & -12906 \\
    Lithuania & OMXV  & -7854 & \textbf{-15804} & -15566 & -14836 & -15789 & -15563 & -15782 & -15784 & -15793 \\
    Luxembourg & LuxX Index & -18766 & -19931 & -20027 & -19979 & -20039 & -20036 & \textbf{-20042} & -20038 & -20033 \\
    Malaysia & FTSE Bursa Malaysia KLCI & -17612 & -18216 & -18215 & -18174 & -18208 & -18211 & \textbf{-18217} & -18216 & -18194 \\
    Mauritius & SEMDEX & -10379 & -10823 & -10844 & -10832 & -10845 & 20553 & \textbf{-10855} & -10855 & -10837 \\
    Mexico & IPC   & -16943 & -17427 & \textbf{-17436} & -17416 & -17429 & -17434 & -17436 & -17436 & -17411 \\
    Morocco & MASI  & -15537 & -16080 & -16171 & -16169 & -16170 & -16171 & \textbf{-16190} & -16190 & -16165 \\
    Namibia & NSX Overall Index & -18172 & -19854 & -19908 & -20073 & \textbf{-20109} & -19921 & -20057 & -20049 & -20074 \\
    Netherlands & AEX   & -21611 & \textbf{-22413} & -22404 & -22347 & -22401 & -22396 & -22405 & -22404 & -22381 \\
    New Zealand & NZX 50 Index & -20975 & \textbf{-22024} & -21980 & -21911 & -22010 & -21958 & -22012 & -22010 & -21986 \\
    Norway & OBX Index & -18507 & -19024 & -19019 & -18963 & -19021 & -19011 & -19035 & \textbf{-19035} & -19006 \\
    Oman  & MSM 30 & -8908 & -9111 & -9131 & -9411 & -9364 & -8882 & -9477 & -9473 & \textbf{-9493} \\
    Pakistan & KSE 100 Index & -14254 & -14672 & \textbf{-14675} & -14670 & -14668 & -14675 & -14675 & -14674 & -14648 \\
    Peru  & S\&P Lima General Index & -19709 & -20111 & -20141 & \textbf{-20196} & -20189 & -20150 & -20174 & -20173 & -20142 \\
    Philippines & PSEi Composite & -14761 & -15351 & -15396 & \textbf{-15495} & -15489 & -15407 & -15475 & -15472 & -15467 \\
    Poland & WIG   & -20917 & \textbf{-21400} & -21397 & -21365 & -21391 & -21392 & -21396 & -21396 & -21368 \\
    Portugal & PSI 20 & -21187 & \textbf{-21868} & -21851 & -21798 & -21854 & -21839 & -21861 & -21861 & -21838 \\
    Qatar & QE 20 Index & -10029 & \textbf{-10702} & -10672 & -10626 & -10691 & -10657 & -10697 & -10696 & -10672 \\
    Romania & BET 10 & -23934 & -24787 & -24812 & -24734 & -24826 & -24795 & \textbf{-24835} & -24834 & -24805 \\
    Russia & MICEX & -20378 & \textbf{-20672} & -20670 & -20655 & -20662 & -20668 & -20664 & -20665 & -20639 \\
    Russia & RTSI  & -18509 & \textbf{-18999} & -18984 & -18955 & -18984 & -18976 & -18991 & -18990 & -18980 \\
    Saudi Arabia & Tadawul All Share & -11094 & -11571 & -11574 & -11584 & -11567 & -11573 & -11569 & -11568 & \textbf{-11674} \\
    Serbia & BELEX & -12031 & -12214 & \textbf{-12220} & -12218 & -12214 & -11996 & -12218 & -12219 & -12196 \\
    Singapore & STI Index & -22029 & -22858 & -22857 & -22766 & -22856 & -22849 & -22874 & \textbf{-22875} & -22847 \\
    South Africa & FTSE/JSE All-Share Index & -20656 & \textbf{-21148} & -21142 & -21102 & -21137 & -21136 & -21144 & -21144 & -21116 \\
    South Korea & KOSPI & -16087 & -16920 & -16920 & -16831 & -16914 & -16917 & \textbf{-16923} & -16922 & -16895 \\
    Spain & IBEX 35 & -22726 & -23362 & -23413 & -23368 & -23431 & -23398 & \textbf{-23436} & -23435 & -23416 \\
    Sri Lanka & CSE All-Share & -15182 & -15313 & \textbf{-15313} & -15307 & -15307 & -15312 & -15309 & -15308 & -15285 \\
    Sweden & OMXS30 & -21207 & \textbf{-21870} & -21859 & -21804 & -21857 & -21851 & -21863 & -21862 & -21837 \\
    Switzerland & SMI   & -21764 & \textbf{-22406} & -22397 & -22340 & -22399 & -22387 & -22406 & -22405 & -22377 \\
    Taiwan & Taiwan Weighted & -11261 & -11745 & -11742 & -11688 & -11737 & -11740 & -11753 & \textbf{-11755} & -11729 \\
    Thailand & SET   & -16678 & \textbf{-16944} & -16942 & -16930 & -16934 & -16940 & -16940 & -16940 & -16910 \\
    Tunesia & Tunindex & -11934 & -11986 & -11983 & \textbf{-11987} & -11980 & -11984 & -11974 & -11977 & -11970 \\
    Turkey & BIST 100 & -22726 & -23365 & -23377 & -23382 & -23379 & -23379 & -23412 & \textbf{-23412} & -23387 \\
    United Arab Emirates & DFM   & -8429 & \textbf{-8567} & -8561 & -8551 & -8557 & -8558 & -8564 & -8565 & -8537 \\
    United Arab Emirates & Abu Dhabi & -10811 & -11015 & -11013 & -11000 & -11013 & -11008 & \textbf{-11019} & -11018 & -10997 \\
    USA   & Dow 30 & -20302 & -21081 & -21184 & -21215 & \textbf{-21217} & -21189 & -21208 & -21206 & -21203 \\
    USA   & S\&P 500 & -20081 & -20808 & -20842 & -20929 & -20921 & -20848 & \textbf{-20932} & -20931 & -20923 \\
    USA   & Nasdaq & -18630 & -19534 & -19556 & -19597 & -19585 & -19559 & -19601 & -19600 & \textbf{-19612} \\
    Venezuela & IBC   & -4658 & -5175 & -5220 & -5285 & \textbf{-5304} & -5233 & -5292 & -5289 & -5288 \\
    Vietnam & HNX 30 & -10943 & \textbf{-10996} & -10990 & -10987 & -10982 & -10990 & -10986 & -10985 & -10963 \\
    Zambia & All Share Index & -2109 & -2419 & -2375 & -2416 & -2407 & -2362 & -2453 & -2467 & \textbf{-2971} \\
    \hline
    \end{tabular}%
				\end{tiny}
		  \caption{BIC of fitted distributions for hourly log-returns, from 11/02/2016 1pm until 11/02/2017 12pm.}
  \label{tab:hBIC}%
\end{table}%

For a start, Tables~\ref{tab:hKS}--\ref{tab:hBIC} present the KS, AD, AIC and BIC statistics. In them we see that the KS select the 3St most often, the AD overwhelmingly selects as well the 3St, the AIC selects the 2St12 most of the time but closely followed by the 3St and 2St39, and the BIC selects most often the simple Student's $t$ distribution, but the 2St12 and 2St39 are selected a number of times as well.
The Table~\ref{tab:minshourly} summarizes how often each distribution has the lowest KS, AD, AIC or BIC statistic among the models. {}We see that the 3St is selected most often by both of the criteria KS and AD, but the Student's $t$ is selected most often by the BIC. This points out to the fact that the 3St is also an excellent model for high-frequency returns (at least hourly returns). If one does not want to deal with a relatively complicated model, the alternative of the single Student's $t$ distribution could describe the data globally, but the reader should we warned that the tails might be poorly described then (the AD statistic never selects the Student's $t$ distribution in all of our samples).

Unsurprisingly, the 3St yields a better fit in terms of KS and AD statistics than the other distributions considered because it has 8 parameters instead of (at most) 5. However, we mention two additional aspects why mixtures of Student's $t$ distributions are appealing. First, if we drop the 3St and only consider the remaining distributions, the 2St12 and 2St39 would outperform the GH distribution (all have 5 parameters) for hourly data. For daily data, the 2St12 would be second best after the GH distribution. This means that these 2-mixtures are comparable to the GH distribution. Second, if we include (the herein omitted) results of 2- or 3-mixtures of normal distributions and compare them directly with the Student's $t$ mixture counterparts we would observe that most often the Student's $t$ mixtures outperform the normal mixtures for all criteria.

\begin{table}[htbp]
  \centering
		\begin{tabular}{rrrrrrrrrr}
    \hline
 &    N     & St    & NIG  & V$\Gamma$  & GH    & Meix & 2St12 & 2St39 & 3St \\
    \hline
    KS  & 0 & 0 & 2  &  0 & 1 & 3 & 6 & 5 & 61 \\
    AD  & 0 &  0 &  0 & 0 & 0 & 0 & 1 & 3  & 74 \\
    AIC  & 0 & 9 & 8  & 2 & 5 & 2 & 20 & 14  & 18 \\
    BIC  & 0 &  27 &  5 & 4 & 3 & 2 &  18 &  12 & 7 \\
    \hline    \end{tabular}%
		  \caption{Number of lowest statistics per distribution, for hourly returns.}
  \label{tab:minshourly}%
\end{table}%

\begin{figure}
	\centering
\begin{tabular}{ccc}
  {\includegraphics[width=.33\textwidth, height=5cm]{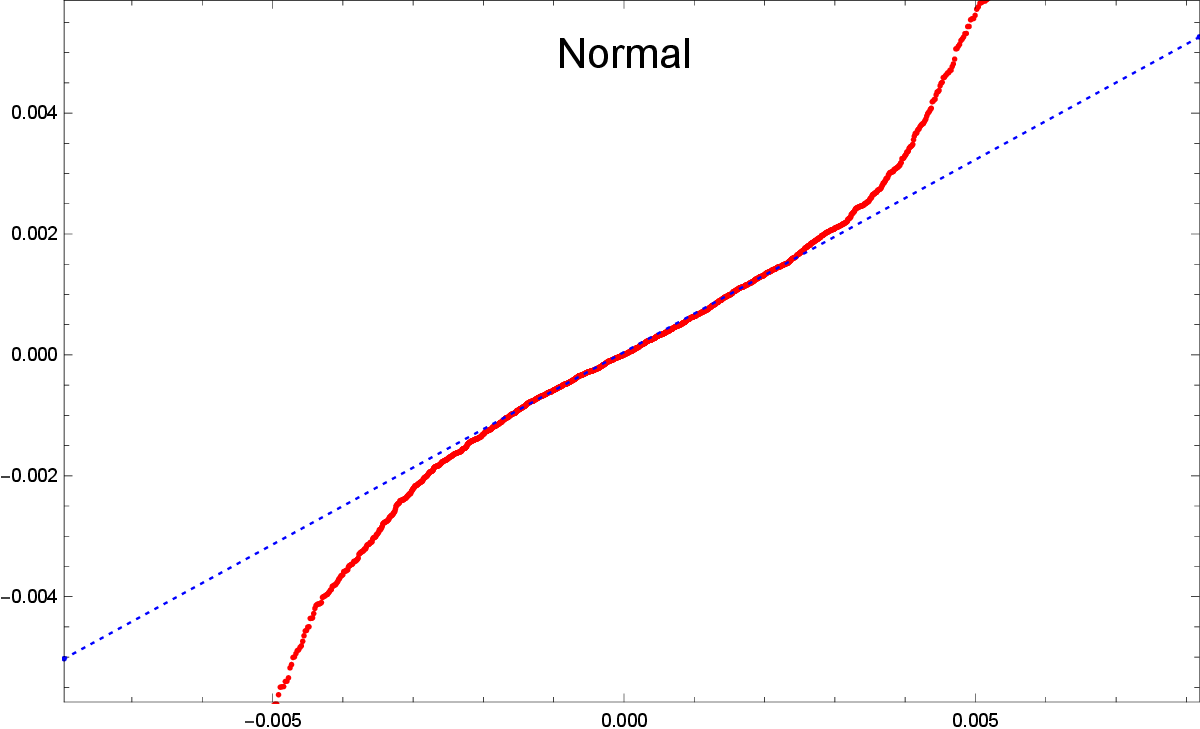}} &
  {\includegraphics[width=.33\textwidth,
  height=5cm]{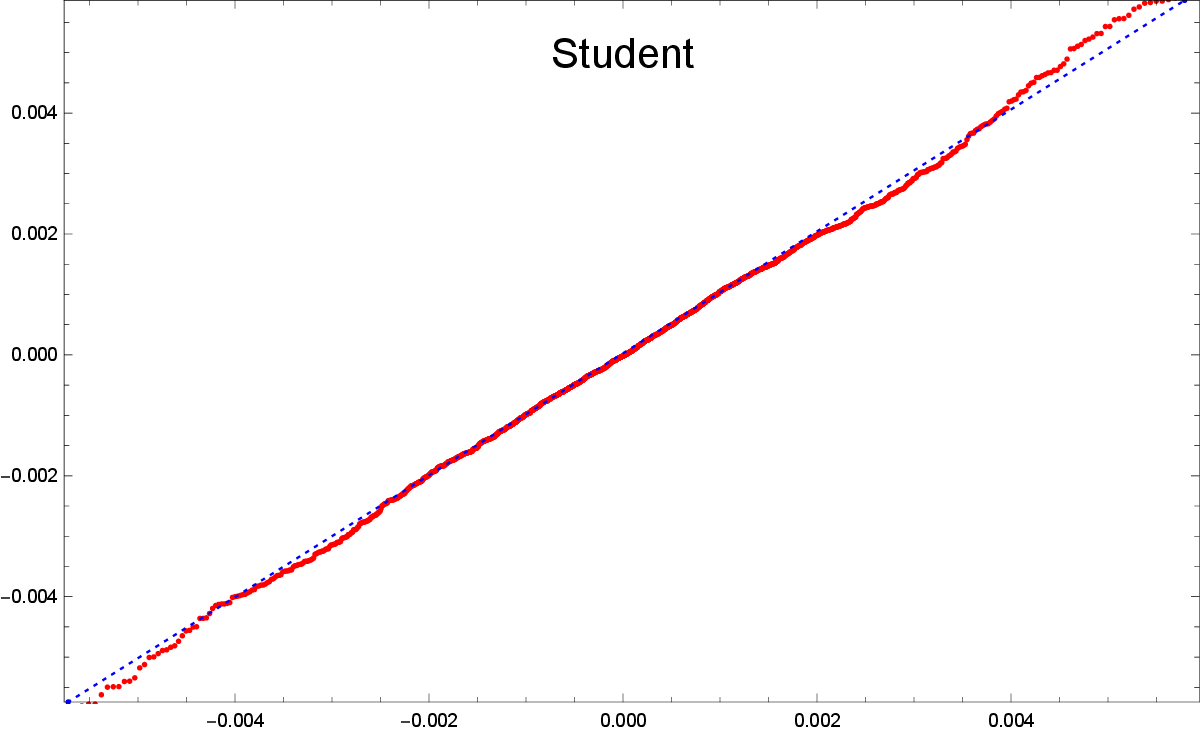}} &
  {\includegraphics[width=.33\textwidth,
  height=5cm]{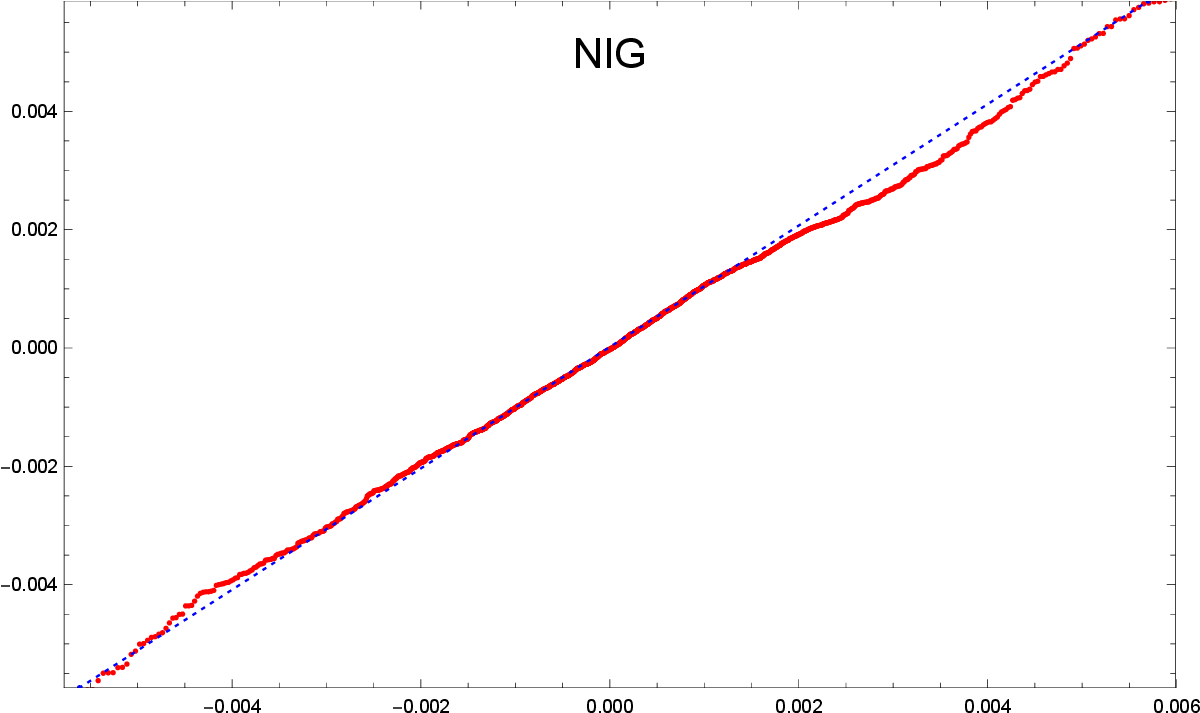}}\\
  {\includegraphics[width=.33\textwidth, height=5cm]{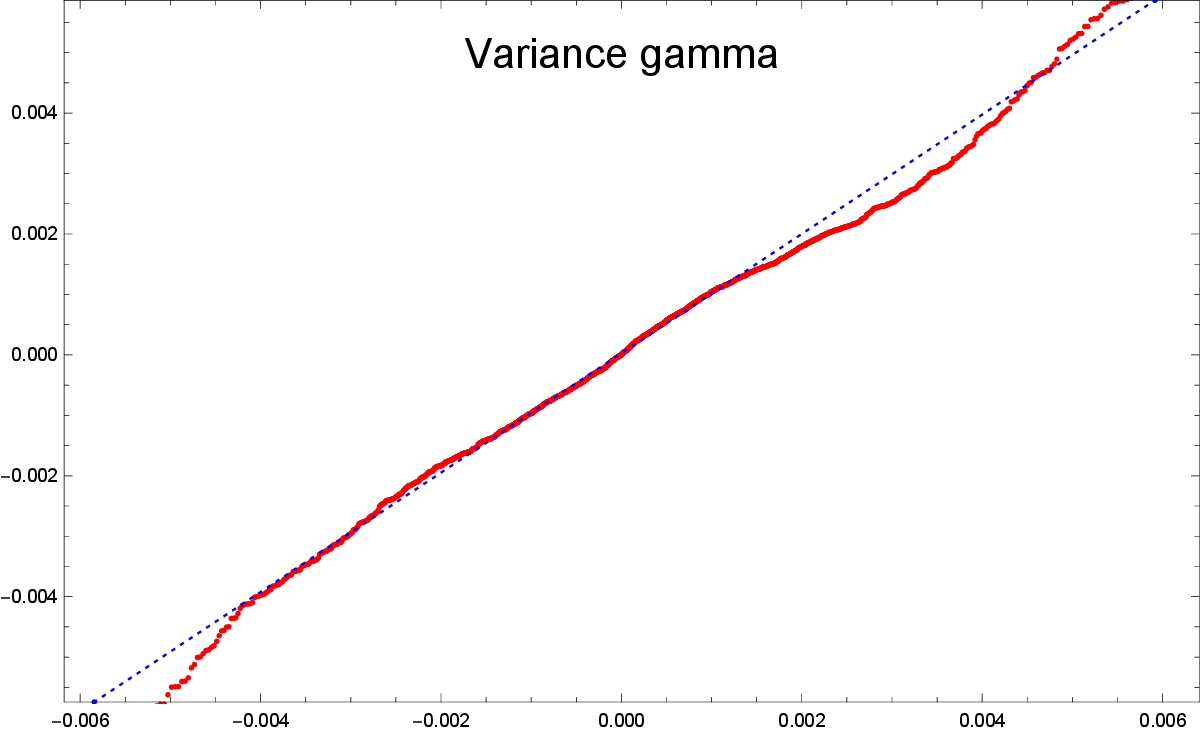}} &
  {\includegraphics[width=.33\textwidth,
  height=5cm]{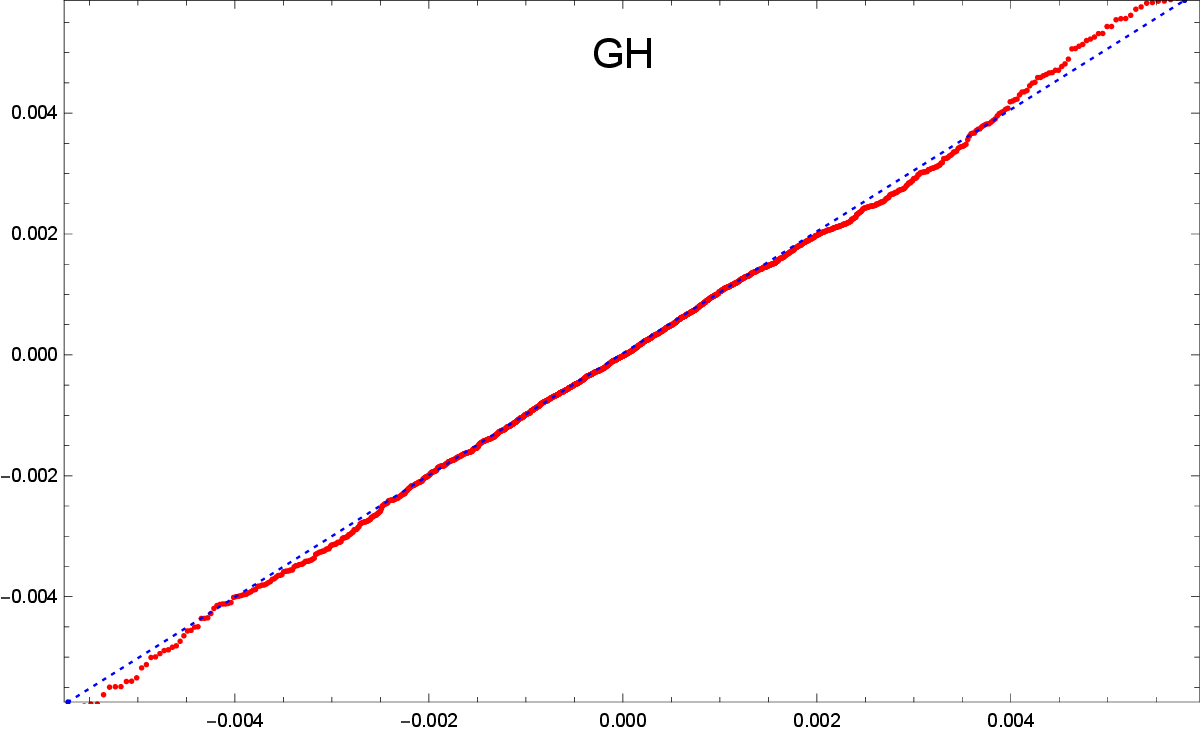}} &
  {\includegraphics[width=.33\textwidth,
  height=5cm]{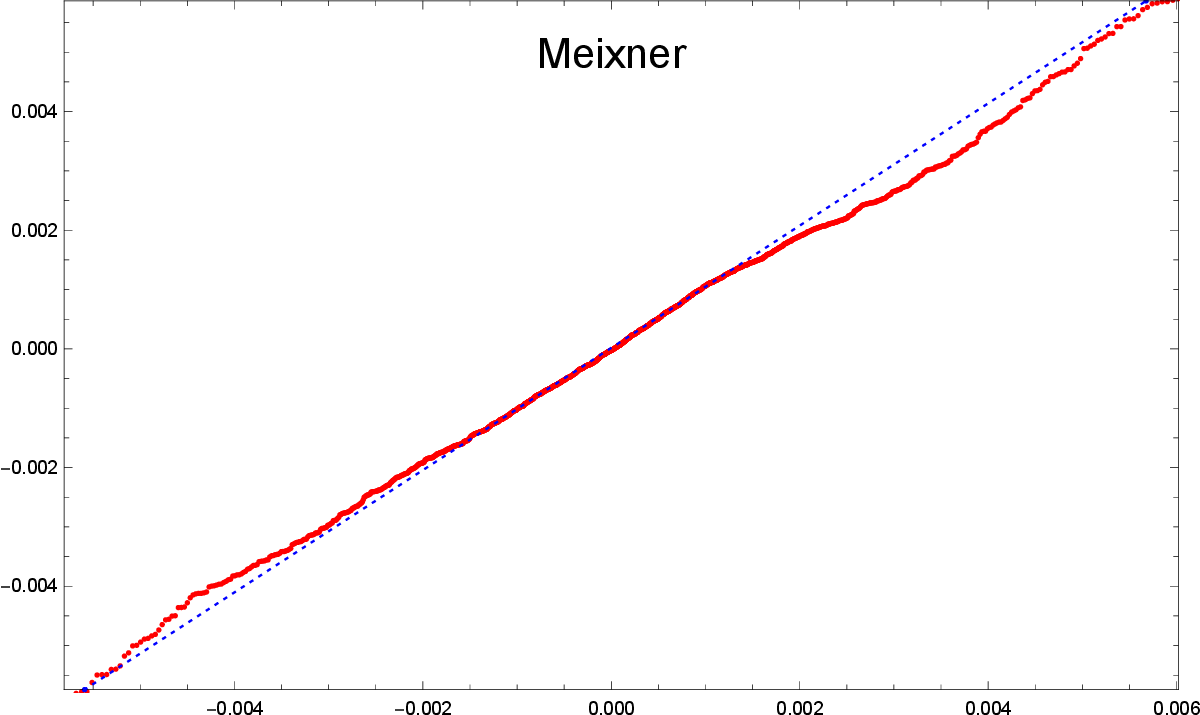}}\\
  {\includegraphics[width=.33\textwidth, height=5cm]{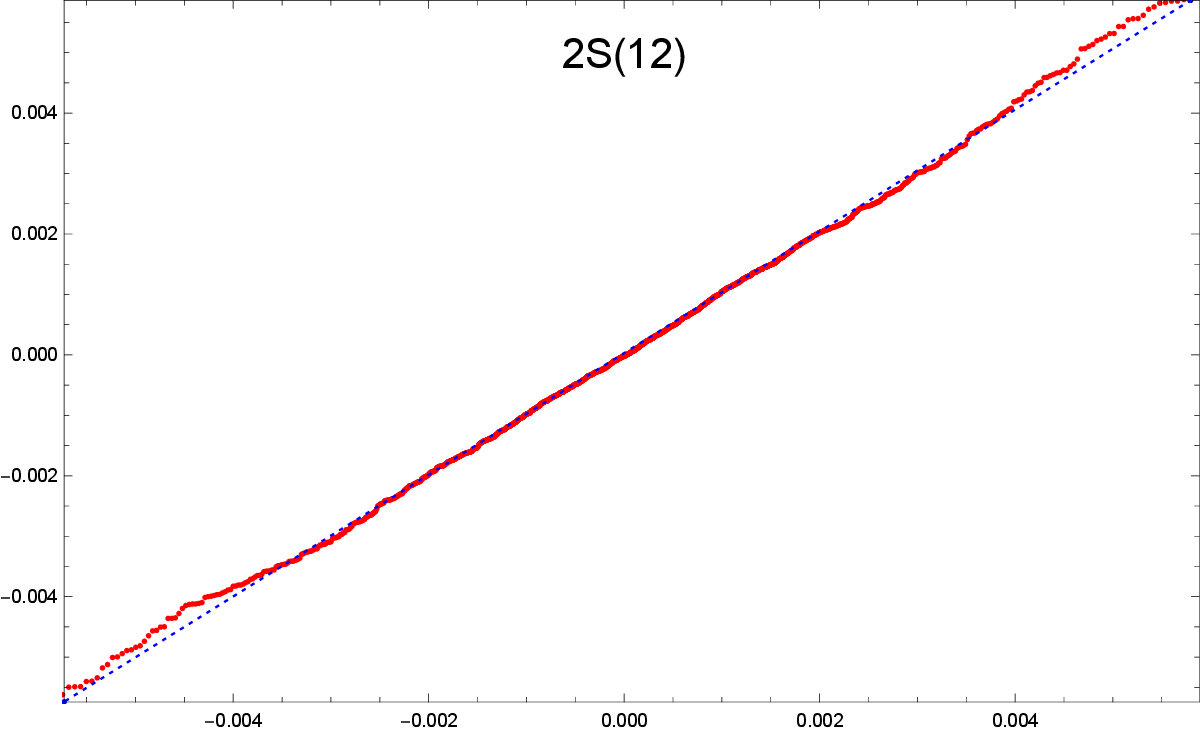}} &
  {\includegraphics[width=.33\textwidth,
  height=5cm]{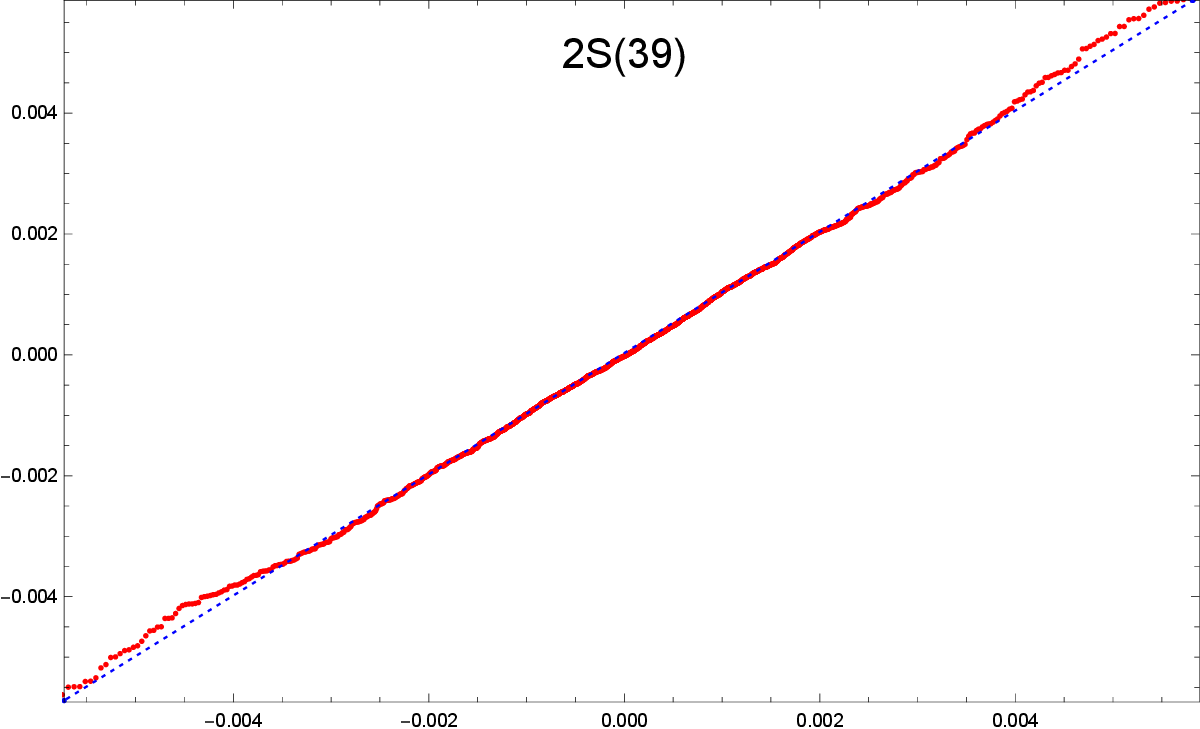}} &
  {\includegraphics[width=.33\textwidth,
  height=5cm]{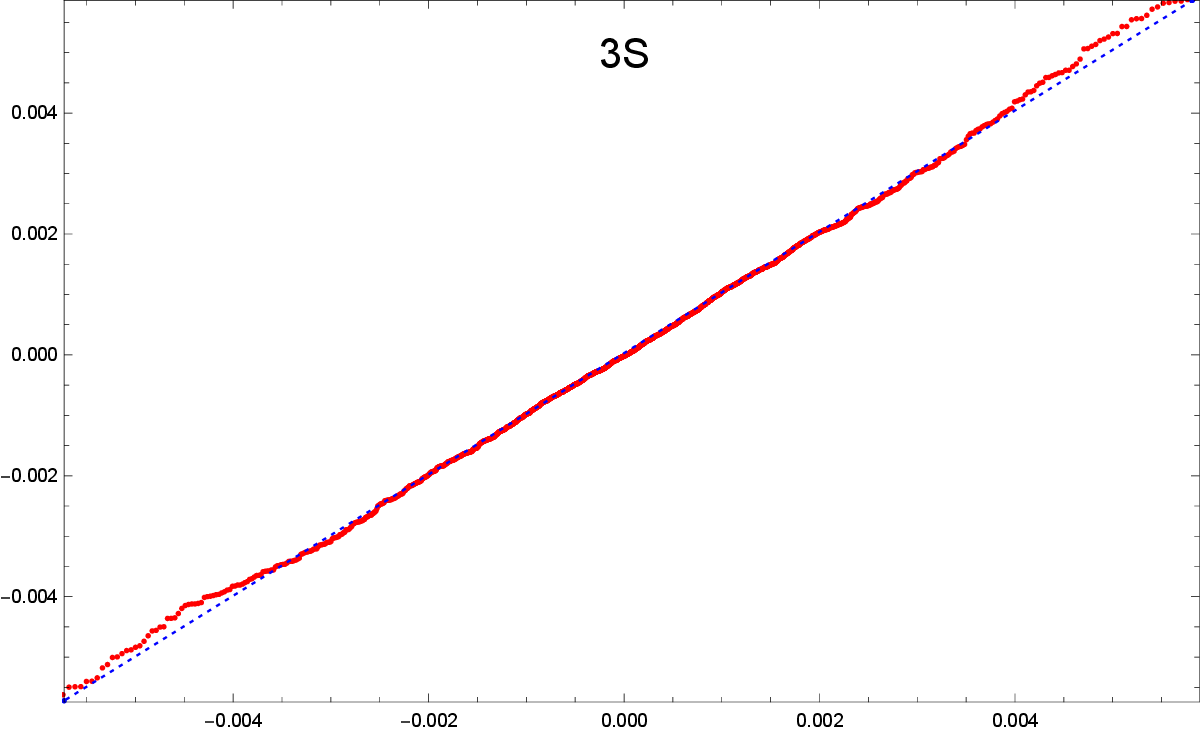}}\\
  \end{tabular}
  \caption{QQ plots of empirical quantiles for hourly DAX log-returns from 11/02/2016 1pm until 11/02/2017 12pm versus model distributions.}
  \label{pic:qqDAXh}
\end{figure}

For the qualitative goodness-of-fit,
Figure~\ref{pic:qqDAXh} shows QQ plots for hourly returns for the DAX example against model distributions. The fits are generally very good, and those that contain the Student's $t$ distribution in their definition are visually excellent.

\section{Further assessment of the 3St distribution}\label{assessment3St}

In this section, we perform further assessments of the 3St distribution which was most often the best fit. We will:
\begin{itemize}
\item[i)] Investigate the relation between each  component of the mixture 3St and the market conditions of the studied equity indices.
\item[ii)] Analyze the contribution of the different components to the overall goodness-of-fit of the 3St. For that, we plot the residues of the empirical and modelled frequencies of the data. Additionally, we perform Chi-square tests to assess the goodness-of-fit when the full 3St is considered, two of its components are considered, or only one of its components is considered.
\end{itemize}

For the first aim, we compute the
posterior probabilities $\tau_1(x),\tau_2(x),\tau_3(x)$ defined above
for all log-returns $x$ in each sample. We do so to detect whether different market conditions (e.g., periods of high and low absolute value log-returns) lead to a pattern in the posterior probabilities.

\begin{figure}
	\centering
\begin{tabular}{cc}
{\includegraphics[width=.5\textwidth, height=5cm]{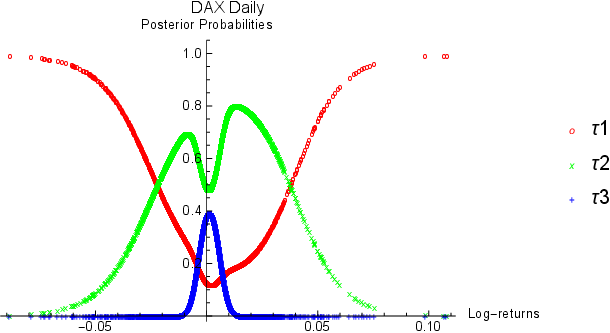}}&
  {\includegraphics[width=.5\textwidth,
  height=5cm]{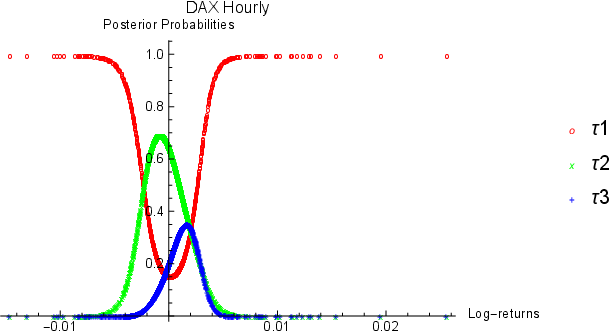}}\\
  \end{tabular}
  \caption{Posterior probability plots according to the estimated 3St of empirical data for daily DAX log-returns from 01/03/1997 until 11/02/2017 and hourly DAX log-returns from 11/02/2016 1pm until 11/02/2017 12pm.}
  \label{postprob}
\end{figure}

Figure~\ref{postprob} shows posterior probabilities
for the example of the German DAX index. Figures for the other indices are contained in the supplementary material. For daily data, we generally observe that the $\tau_1(x)$ (related with a St with $\nu_1=4$ degrees of freedom) takes appreciable values in the tails of the log-returns. $\tau_2(x)$ (related with a St with $\nu_2=12$ degrees of freedom) describes the intermediate or moderate values between the extreme values and the mode of the data, and $\tau_3(x)$ (related with a St with $\nu_3=39$ degrees of freedom, very similar to a normal distribution) describes mostly the more frequent values around the mode of the data.
For hourly data, a similar pattern emerges but
there are some exceptions to this behaviour that can be observed in the supplementary graphs. Thus, the first component of the 3St models well the tails of the data, the second models mainly the intermediate values, and the third the most likely and small (in absolute value) log-returns, independently of the moment when each specific log-return takes place.

\begin{figure}
	\centering
\begin{tabular}{cc}
{\includegraphics[width=.5\textwidth, height=5cm]{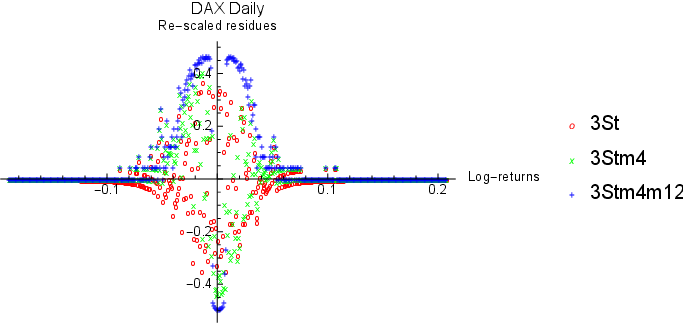}}&
  {\includegraphics[width=.5\textwidth,
  height=5cm]{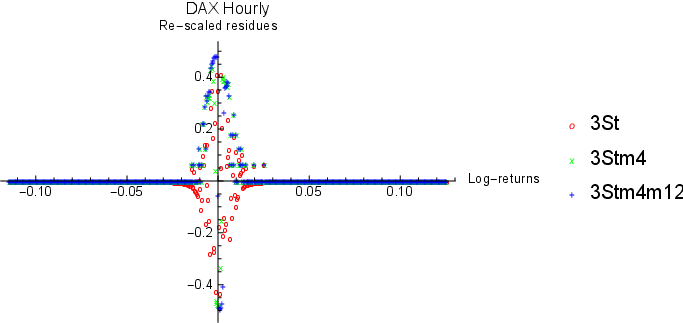}}\\
  \end{tabular}
  \caption{Re-scaled residues according to the 3St, 3Stm4, 3Stm4m12 of empirical data for daily DAX log-returns from 01/03/1997 until 11/02/2017 and hourly DAX log-returns from 11/02/2016 1pm until 11/02/2017 12pm.}
  \label{residues}
\end{figure}

For the second aim, we compute the differences (or residues) between the empirical frequencies of the log-returns and the predicted frequencies, using the full 3St, the 3St without the $\nu_1=4$ component (and call it 3Stm4), and the 3St without the $\nu_1=4$ and $\nu_2=12$ components (3Stm4m12). The latter one is essentially the St distribution in the mixture with $\nu_3=39$ degrees of freedom. We use  500 bins to classify the data. When subtracting the components we have not re-estimated the remaining components because we want to assess how much each component adds to the overall goodness-of-fit.
In order to show a more useful information about the differences in the residues, we perform the following re-scaling:\footnote{We thank an anonymous referee for this nice suggestion.}
First, we select, for each sample, a constant 
$$
a=\frac{1}{40}\max_i(|{\rm res}_{\rm 3Stm4m12}(i)|)
$$
and then plot jointly
\begin{eqnarray}
&&\frac{1}{\pi}\arctan\left(\frac{1}{a}{\rm res}_{\rm 3St}(i)\right)\nonumber\\
&&\frac{1}{\pi}\arctan\left(\frac{1}{a}{\rm res}_{\rm 3Stm4}(i)\right)\nonumber\\
&&\frac{1}{\pi}\arctan\left(\frac{1}{a}{\rm res}_{\rm 3Stm4m12}(i)\right)\nonumber
\end{eqnarray}
Trial-and-error has shown us that the above choice of $a$ serves well in most cases so that the structure in the small residuals can be easily observed.
Figure~\ref{residues} plots the re-scaled residues for the German DAX example. The plots for the other indices are contained in the supplementary material.

We observe that the re-scaled residues for the 3St are generally smaller or comparable to those of the 3Stm4, and both of them, smaller than those of the 3Stm4m12 for all samples, both daily and hourly.

Additionally, we perform Chi-square tests of goodness-of-fit for the models 3St, 3Stm4, 3Stm4m12 to investigate how much the different components contribute to the overall fit. The results are in the folder of supplementary materials (\lq\lq Excel sheets\rq\rq\,\ in the sheets \lq\lq Results Chi-square test\rq\rq), both for daily and hourly data sets. The summary of the results is as follows: For daily data sets, the 3St is not rejected 100\% of the cases, the 3Stm4 is not rejected 68\% of the cases, and the 3Stm4m12 is always rejected. For hourly data sets, the 3St is not rejected 100\% of the instances, the 3Stm4 is not rejected 37\% of the samples and the 3Stm4m12 is not rejected only 1.3\% of the cases.

\section{Conclusion}\label{sec:conclusionappl}
This paper provides a comprehensive comparison of models for equity index returns for both traditional and exotic markets. We analyze two samples, namely, daily returns over an almost 20-year horizon and hourly returns for the last year. We find
that to consider mixtures of two and above all three Student's $t$ distributions (with numbers of degrees of freedom fixed a priori), offer an excellent performance with respect to three out of four statistical criteria we use, namely the KS and AD statistics, and AIC criterion. This result is robust to the consideration of daily and hourly log-returns for a wide family of stock indices.
We discuss which component in the mixture 3St models extreme/moderate/small log-returns of the studied equity indices.

Our results point out to an excellent performance of the 3St model, which is moderately complex (8 parameters) but its descriptive power, above all at the tails, makes it to be a perfect candidate for subsequent studies of Risk measures, like Value-at-Risk and related quantities. We hope to address this issue in other research works.

\section*{Author contributions}
Till Massing: Conceptualization, data curation, formal analysis, funding acquisition, investigation, methodology, software, supervision, validation, visualization, writing-original draft, writing-review \& editing. Arturo Ramos: Conceptualization, data curation, formal analysis, funding acquisition, investigation, methodology, resources, software, validation, visualization, writing-original draft, writing-review \& editing.

\section*{Competing interests statement}
The authors declare to have no competing interests concerning the research carried out in this article.





\section*{Acknowledgments}
We thank an anonymous reviewer for very helpful suggestions. The work of Till Massing has been supported by \emph{Deutsche Forschungsgemeinschaft}, (Germany) via SFB 823. The work of Arturo Ramos has been supported by the Spanish \emph{Ministerio de Econom\'{\i}a y Competitividad} (ECO2017-82246-P) and by Aragon Government (ADETRE Reference Group).

\bibliographystyle{apalike}
\bibliography{bibliography}

\vfill\eject

\vfill\eject

\end{document}